\pdfoutput=1
\documentclass[manuscript,screen,nonacm]{acmart}

\usepackage{multirow}
\usepackage{array}
\newcolumntype{P}[1]{>{\raggedright\arraybackslash}p{#1}}
\usepackage{subcaption}
\usepackage{algorithm}
\usepackage{algorithmic}
\usepackage{pifont}
\usepackage[most]{tcolorbox}
\usepackage{tikz}
\usetikzlibrary{positioning,arrows.meta,fit,backgrounds,calc}
\newcommand{\cmark}{\ding{51}}
\newcommand{\xmark}{\ding{55}}

\setcopyright{rightsretained}
\renewcommand\footnotetextcopyrightpermission[1]{}

\tcbset{breakable}
\begin{document}

\title[Noise-Aware Testing of NISQ Quantum Programs]{Noise-Aware Testing of NISQ Quantum Programs: Self-Baseline Oracle with a Controlled False-Positive Rate}

\author{Mir Mohammad Yousuf}
\orcid{0009-0001-7453-3358}
\email{yousuf\_2022phaite006@nitsri.ac.in}
\affiliation{%
  \institution{National Institute of Technology Srinagar}
  \department{Department of Information Technology}
  \city{Srinagar}
  \state{Jammu and Kashmir}
  \country{India}}

\author{Shabir Ahmad Sofi}
\orcid{0000-0002-8740-2875}
\email{shabir@nitsri.ac.in}
\affiliation{%
  \institution{National Institute of Technology Srinagar}
  \department{Department of Information Technology}
  \city{Srinagar}
  \state{Jammu and Kashmir}
  \country{India}}

\renewcommand{\shortauthors}{Yousuf and Sofi}

\begin{abstract}
Testing quantum programs on Noisy Intermediate-Scale Quantum (NISQ) devices faces a test-oracle problem: hardware noise and finite-shot sampling make the runs of a correct program vary as much as those of a faulty one, so oracles with a fixed acceptance threshold produce false positives that depend on the program and the backend. We present a noise-aware testing framework built on \emph{self-baseline calibration}. In regression and mutation testing, each run of the program under test is compared with repeated executions of a trusted reference program on the same backend, and the acceptance threshold is the $(1-\alpha)$ quantile of the finite-shot null distribution of the Jensen--Shannon divergence, so the false-positive rate of the resulting \emph{noise-aware oracle} is a nominal level set in advance. A noncentral chi-square model predicts its detection power, a homogeneity test and a two-run sentinel detect drift, a mutation engine labels every mutant exactly by statevector simulation within a three-class noise-aware taxonomy, and zero-noise extrapolation attributes rejections to faults or noise without a noiseless simulator. We evaluate the framework in simulation on nine core benchmarks, 32 QASMBench programs, ten real bugs from Bugs4Q, five depolarizing and two coherent-error noise regimes and three device snapshots, and on an IBM Heron device. First, the oracle controls false positives on average over calibration windows: $0.0011$ at $\alpha = 10^{-3}$ over 9{,}000 correct runs ($0.0013$ on the QASMBench programs), against 36--74\% for fixed thresholds, 8.9\% for a QOIN-style filter and 10.8\% for a threshold calibrated globally on the device's noise model; under the evaluated conditions, the nominal level was approximately maintained with at least 16 shots per observed outcome, whereas below 8 it was not reliably controlled; the intermediate range was not sampled. Second, it detects faults without sacrificing this control: it attains the highest F1 of all oracles on non-equivalent injected faults ($0.941$), kills 93.8\% of the semantic mutants and detects 96.4\% of the runs of observable real bugs; 24--70\% of draws of common gate-insertion operators are equivalent mutants, and semantic mutants separate strongly from the two semantics-preserving classes, which separate from each other more weakly and depending on the noise level. Third, attribution is a conditional diagnostic: at low and medium noise it labels 98--99\% of the rejected bugs and 90\% of the rejected noise effects correctly, and it fails for noise effects at high noise. On the IBM device, using about 14 minutes of QPU time, the oracle rejected none of 82 correct runs of 17 programs and detected 120 of 121 faults and semantic mutants, and over 26 hours a two-run sentinel reduced the false positives caused by drift from 27.4\% to 1.8\% of the protected runs, showing that drift must be detected before testing.
\end{abstract}



\maketitle
\hypersetup{pdfauthor={Mir Mohammad Yousuf and Shabir Ahmad Sofi}}

\section{Introduction}
\label{sec:introduction}

A quantum program can be correct and still fail its test. On today's Noisy Intermediate-Scale Quantum (NISQ) hardware, every gate is imperfect, every measurement is stochastic, and the resulting run-to-run variation can be statistically indistinguishable from a software fault~\cite{preskill2018quantum,bharti2022noisy,yousuf2026systematic,pan2023understanding}. This is the normal operating condition of quantum software, and it challenges an assumption that has long underpinned software testing: that a correct program produces a recognizable output, subject at most to a small and uniform tolerance. Traditional test oracles interpret deviations from an expected output as evidence of defects; on NISQ hardware, legitimate hardware-induced variation produces the same observable behavior, and empirical studies report failures that arise from hardware behavior rather than implementation errors~\cite{paltenghi2024survey,pan2023understanding}. The central challenge is therefore deciding whether an observed execution should be regarded as correct.

The oracle problem is harder in quantum computing than in classical software: the no-cloning theorem prevents complete inspection of quantum states, measurement collapses the state observed, and the exponential state space precludes exhaustive verification. Noise adds a further layer: the variability of an execution becomes a property of the backend, so correctness decisions must separate deviations caused by faults from those caused by device noise and finite-shot sampling. Yet practitioners often resort to deterministic assertions, fixed seeds and approximate equality~\cite{ramalho2025wild}, and evaluation practices remain largely inherited from classical software engineering~\cite{li2026methodological}.

Existing research has addressed this problem through three broad classes of correctness oracles. The first relies on \emph{fixed-threshold} decision rules, where executions are accepted whenever the observed divergence remains below a predetermined global tolerance. Such thresholds inevitably become unsound once hardware noise exceeds the chosen limit; here and throughout, we call an oracle \emph{sound} in a setting if it rarely rejects correct runs there (at its nominal level for an oracle that has one, and at most $5\%$ otherwise). The second class adopts \emph{noise-filtering} strategies, exemplified by approaches such as QOIN~\cite{muqeet2024qoin}, which estimate hardware noise and compensate for it before comparing program behavior with the ideal noiseless output. These methods are conceptually related to quantum error mitigation~\cite{cai2023quantum,temme2017error} but require an explicit or learned representation of the underlying noise characteristics, making them dependent on particular hardware configurations. A third line of work employs \emph{metamorphic}, \emph{property-based} or \emph{statistical} testing formulations~\cite{barrera2021state,assertsq2025,morphq,huang2019statistical,pontolillo2025ideal}, reducing dependence on ideal outputs by expressing correctness through programmer-specified relations or statistical properties. While effective in some settings, these approaches shift the burden of correctness from the program itself to externally supplied specifications.

Fixed-threshold and noise-filtering techniques evaluate executions against a reference the device cannot realize, the noiseless ideal output; metamorphic approaches replace it with specifications that the developer must supply. None constructs the oracle from the behavior of a trusted version of the program on the target backend, and none of the threshold-based oracles states the false-positive rate its threshold delivers under noise.

This paper adopts a different perspective for the setting in which a trusted reference version of the program is available, which is the setting of mutation analysis and of regression testing: we decide whether a program under test $P'$ behaves like the reference $P$ \emph{on the same backend}. Instead of comparing every execution against an ideal reference that the hardware cannot faithfully produce, we compare it against a baseline that $P$ establishes about itself through repeated execution on that backend. We refer to this principle as \emph{self-baseline calibration}. Its key observation is that two correct executions are independent multinomial samples of one output distribution, so the distribution of their divergence under the hypothesis ``$P'$ behaves like $P$'' is known from finite-shot statistics. The resulting \emph{noise-aware oracle} accepts a run whenever its divergence from the self-baseline lies below the $(1-\alpha)$ quantile of that null distribution. The false-positive rate is therefore a nominal level $\alpha$ chosen by the tester, detection power follows from a noncentral chi-square model, the calibration runs additionally serve to detect drift through a homogeneity test, and a short sentinel before each test batch decides when to recalibrate. The oracle requires neither the ideal output, nor a parametric noise model, nor a learned noise representation; it does require the trusted reference program.

Across nine core benchmarks and five noise regimes, the observed false-positive rate (FPR) of the noise-aware oracle is $0.0011$ at $\alpha = 10^{-3}$ over 9{,}000 correct runs ($0.0013$ on $32$ QASMBench programs), whereas fixed thresholds reach 36--74\%, a QOIN-style filter 8.9\% and a threshold calibrated globally on the device's noise model 10.8\%. On non-equivalent injected faults, the oracle attains F1 $0.941$; it rejects 93.8\% of semantic mutants and detects $96.4\%$ of the runs of observable real bugs from Bugs4Q. On an IBM Heron device, it rejected none of $82$ correct runs and detected $120$ of $121$ faults and semantic mutants. The evaluation rests on exact semantic labels: every mutant and fault is classified as equivalent or not by statevector simulation before any noisy execution, so detection is never credited for, or penalized on, programs indistinguishable from the original.

\subsection{Research gap and problem formulation}
\label{subsec:problem-statement}

Four methodological limitations remain. First, quantum mutation testing largely inherits classical, noise-oblivious semantics, so a killed mutant may reflect hardware variation rather than a fault, and the rate at which quantum operators produce equivalent mutants, a classical concern~\cite{papadakis2018mutation,madeyski2014overcoming}, has not been quantified. Second, oracles rely on fixed thresholds or hardware-specific noise representations, neither sound across heterogeneous noise, and none sets its false-positive rate in advance or accounts for the false negatives it trades for soundness. Third, mutation taxonomies introduce categories without establishing that they are statistically distinguishable under noise. Fourth, failure attribution has seen little evaluation against labeled ground truth and typically relies on a noiseless simulator, which is unavailable on hardware. We therefore investigate the following problem:

\emph{How can a noise-aware testing framework for NISQ quantum programs provide a correctness oracle with a controlled false-positive rate under heterogeneous noise, while supporting mutation testing with validated mutants, an empirically grounded mutation taxonomy, quantified fault-detection capability, and attribution of observed failures to software defects or hardware noise?}

We decompose it into five research questions, each answered on simulated noise, device snapshots and IBM Quantum hardware, in the order in which a tester relies on the framework: the oracle must be sound (RQ1); mutation analysis depends on its verdicts (RQ2); attribution explains ambiguous verdicts (RQ3); sensitivity quantifies what soundness costs (RQ4); and drift and cost determine practicality (RQ5).

\begin{description}
\item[\textbf{RQ1.}] \emph{Does the noise-aware oracle control its false-positive rate at the nominal level across heterogeneous noise conditions and device noise models, and how does it compare with fixed-threshold, envelope-based, noise-filtering and noise-model-calibrated oracles?}
\item[\textbf{RQ2.}] \emph{How often do quantum mutation operators produce equivalent mutants, is the proposed noise-aware mutation taxonomy empirically supported, and to what extent does mutation testing under the noise-aware oracle detect semantic mutants?}
\item[\textbf{RQ3.}] \emph{When the oracle reports a deviation, can the failure be attributed to a software fault or to hardware noise without a noiseless simulator?}
\item[\textbf{RQ4.}] \emph{What fault-detection capability (recall, precision, F1) does the noise-aware oracle achieve, how well does the analytic power model predict it, and how does it scale with circuit size?}
\item[\textbf{RQ5.}] \emph{How does the oracle behave under drift between calibration and testing, how should recalibration be triggered, and what does calibration cost in executions?}
\end{description}

\subsection{Contributions}

This paper makes four contributions.

\begin{enumerate}
\item \textbf{A noise-aware oracle with a nominal false-positive level.} We formulate the NISQ test oracle problem in the differential setting as a two-sample test against a self-baseline on the same backend, and design the oracle around it: the acceptance threshold is the $(1-\alpha)$ quantile of the finite-shot null distribution of the Jensen--Shannon divergence, estimated by parametric bootstrap with a correction for unobserved outcomes; pooling $K$ calibration runs raises sensitivity, with a single-run variant for tight execution budgets; a homogeneity test checks that the baseline is valid; a noncentral chi-square model predicts recall and the shot budget for a target power; and a sentinel policy decides when to recalibrate. The underlying test statistics are classical~\cite{read1988goodness}; the contribution is the oracle design and the evidence that fixed, filtered and globally calibrated thresholds cannot keep the false-positive rate at a level set in advance across programs and noise levels.

\item \textbf{A noise-aware mutation testing methodology with exact semantic labels.} We provide a mutation engine that places operators before measurement and labels every mutant as equivalent or not on the tested inputs by statevector simulation, a quantification of equivalent-mutant rates for quantum mutation operators and fault injectors, and a three-class mutation taxonomy (semantic, noise-amplifying, noise-neutral) validated with stated hypotheses, tests that account for shared calibrations, effect sizes and a dose-response model of noise amplification.

\item \textbf{A hardware-executable bug--noise attribution procedure.} We attribute a reported deviation to a fault or to noise by zero-noise extrapolation of the output distributions of both programs using global unitary folding, and evaluate it against exact, non-circular labels.

\item \textbf{A comprehensive empirical evaluation.} We evaluate the framework on nine core benchmarks and $32$ QASMBench programs, ten real bugs from Bugs4Q, five depolarizing and two coherent-error noise regimes, three device-snapshot backends, circuits of up to 16 qubits and IBM Quantum hardware, against fixed-threshold, envelope-based (statistical process control), QOIN-style, noise-model-calibrated and re-execution baselines, reporting full confusion matrices, cluster-robust confidence intervals and variance-based sensitivity indices, and characterize where the approach stops working.
\end{enumerate}

The underlying principle, deriving the acceptance threshold from a system's own observed variability, may also apply to other software observed through a noisy channel, such as flaky tests in continuous integration~\cite{luo2014empirical,parry2021survey,bell2018deflaker} and machine-learning inference under numerical nondeterminism~\cite{zhang2020machine}. Every experiment and quantitative claim in this paper, however, concerns quantum programs on NISQ hardware or models of it; Section~\ref{subsec:beyond-quantum} discusses this scope.

\paragraph{Paper organization.}
Section~\ref{sec:related} reviews related work, Section~\ref{sec:approach} presents the framework, Section~\ref{sec:design} the study design, and Section~\ref{sec:results} the results. Section~\ref{sec:discussion} discusses implications, Section~\ref{sec:threats} threats to validity, and Section~\ref{sec:conclusion} concludes.

\section{Background and Related Work}
\label{sec:related}

Existing quantum software testing techniques respond to NISQ noise in three ways: they \emph{filter} it out via error mitigation or learned noise models, \emph{tolerate} it with a fixed divergence threshold, or \emph{restructure} the correctness check via metamorphic relations, assertions, property-based testing or richer measurement specifications~\cite{barrera2021state,paltenghi2024survey,zhao2020qse}.

\paragraph{Terminology.}
\emph{Noise} denotes all stochastic deviations of a physical execution from the ideal unitary semantics of a program: gate, decoherence and readout errors and their drift. It is distinct from \emph{finite-shot sampling variability}, which arises even on a noiseless simulator. A \emph{fault} (or bug) is a defect that changes the program's ideal semantics.

\subsection{Quantum Software Testing Tools and Adequacy}
\label{subsec:related-tools}

Assertion-based approaches such as AssertsQ~\cite{assertsq2025} verify properties of quantum states; MorphQ~\cite{morphq} and QDiff~\cite{wang2021qdiff} apply metamorphic and differential testing~\cite{segura2016survey}; QMutPy~\cite{qmutpy,fortunato2022mutation} and Muskit~\cite{muskit} adapt mutation analysis to Qiskit programs; and Pontolillo et al.~\cite{pontolillo2025ideal} adapt property-based testing to noisy quantum computers. Most of these techniques assume near-ideal execution or rely on manually selected thresholds whose robustness under heterogeneous noise has not been established. Test-generation frameworks such as Quito~\cite{wang2022quito}, QuCAT~\cite{wang2024qucat} and QuSBT~\cite{wang2022qusbt} address input generation rather than oracle soundness, mutation adequacy or failure attribution.

Coverage criteria for quantum programs include input and output coverage~\cite{ali2021assessing} and gate branch coverage~\cite{fortunato2024gate}; they measure structural and behavioral exercise but do not address whether deviations arise from faults or noise, and we treat them as complementary to the adequacy metrics introduced here.

\subsection{Mutation Testing for Quantum Programs}
\label{subsec:related-mutation}

Quantum mutation tools such as QMutPy and Muskit define quantum-specific operators (gate insertion, deletion, replacement and measurement manipulation) and report mutation scores, treating observable deviations as kills~\cite{qmutpy,muskit}. They inherit classical assumptions: behavioral differences are interpreted as faults, although under NISQ execution noise can itself induce deviations or mask genuine faults. The closest work to ours is the empirical study of Fortz et al.~\cite{fortz2026robust}, which examines mutation analysis of 41 quantum programs under noisy simulators of three IBM devices. They report that noise substantially changes the behavioral distance between programs and their mutants, making equivalent mutants harder to separate from genuine faults; that density-matrix metrics discriminate best but are unavailable on hardware, whereas output-distribution metrics reach up to 73\% accuracy; that thresholds set for the noise profile outperform noiseless thresholds; and that noise effects depend more on the algorithm and circuit than on the mutation type. Our work is complementary. Fortz et al.\ compare distance metrics and threshold-selection strategies; we derive the threshold of an output-distribution oracle from finite-shot statistics so that its false-positive rate is set in advance, label mutant equivalence exactly so that detection is measured only on non-equivalent mutants, and validate a mutation taxonomy and an attribution procedure on top of that oracle. Their observation that noise effects depend more on the circuit than on the mutation type is consistent with our finding that the two semantic operator families are not statistically separable (Section~\ref{subsec:rq2-results}). We also evaluate their threshold construction as a baseline (Section~\ref{subsec:baselines}).

Pan et al.~\cite{pan2023understanding} analyze the influence of noise on program behavior, and Virani et al.~\cite{virani2025distinguishing} propose statistical techniques for distinguishing bug-induced failures from noise; neither redefines mutation adequacy for NISQ execution.

\subsection{Noise-Aware Testing and Oracles}
\label{subsec:related-noise-aware}

QOIN~\cite{muqeet2024qoin} learns backend- and circuit-specific noise characteristics by transfer learning and filters noise from execution results before applying an existing oracle. Such approaches recover an approximation of the ideal behavior but preserve the underlying oracle semantics.

A complementary line of research investigates improvements to distribution-based correctness oracles by redesigning the measurements used to characterize quantum program behavior. Oldfield et al.~\cite{oldfield2025faster} observe that treating quantum programs purely as classical output distributions suffers from two fundamental limitations: the exponential growth of the output space and the inability of computational-basis measurements to detect certain classes of quantum faults, such as relative phase errors. They address these limitations through reduced program specifications and projective measurements in multiple bases. Related work on assertion-based verification~\cite{oldfield2025bloch,assertsq2025} and specification-aware test generation~\cite{ye2023quratest} similarly focuses on improving fault detectability by enriching program specifications or measurement strategies. Bensoussan et al.~\cite{bensoussan2026toward} propose live noise fingerprinting: device noise is characterized continuously from empirical observations with classical shadow tomography, and fingerprints of different platforms are compared to flag programs likely to behave differently across them. Noise fingerprints characterize the \emph{device}; our oracle characterizes the variability of a \emph{program} on a device, and the drift test of Section~\ref{subsubsec:drift} detects when that variability changes. The two are natural complements for deciding when recalibration is needed.

Less attention has been given to oracles that judge correctness directly under noisy execution, with a false-positive level set in advance, without an ideal reference or explicit noise mitigation.

\subsection{Statistical Foundations: From Process Control to Self-Baseline Calibration}
\label{subsec:related-spc}

Estimating a process mean and dispersion from repeated observations and flagging deviations beyond a $\mu + k\sigma$ control limit forms the basis of Shewhart statistical process control (SPC)~\cite{montgomery2020introduction}, while detecting distributional changes over time has been extensively studied in sequential change detection and CUSUM methods~\cite{page1954continuous,basseville1993detection}. Similar baselining principles appear in classical flaky-test analysis and performance regression testing, where current executions are compared against historical execution baselines. A control-limit envelope $\tau = \max(\mu + k\sigma, \tau_{\min})$ over pairwise divergences of repeated runs is the natural transfer of this idea to quantum output distributions, and we include it as a baseline (the \emph{SPC envelope}). It has two weaknesses in this setting: the pairwise divergences of $K$ runs are dependent and skewed, so $k$ carries no coverage guarantee, and a floor $\tau_{\min}$ is needed to prevent degenerate limits on concentrated outputs, which then dominates the decision. Our oracle instead uses the known sampling distribution of divergence statistics between multinomial samples: the Jensen--Shannon divergence is a member of the power-divergence family whose scaled form is asymptotically chi-square distributed under the null hypothesis of equal distributions~\cite{read1988goodness,agresti2013categorical}, so a threshold at a nominal false-positive level can be computed for each program and backend.

The statistical machinery is therefore classical, and we do not claim a new test statistic. Statistical tests have been used in quantum software testing before: statistical assertions check properties of intermediate states with chi-square tests~\cite{huang2019statistical}, and property-based testing has been adapted to noisy hardware~\cite{pontolillo2025ideal}. These approaches test a program against a \emph{specification} of its expected behavior. What has been missing is a differential oracle that tests a program against the \emph{noisy behavior of a trusted version of itself on the same backend}, together with an account of what such an oracle needs to be usable in a testing process: a threshold whose false-positive rate is chosen rather than tuned, a sensitivity that can be planned, a check that the baseline is still valid, labels that separate detectable from undetectable mutants, and a way to explain the rejections that are caused by noise. Our contribution is this oracle design and its empirical characterization, including the evidence that the thresholds in current use, fixed, filtered or calibrated globally on a noise model, cannot keep the false-positive rate at a level set in advance across programs and noise levels.

\subsection{Methodological Gaps}
\label{subsec:motivation}

Under noise, a killed mutant no longer implies test effectiveness: noise may amplify small syntactic changes so that a correct program or an equivalent mutant fails, or mask genuine faults, yet QMutPy and Muskit evaluate mutants with noise-oblivious oracles~\cite{qmutpy,muskit}, which may overestimate adequacy~\cite{barrera2021state,paltenghi2024survey}. Noise-aware simulation and benchmarking studies confirm that noise dominates observable behavior as depth, entanglement and gate density grow, even for correct programs~\cite{wang2025accelerating,grurl2022noise,li2023qasmbench,pan2023understanding}. Distribution-based oracles typically use a manually selected threshold that is fixed across programs and backends~\cite{assertsq2025,morphq}, and no fixed threshold states the false-positive rate it delivers. Mutation taxonomies are motivated by operator semantics but not evaluated for statistically distinguishable behavior under noise, and the equivalent-mutant problem, undecidable in general for classical software~\cite{madeyski2014overcoming,papadakis2018mutation}, is decidable for quantum programs of moderate width by statevector simulation, which existing tools do not exploit. Failure attribution matters for debugging, fault localization and repair~\cite{sato2024locating,pontolillo2024delta,li2024automatic,tan2025hornbro}, which assume that failures are genuine defects, but attribution mechanisms~\cite{virani2025distinguishing} are evaluated on limited cases, and those that need a noiseless simulation cannot run on hardware beyond classical simulation. Methodological analyses also report inconsistent experimental design, oracle selection and noise handling across empirical studies~\cite{li2026methodological}.

Table~\ref{tab:rw-comparison} summarizes representative approaches across four dimensions: noise-aware oracles, noise-sensitive mutation models, adequacy metrics for noisy execution and failure attribution. Existing approaches address one or two in isolation; to our knowledge, none integrates them in one methodology evaluated under a unified protocol.

\begin{table}[t]
\centering
\caption{Comparison of representative quantum software testing approaches. A \cmark\ marks a dimension the approach treats as a first-class concern: \emph{noise-aware oracle}, the correctness decision accounts for hardware noise; \emph{noise-sensitive mutations}, mutation operators categorized by their interaction with noise; \emph{noise-aware metrics}, adequacy metrics defined under noisy execution; \emph{failure attribution}, a mechanism separating bug-induced from noise-induced failures.}
\label{tab:rw-comparison}
\small
\begin{tabular}{lcccc}
\toprule
\textbf{Work} & \textbf{Noise-aware} & \textbf{Noise-sensitive} & \textbf{Noise-aware} & \textbf{Failure} \\
& \textbf{oracle} & \textbf{mutations} & \textbf{metrics} & \textbf{attribution} \\
\midrule
AssertsQ~\cite{assertsq2025}        & \cmark & \xmark & \xmark & \xmark \\
MorphQ~\cite{morphq}                & \xmark & \xmark & \xmark & \xmark \\
QDiff~\cite{wang2021qdiff}          & \xmark & \xmark & \xmark & \xmark \\
QMutPy~\cite{qmutpy}                & \xmark & \xmark & \xmark & \xmark \\
Muskit~\cite{muskit}                & \xmark & \xmark & \xmark & \xmark \\
QOIN~\cite{muqeet2024qoin}          & \cmark & \xmark & \xmark & \xmark \\
Pontolillo et al.~\cite{pontolillo2025ideal} & \cmark & \xmark & \xmark & \xmark \\
Fortz et al.~\cite{fortz2026robust} & \cmark & \xmark & \xmark & \xmark \\
Virani et al.~\cite{virani2025distinguishing} & \xmark & \xmark & \xmark & \cmark \\
\midrule
\textbf{This work}                  & \cmark & \cmark & \cmark & \cmark \\
\bottomrule
\end{tabular}
\end{table}

\section{The Noise-Aware Testing Framework}
\label{sec:approach}

\begin{figure}[t]
\centering
\includegraphics[width=\linewidth]{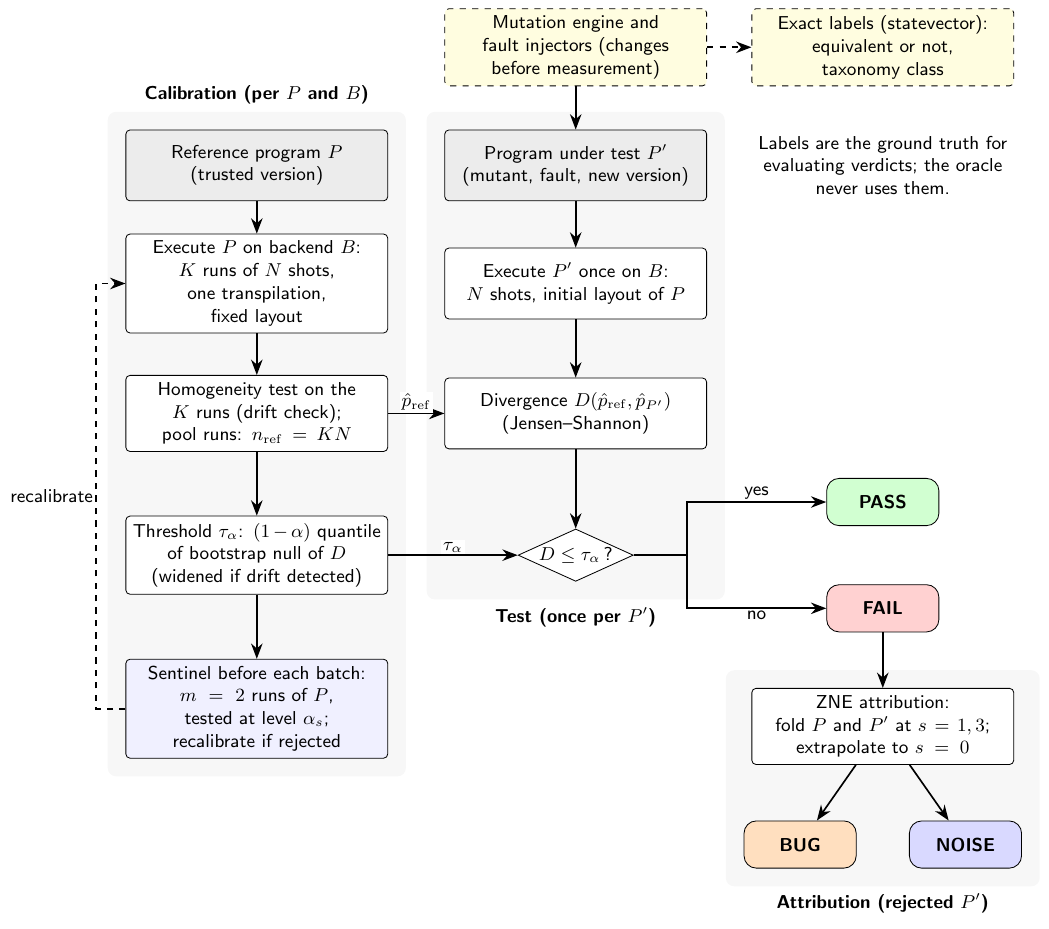}
\Description{Flowchart in three columns. Calibration: the reference program is executed K times, a homogeneity test checks the runs for drift, the runs are pooled and a bootstrap threshold is computed; a sentinel before each batch triggers recalibration. Test: the program under test is executed once and its Jensen-Shannon divergence from the pooled reference is compared with the threshold, giving PASS or FAIL. Attribution: a failed program is folded at noise scales 1 and 3 and extrapolated to zero noise, giving BUG or NOISE. Exact statevector labels are used only to evaluate the verdicts.}
\caption{Workflow of the framework. The reference program $P$ is executed $K$ times on the target backend; the pooled counts form the self-baseline reference, the homogeneity test checks the calibration runs for drift, and the null-quantile threshold $\tau_\alpha$ is computed from the reference counts. Each program under test $P'$ (a mutant, an injected fault, or a new version) is executed once and accepted if its divergence from the reference does not exceed $\tau_\alpha$. Rejected programs can be passed to zero-noise-extrapolation attribution, which labels the deviation as a fault or as a noise effect. Before each batch of tests, a sentinel of two executions of $P$ checks that the reference is still valid and triggers recalibration if it is not (Section~\ref{subsec:oracle}). Exact semantic labels (statevector) are used only to construct and evaluate mutants and faults, never by the oracle itself.}
\label{fig:workflow}
\end{figure}

The framework has three components (Figure~\ref{fig:workflow}): a \emph{noise-aware oracle} whose threshold is derived from the finite-shot sampling distribution of the divergence between repeated executions of the reference program (Section~\ref{subsec:oracle}), a \emph{mutation engine} whose mutants carry exact semantic labels within a three-class taxonomy (Section~\ref{subsec:mutation}), and a \emph{bug--noise attribution} procedure based on zero-noise extrapolation (Section~\ref{subsec:attribution}). A power model (Section~\ref{subsec:power}) and adequacy metrics (Section~\ref{subsec:mutation-metrics}) complete it.

\subsection{Setting and Problem Statement}
\label{subsec:framework-overview}

\emph{Self-baseline calibration} is the general principle: judging a program's behavior against a baseline established by repeated execution of a trusted reference on the same backend, rather than against a fixed tolerance or an ideal output. This paper instantiates it with Jensen--Shannon divergences and their finite-shot null distribution; the \emph{noise-aware oracle} is the resulting artifact. The oracle is \emph{differential}: it decides whether a program under test $P'$ behaves like a trusted reference $P$ on the same backend $\mathcal{B}$, as in mutation analysis ($P'$ a mutant) and regression testing ($P$ the last accepted version). The correctness of $P$ itself must be established by other means, such as specification-based or property-based testing on small instances.

\begin{definition}[NISQ testing problem]
\label{def:nisq-testing}
Given a reference program $P$, a program under test $P'$ and a noisy backend $\mathcal{B}$, the NISQ testing problem is to decide whether the output distribution of $P'$ on $\mathcal{B}$ differs from that of $P$ on $\mathcal{B}$ by more than finite-shot sampling explains, and, if it does, whether the difference is caused by a change of the program's ideal semantics (a \emph{fault}) or only by its interaction with hardware noise (a \emph{noise effect}). The attribution procedure of Section~\ref{subsec:attribution} returns one of \{\textsc{Bug}, \textsc{Noise}\}.
\end{definition}

The framework accepts a Qiskit circuit ending in computational-basis measurement of some or all qubits (variational parameters fixed), a backend (a noisy simulator, a Qiskit fake backend or an IBM Quantum device), and a configuration (shots $N$, calibration size $K$, level $\alpha$, seed).

\paragraph{Observability limitation.}
A change that leaves the measured distribution unchanged for the tested input cannot be detected by any threshold on that distribution; relative phase errors immediately before measurement are the canonical example. Equivalence and detection are therefore defined with respect to the measured output distribution, and injected faults that change the unitary but not the measured output are reported separately. Richer measurement specifications~\cite{oldfield2025faster} address this limitation and are orthogonal to the threshold construction.

\subsection{Noise-Aware Oracle}
\label{subsec:oracle}

\paragraph{Divergence statistic.}
Let $p$ and $q$ be empirical output distributions over measured bitstrings. We use the base-2 Jensen--Shannon divergence~\cite{lin1991divergence}
\begin{equation}
D(p, q) = \tfrac{1}{2}\, \mathrm{KL}\!\left(p \,\middle\|\, m\right) + \tfrac{1}{2}\, \mathrm{KL}\!\left(q \,\middle\|\, m\right), \qquad m = \tfrac{1}{2}(p + q),
\end{equation}
which is symmetric and bounded in $[0, 1]$. The two distributions are aligned over the union of observed bitstrings, additively smoothed with $\varepsilon = 10^{-12}$, and renormalized before $D$ is computed. We state this explicitly because the Jensen--Shannon \emph{distance} $\sqrt{D}$ and the divergence $D$ are frequently conflated; all thresholds in this paper are on the divergence scale.

\paragraph{Null distribution of the divergence.}
Let the reference be estimated from $n_{\mathrm{ref}}$ shots and the program under test from $n_{\mathrm{test}}$ shots. Under the null hypothesis $H_0$: ``$P'$ has the same output distribution $\pi$ as $P$ on $\mathcal{B}$'', the two count vectors are independent multinomial samples of $\pi$. A second-order expansion of $D$ around $\pi$ gives $D \ln 2 \approx \frac{1}{8}\sum_x (\hat p_x - \hat q_x)^2/\pi_x$, and the multinomial central limit theorem gives the Pearson-type limit
\begin{equation}
T \;=\; \frac{8 \ln 2 \; D(\hat p, \hat q)}{1/n_{\mathrm{ref}} + 1/n_{\mathrm{test}}} \;\xrightarrow{\;d\;}\; \chi^2_{d-1},
\label{eq:null}
\end{equation}
where $d$ is the number of outcomes with $\pi_x > 0$. This is the two-sample instance of the classical result for power-divergence statistics~\cite{read1988goodness,agresti2013categorical}. Two consequences matter for testing. First, the divergence between two \emph{correct} runs does not depend on how noisy the backend is, only on how many outcomes the noisy distribution occupies and on the shot budget: noise matters through $d$. Second, a threshold at the nominal level $\alpha$ follows directly, valid asymptotically:
\begin{equation}
\tau_\alpha^{\chi^2} \;=\; \chi^2_{d-1,\,1-\alpha}\;\frac{1/n_{\mathrm{ref}} + 1/n_{\mathrm{test}}}{8 \ln 2}.
\label{eq:chi2-threshold}
\end{equation}

\paragraph{Bootstrap threshold.}
Equation~\eqref{eq:chi2-threshold} requires $d$ and assumes large counts per outcome. The oracle therefore uses a parametric bootstrap~\cite{efron1993introduction} from the reference counts. The plug-in estimate $\hat\pi$ is corrected for outcomes not yet observed: with $f_1$ and $f_2$ the numbers of outcomes observed once and twice in $n$ reference shots, $\hat S_0 = f_1^2/(2 f_2)$ unseen outcomes (bias-corrected to $f_1(f_1 - 1)/2$ when $f_2 = 0$, and capped by the number of possible outcomes)~\cite{chao1984nonparametric} share the Good--Turing unseen mass $f_1/n$~\cite{good1953population}, and the observed probabilities are scaled by $1 - f_1/n$. We draw $B$ pairs ($B = 4000$ in the main false-positive evaluation of RQ1 and in the hardware analyses, $2000$ or $3000$ in the other experiments; Section~\ref{subsec:params}) of multinomial samples of sizes $(n_{\mathrm{ref}}, n_{\mathrm{test}})$ from $\hat\pi$, compute $D$ for each pair, and set $\tau_\alpha$ to the empirical $(1-\alpha)$ quantile. Equation~\eqref{eq:chi2-threshold} is reported alongside as a closed-form check.

\paragraph{Pooled reference.}
The reference is the pool of $K$ calibration executions of $P$, so $n_{\mathrm{ref}} = K N$ and the term $1/n_{\mathrm{ref}}$ shrinks with $K$. Pooling yields a more precise reference, which lowers the threshold and raises power (Section~\ref{subsec:power}), and several independent runs, which make the drift test below possible. We call the oracle with a pooled reference ($K = 8$ by default) the \emph{noise-aware oracle} (NAO), and its single-reference variant ($K = 1$) NAO-1. NAO-1 forgoes both benefits for a calibration cost of one execution and is intended for stable, compiled-once backends; NAO is the recommended default, and NAO-1 is evaluated alongside it to measure the contribution of pooling and the drift test.

\paragraph{Drift detection.}
\label{subsubsec:drift}
Equation~\eqref{eq:null} holds only if the backend is stationary while $P$ and $P'$ are executed. Under stationarity, the $K$ calibration runs are multinomial samples of one distribution, so the oracle applies a Pearson chi-square test of homogeneity to their $K \times d$ contingency table, merging outcomes with expected count below five per run~\cite{agresti2013categorical}. If homogeneity is rejected at $\alpha_{\mathrm{drift}} = 0.01$ (for example, because the noise drifted or the circuit was recompiled onto other qubits), the oracle reports drift and widens the threshold to $\max(\tau_\alpha,\ \bar\delta + k\, s_\delta)$, where $\bar\delta$ and $s_\delta$ are the mean and standard deviation of the pairwise divergences among the calibration runs and $k = 4$. Drift \emph{after} calibration is checked by a \emph{sentinel}: before a batch of tests, $P$ is executed $m$ times, and the pooled runs are compared with the reference by the oracle's own two-sample test at level $\alpha_s$; if it rejects, the reference is recalibrated. Because a false recalibration costs only $K$ executions, $\alpha_s$ can be less strict than $\alpha$, which raises the sentinel's power. RQ5 evaluates $m \in \{1, 2, 4\}$ and $\alpha_s \in \{10^{-3}, 10^{-2}\}$; we use $m = 2$ and $\alpha_s = 10^{-2}$.

Algorithm~\ref{alg:oracle} summarizes the oracle.

\begin{algorithm}[t]
\caption{Noise-aware oracle (NAO)}
\label{alg:oracle}
\begin{algorithmic}[1]
\REQUIRE reference program $P$, backend $\mathcal{B}$, shots $N$, calibration size $K$, level $\alpha$, bootstrap size $B$, drift level $\alpha_{\mathrm{drift}}$ and widening factor $k$, sentinel size $m$ and level $\alpha_s$; program under test $P'$
\ENSURE verdict (\textsc{Pass}/\textsc{Fail}), divergence $D$, threshold $\tau$, drift flag
\IF{reference not cached}
    \STATE execute $P$ on $\mathcal{B}$ $K$ times with $N$ shots $\rightarrow c_1,\dots,c_K$ \COMMENT{one transpilation, fixed layout}
    \STATE $c_{\mathrm{ref}} \gets \sum_i c_i$;\ $\hat\pi \gets$ plug-in of $c_{\mathrm{ref}}$ with unseen-outcome correction
    \STATE $\tau \gets (1-\alpha)$ quantile of $D$ over $B$ bootstrap pairs of sizes $(KN, N)$ drawn from $\hat\pi$
    \STATE drift $\gets$ false
    \IF{homogeneity test on $c_1,\dots,c_K$ rejects at $\alpha_{\mathrm{drift}}$}
        \STATE $\tau \gets \max(\tau,\ \bar\delta + k\, s_\delta)$;\ drift $\gets$ true
    \ENDIF
\ENDIF
\IF{a new batch of tests begins}
    \STATE execute $P$ on $\mathcal{B}$ $m$ times with $N$ shots $\rightarrow s_1,\dots,s_m$;\ $s \gets \sum_j s_j$ \COMMENT{sentinel}
    \STATE $\tau_s \gets$ threshold of lines 4--8 with level $\alpha_s$ and test size $mN$, i.e.\ pairs of sizes $(KN, mN)$
    \IF{$D(c_{\mathrm{ref}}, s) > \tau_s$}
        \STATE discard the cached reference and repeat lines 2--8 \COMMENT{recalibrate}
    \ENDIF
\ENDIF
\STATE execute $P'$ on $\mathcal{B}$ with $N$ shots, on the initial layout of $P$ $\rightarrow c'$
\RETURN $\big(D(c_{\mathrm{ref}}, c') \le \tau,\ D(c_{\mathrm{ref}}, c'),\ \tau,\ \text{drift}\big)$
\end{algorithmic}
\end{algorithm}

\paragraph{What the level means.}
\label{subsubsec:level}
The level $\alpha$ is \emph{nominal}, not an unconditional finite-sample guarantee. The rejection probability of a correct program equals $\alpha$ only to the extent that four approximations hold: the plug-in estimate $\hat\pi$, with its unseen-outcome correction, represents the support and shape of $\pi$; the bootstrap quantile is close to the true null quantile (at $\alpha = 10^{-3}$ and $B = 4000$, only about four draws lie beyond it, so the level holds on average over windows, not in each window); the calibration and test runs are samples of one distribution (stationary backend, identical compilation); and there are enough shots per outcome to estimate the null from the reference. The first two are statistical approximations whose accuracy RQ1 measures; the last two are assumptions about the test setting, of which the drift test and the sentinel check stationarity. Our claim is therefore empirical: under the conditions evaluated in RQ1, the observed false-positive rate is consistent with $\alpha$ where these assumptions hold, and RQ1 reports where it is not.

\begin{definition}[Noise-aware oracle]
\label{def:oracle}
For a reference program $P$, a program under test $P'$, a backend $\mathcal{B}$ and a level $\alpha$, the noise-aware oracle returns \textsc{Pass} if and only if $D(\hat p_{\mathrm{ref}}, \hat p_{P'}) \le \tau_\alpha(P, \mathcal{B})$, where $\hat p_{\mathrm{ref}}$ is the pooled calibration distribution of $P$ on $\mathcal{B}$, $\hat p_{P'}$ the empirical distribution of one execution of $P'$ on $\mathcal{B}$, and $\tau_\alpha$ the threshold of Algorithm~\ref{alg:oracle}. If $P'$ has the same output distribution as $P$ and the backend is stationary, the oracle rejects with probability approximately $\alpha$; the accuracy of the approximation depends on the number of shots per outcome (see \emph{What the level means} above).
\end{definition}

\subsection{Detection Power and Shot Budget}
\label{subsec:power}

If $P'$ differs from $P$, the scaled statistic $T$ of Equation~\eqref{eq:null} is approximately noncentral chi-square with $d-1$ degrees of freedom and noncentrality $\lambda = 8 \ln 2\; \delta / (1/n_{\mathrm{ref}} + 1/n_{\mathrm{test}})$, where $\delta$ is the divergence between the two true (noisy) output distributions~\cite{agresti2013categorical}. The probability that the oracle detects the difference is therefore
\begin{equation}
\mathrm{power}(\delta) \;=\; \Pr\!\left[\chi^2_{d-1}(\lambda) > \chi^2_{d-1,\,1-\alpha}\right].
\label{eq:power}
\end{equation}
Equation~\eqref{eq:power} makes the soundness--sensitivity trade-off explicit and quantitative. It yields the \emph{minimum detectable divergence} at a given shot budget and power, and conversely the shot budget needed to detect a fault of a given size. It also shows how noise reduces sensitivity: noise both shrinks $\delta$ (it pulls the faulty and the correct distributions towards the maximally mixed distribution) and increases $d$.

\subsection{Noise-Aware Mutation Testing}
\label{subsec:mutation}

\paragraph{Placement and exact labels.}
Operators that insert gates place them at a uniformly random position before the first measurement; operators that remove or rewrite gates act on gates before the first measurement. Gates placed after the final measurement cannot influence the measured output, and Section~\ref{subsec:rq2-results} shows that such placement yields equivalent mutants in every case. For each generated mutant, the framework computes the exact ideal output distribution of the original and of the mutant by statevector simulation and records their divergence $D_{\mathrm{ideal}}$; the mutant is \emph{equivalent} if $D_{\mathrm{ideal}} < 10^{-10}$ for the tested input. By default the input is $|0\cdots0\rangle$, the initial state of every benchmark; the framework can prepend preparation of arbitrary computational-basis inputs, and Section~\ref{subsec:rq2-results} reports equivalence over four inputs. The exact labels require simulating the circuit, which is feasible for the program widths at which mutation analysis is performed in this paper; at larger widths they can be computed on the reference version or on reduced instances.

\paragraph{Taxonomy.}
Definition~\ref{def:taxonomy} organizes mutants by two properties: whether the mutation changes the program's ideal semantics, and how it changes the program's physical execution.

\begin{definition}[Noise-aware mutation taxonomy]
\label{def:taxonomy}
A mutant $P'$ of a program $P$ belongs to one of three classes:
\begin{itemize}
\item \textbf{Semantic} ($\mathcal{M}_S$): $D_{\mathrm{ideal}}(P, P') > 0$. The mutation changes the ideal output. Two operator families generate semantic mutants: \emph{additive} (gates are inserted) and \emph{subtractive} (gates are removed).
\item \textbf{Noise-amplifying} ($\mathcal{M}_A$): $D_{\mathrm{ideal}}(P, P') = 0$ and the circuit of $P'$ compiled for the target backend contains more physical operations than that of $P$. The ideal output is unchanged, but the mutant is exposed to more gate error than the original; whether its output distribution changes measurably depends on the program, the noise and the shot budget (Section~\ref{subsec:rq2-results}).
\item \textbf{Noise-neutral} ($\mathcal{M}_N$): $D_{\mathrm{ideal}}(P, P') = 0$ and the circuit of $P'$ compiled for the target backend contains no more physical operations than that of $P$; gates may be reordered or re-expressed.
\end{itemize}
\end{definition}

The semantic criterion is a property of the program; the physical criterion is a property of the program \emph{and} the compilation target, because which operations are physical (rather than virtual, such as $R_z$ on IBM devices) and how a circuit is routed depend on the target. The operators of Table~\ref{tab:mutation-operators} are designed for one class each: the noise-amplifying operators insert gates whose product is the identity, protected from cancellation by barriers, and the noise-neutral operators reorder commuting gates or re-express a gate in an equivalent form, for example $H \cdot CZ \cdot H$, which compiles back to a CNOT on a device with a native CNOT. Whether a mutant meets the physical criterion of its operator's class on a given target is verified by comparing the gate counts of the compiled mutant and original (Section~\ref{subsec:rq2-results}).

\begin{figure}[t]
\centering
\includegraphics[width=\linewidth]{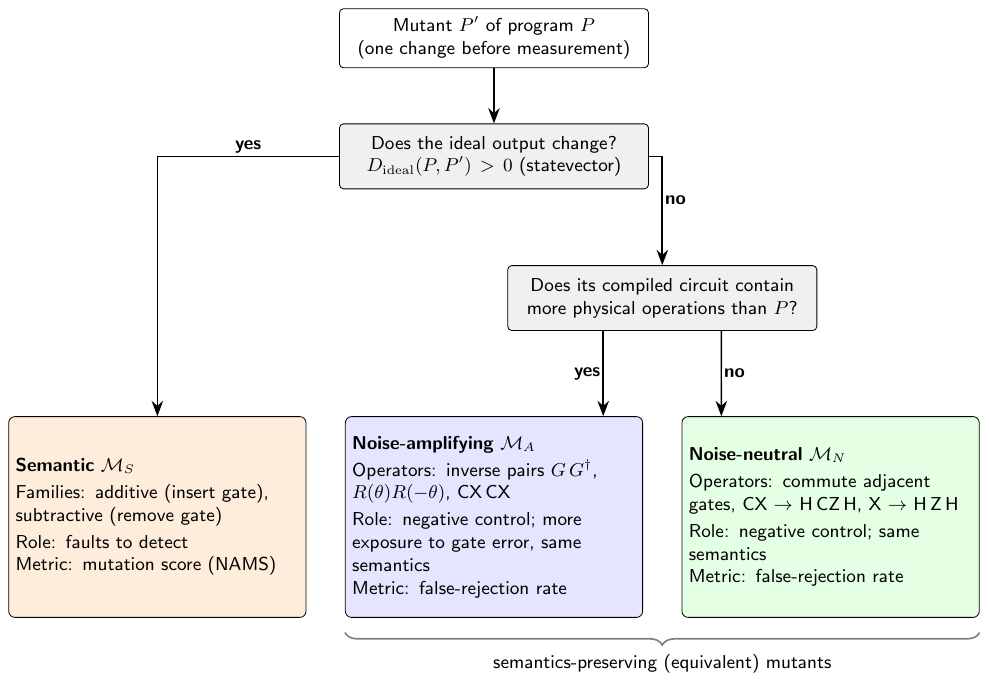}
\Description{Decision diagram. A mutant that changes the ideal output is semantic. A mutant that preserves the ideal output is noise-amplifying if its compiled circuit contains more physical operations than the original, and noise-neutral otherwise. Each class box lists its operators, its role and its metric.}
\caption{The three mutation classes. The semantic class contains the faults to be detected; the two semantics-preserving classes are negative controls. Noise-amplifying mutants preserve the ideal output but add physical gate error, so a sensitive oracle may observe a genuine distributional difference; attribution decides whether such a difference is a noise effect. Section~\ref{subsec:rq2-results} tests whether the three classes are empirically distinguishable.}
\label{fig:taxonomy}
\end{figure}

Semantic mutants are the faults a test suite should detect. Noise-amplifying and noise-neutral mutants are \emph{negative controls}: an oracle that rejects them reports a fault that does not exist. A noise-amplifying mutant is a particularly informative control, because a sensitive oracle can correctly detect that its output distribution differs from the original's (it is noisier) although its semantics are unchanged; such rejections are the cases that attribution must resolve. Table~\ref{tab:mutation-operators} lists the operators and Figure~\ref{fig:taxonomy} summarizes the taxonomy.

\begin{table}[t]
\centering
\caption{Mutation operators by taxonomy class. Insertions are placed at a uniformly random position before the first measurement; inserted pairs are separated by barriers so that the transpiler cannot cancel them.}
\label{tab:mutation-operators}
\small
\begin{tabular}{lll}
\toprule
\textbf{Class (family)} & \textbf{Operator} & \textbf{Description} \\
\midrule
\multirow{4}{*}{Semantic (additive)} & \texttt{insert\_cnot} & insert a CNOT between two distinct random qubits \\
& \texttt{insert\_t} & insert a T gate on a random qubit \\
& \texttt{insert\_x} & insert an X gate on a random qubit \\
& \texttt{insert\_rotation} & insert $R_{x/y/z}(\theta)$, $|\theta| \sim \mathcal{U}[\pi/8, \pi/2]$ \\
\midrule
\multirow{3}{*}{Semantic (subtractive)} & \texttt{remove\_gate} & remove a randomly selected gate \\
& \texttt{remove\_1q\_gate} & remove a randomly selected single-qubit gate \\
& \texttt{rotation\_to\_identity} & remove a randomly selected rotation gate \\
\midrule
\multirow{3}{*}{Noise-amplifying} & \texttt{inverse\_pair\_1q} & insert $G\,G^\dagger$, $G \in \{H, X, Y, Z, S, T\}$ \\
& \texttt{cx\_pair} & insert two identical CNOTs \\
& \texttt{rotation\_pair} & insert $R(\theta)\,R(-\theta)$, $R \in \{R_x, R_y\}$ \\
\midrule
\multirow{3}{*}{Noise-neutral} & \texttt{commute\_swap} & swap two adjacent gates acting on disjoint qubits \\
& \texttt{cx\_to\_h\_cz\_h} & replace a CNOT by $H \cdot CZ \cdot H$ on its target \\
& \texttt{x\_to\_h\_z\_h} & replace an X gate by $H \cdot Z \cdot H$ in place \\
\bottomrule
\end{tabular}
\end{table}

\paragraph{Generation.}
The engine samples an operator uniformly within the requested family and applies it to a fresh copy of the original (one mutation per mutant), driven by a seeded generator; an inapplicable operator is redrawn, so every mutant is labeled with the operator that produced it. For semantic families, the engine redraws (up to $60$ times) until it obtains a non-equivalent mutant and records the equivalent draws it rejected; for the semantics-preserving families, it verifies that every mutant is equivalent. Every program's engine is seeded with the same value.

\subsection{Mutation Adequacy Metrics}
\label{subsec:mutation-metrics}

\begin{definition}[Noise-aware mutation score]
\label{def:nams}
Within an evaluation cell (one benchmark, one noise level and one oracle calibration), let the original program pass the oracle. The noise-aware mutation score is the fraction of \emph{semantic} mutants rejected by the oracle,
$\mathrm{NAMS} = |\{m \in \mathcal{M}_S : \text{oracle rejects } m\}| \,/\, |\mathcal{M}_S|$.
Cells in which the original program fails the oracle are reported and excluded.
\end{definition}

Because equivalent mutants are excluded from $\mathcal{M}_S$ by construction, NAMS is not deflated by mutants that cannot be killed. The rejection rates of the two semantics-preserving classes are reported as \emph{false-rejection rates}. Detection of injected faults is reported with the full confusion matrix, recall, precision, F1 and false-positive rate, computed only on faults whose ideal measured output on the tested input differs from the original's; rejections of equivalent faults are reported as a separate rate.

\subsection{Bug--Noise Attribution}
\label{subsec:attribution}

When the oracle rejects $P'$, the attribution procedure decides whether the deviation is caused by a change of the ideal semantics (\textsc{Bug}) or by hardware noise (\textsc{Noise}), using only executions on the same backend. It relies on zero-noise extrapolation (ZNE)~\cite{temme2017error,giurgica2020digital}: if the deviation is semantic, it persists as noise is removed; if it is a noise effect, it vanishes.

\begin{enumerate}
\item \emph{Noise scaling.} Both circuits are folded globally, $U \mapsto U\,(U^\dagger U)^{m}$, giving noise scale factors $s = 2m+1 \in \{1, 3\}$ (and $5$ for the quadratic variant); barriers between folds prevent the transpiler from undoing them.
\item \emph{Measurement.} At each scale, $P$ and $P'$ are each executed once with $N$ shots.
\item \emph{Extrapolation.} Every outcome probability of $P$ and of $P'$ is extrapolated to $s = 0$ by Richardson extrapolation (linear through $s = 1, 3$; quadratic through $s = 1, 3, 5$), clipped at zero and renormalized, and the Jensen--Shannon distance $r_0$ between the two extrapolated distributions is computed.
\item \emph{Decision.} A null distribution of $r_0$ is obtained by redrawing, at every scale, both count vectors from their pooled distribution (no difference between $P$ and $P'$), repeating the extrapolation $2000$ times. The case is labeled \textsc{Bug} if $r_0$ exceeds the null $0.99$ quantile and \textsc{Noise} otherwise.
\end{enumerate}

The null of step 4 represents sampling variation only: it assumes that the extrapolated outputs of $P$ and $P'$ coincide. A noise effect satisfies this only to the extent that the extrapolation model holds. If the folded circuits saturate towards the maximally mixed distribution, if $P$ and $P'$ decay at different rates, or if the realized noise scale differs from the nominal factor (for example because a folded circuit is routed differently), a systematic residual remains that the null does not contain, and a noise effect is labeled \textsc{Bug}. The procedure is therefore a conditional diagnostic; RQ3 measures its reliability.

The linear variant costs four executions per case and the quadratic variant six. We also evaluate exponential extrapolation~\cite{giurgica2020digital} under a global depolarizing model $p(s) = u + (p_0 - u) f^s$, with $f$ estimated per circuit from $s = 1, 3$, which removes the bias of a linear fit when the circuits decay at different rates at the price of amplifying sampling noise. No variant requires a noiseless simulator. The ground truth for evaluating attribution is exact: a case is a \textsc{Bug} if and only if the ideal measured outputs of $P$ and $P'$ differ on the tested input.

\section{Study Design}
\label{sec:design}

Table~\ref{tab:rq-design} gives an overview of how each research question is evaluated. Every configuration value stated here is recorded with its results.

\begin{table}[t]
\centering
\caption{Study design per research question. \emph{Cells} are (benchmark, noise level) pairs. $\uparrow$: higher is better; $\downarrow$: lower is better; ``$\approx\alpha$'': the target is the nominal level.}
\label{tab:rq-design}
\footnotesize
\setlength{\tabcolsep}{3pt}
\begin{tabular}{@{}P{0.9cm}P{3.3cm}P{3.3cm}P{3.0cm}P{3.6cm}@{}}
\toprule
\textbf{RQ} & \textbf{Subjects} & \textbf{Environments} & \textbf{Oracles / methods} & \textbf{Metrics} \\
\midrule
RQ1 & 9 benchmarks, correct programs only; 40 calibration windows $\times$ 5 runs per cell; 32 QASMBench programs (extended suite); sampled supports of 8 to 16{,}384 outcomes; bootstrap stability (27 cells, $B$ up to $40{,}000$) & 5 synthetic noise levels; 2 coherent-error levels; 3 device snapshots $\times$ 2 compilation policies; IBM device & NAO, NAO-1 ($\alpha \in \{10^{-3}, 10^{-2}\}$), SPC envelope, fixed thresholds, QOIN-style filter, noise-calibrated global threshold (extended suite) & FPR ($\approx\alpha$, $\downarrow$ for baselines) with cluster-bootstrap CI; dispersion ratio \\
\addlinespace
RQ2 & 12 operators $\times$ 9 benchmarks $\times$ 100 draws (census); 15 mutants per family per cell, 27 cells; amplification dose study ($1{,}620$ mutants); extended suite ($5{,}760$ mutants) & noiseless (census); low, medium, high noise & exact statevector labels; NAO, NAO-1, SPC envelope & equivalent-mutant rate; NAMS $\uparrow$; false-rejection rate $\downarrow$; class contrasts; predicted vs observed rejection \\
\addlinespace
RQ3 & 1{,}062 labeled cases (mutants, faults, unmodified programs), 27 cells; 392 cases on snapshots & low, medium, high noise; 2 device snapshots; IBM device & ZNE (linear, quadratic, exponential); re-execution triage & accuracy $\uparrow$, Cohen's $\kappa$ $\uparrow$, bug precision and recall $\uparrow$, noise recall $\uparrow$ \\
\addlinespace
RQ4 & 50 non-equivalent faults and 1{,}350 correct runs, 45 cells; 163 faults in the extended suite; 10 real bugs (Bugs4Q); GHZ and QAOA up to 16 qubits; global threshold on held-out programs & 5 synthetic noise levels; 2 coherent-error levels; device snapshots; $\eta = 0.005$ (scaling); IBM device & NAO, NAO-1, SPC envelope, fixed thresholds, QOIN-style filter, noise-calibrated global threshold; marginal NAO; ablation (six configurations) & recall $\uparrow$, precision $\uparrow$, F1 $\uparrow$, FPR $\downarrow$; predicted vs observed power \\
\addlinespace
RQ5 & GHZ-3, QAOA-3, Bell-8, QAOA-8; 25 windows per drift level & calibration at $\eta_0$, testing at $\eta_0(1+r)$; IBM device sessions & NAO with recalibration policies (none, sentinel, co-scheduled); homogeneity test & FPR $\downarrow$, drift-detection rate $\uparrow$, executions per campaign $\downarrow$ \\
\bottomrule
\end{tabular}
\end{table}

\subsection{Subject Programs and Inputs}
\label{subsec:benchmarks}

The nine core benchmarks (Table~\ref{tab:benchmarks}) span 2 to 8 qubits: five small benchmarks covering the canonical primitives of prior quantum software testing work, and four larger ones in which noise has more opportunity to corrupt outputs. QAOA-8 is the most demanding (depth 27, 16 entangling gates); Bell-8 and GHZ-8 are shallower but have highly concentrated outputs. For RQ4, a scaling study adds GHZ circuits of 10 to 16 qubits and one-layer QAOA-MaxCut ring circuits of 10 and 12 qubits at medium noise, with eight windows of four correct runs ($32$ per circuit) and one run of each applicable non-equivalent fault of five injectors.

\begin{table}[t]
\centering
\small
\caption{Benchmark circuits. Depth is the logical depth before transpilation, including the measurement layer.}
\label{tab:benchmarks}
\begin{tabular}{lcccll}
\toprule
\textbf{Benchmark} & \textbf{Qubits} & \textbf{Depth} & \textbf{2-qubit gates} & \textbf{Description} & \textbf{Family} \\
\midrule
Bell   & 2 & 3  & 1  & two-qubit entangled state                      & state preparation \\
GHZ-3  & 3 & 4  & 2  & three-qubit GHZ state                          & state preparation \\
QFT-3  & 3 & 7  & 4  & three-qubit quantum Fourier transform          & arithmetic kernel \\
QAOA-3 & 3 & 12 & 6  & one-layer QAOA on a 3-qubit ring               & variational \\
VQE-3  & 3 & 5  & 2  & hardware-efficient ansatz, fixed parameters    & variational \\
Bell-8 & 8 & 9  & 7  & 8-qubit chain-entangled state                  & state preparation \\
GHZ-8  & 8 & 9  & 7  & 8-qubit GHZ state, star topology               & state preparation \\
QAOA-8 & 8 & 27 & 16 & one-layer QAOA-MaxCut on an 8-qubit ring       & variational \\
HWE-6  & 6 & 13 & 10 & two-layer hardware-efficient ansatz, fixed parameters & variational \\
\bottomrule
\end{tabular}
\end{table}

\paragraph{Extended suite.}
RQ1, RQ2 and RQ4 are replicated on $32$ programs from QASMBench~\cite{li2023qasmbench}: every small-scale program with at most $10$ qubits that is a static circuit ending in measurement (excluding mid-circuit measurement, reset, classical control, unparsable programs and QAOA-3, which duplicates a core benchmark). The suite covers arithmetic, error correction, oracle algorithms, phase estimation, linear solvers (HHL), quantum walks, simulation, machine learning (DNN), variational circuits and state preparation, from $2$ to $10$ qubits and up to depth $551$; several programs measure only some qubits. In RQ1 and RQ4, ten windows per program and noise level each judge five correct runs and one run of every applicable fault ($8{,}000$ correct runs; $163$ non-equivalent faults); in RQ2, $15$ mutants per family per program at three noise levels ($5{,}760$ mutants). The noise-calibrated global threshold for this suite is computed from $30$ separate runs of each program.

\paragraph{Inputs.}
All noisy executions start from $|0\cdots0\rangle$. To check that equivalence labels do not depend on this choice, the census of RQ2 also evaluates every mutant and fault on three further random computational-basis states, prepared by prepended X gates; a mutant is equivalent over the four inputs only if its ideal output equals the original's on all of them.

\subsection{Mutants and Injected Faults}
\label{subsec:design-mutants}

\paragraph{Equivalence census (RQ2).}
Each operator of Table~\ref{tab:mutation-operators} is applied 100 times to every benchmark on which it is applicable, without redrawing, and every raw draw is labeled by statevector simulation. Eleven operators are applicable to at least one core benchmark; \texttt{x\_to\_h\_z\_h} is not, because no core benchmark contains an X gate, and is exercised by the extended suite. A placement ablation applies the three insertion operators \texttt{insert\_cnot}, \texttt{insert\_t} and \texttt{insert\_x} $50$ times per benchmark both before and after the final measurement.

\paragraph{Mutation analysis (RQ2).}
For each benchmark, the engine generates $15$ mutants per family (semantic additive, semantic subtractive, noise-amplifying, noise-neutral; seed 42), so each of the 27 cells (nine benchmarks at low, medium and high noise) contains $30$ semantic mutants and $15$ mutants of each semantics-preserving class, $1{,}215$ mutants in total. Each mutant is executed once and judged by all oracles on the same counts.

\paragraph{Noise-amplification dose (RQ2).}
To test whether the rejections of noise-amplifying mutants are explained by the gate error they add, a dose-response study inserts $k$ barrier-protected inverse pairs at independent random positions before measurement: $k \in \{1, 2, 4\}$ CNOT pairs or $k \in \{1, 4, 16\}$ single-qubit pairs, ten mutants per (benchmark, pair type, $k$), nine benchmarks at low, medium and high noise ($1{,}620$ mutant executions). Each mutant is judged by NAO and also executed with $200{,}000$ shots to estimate its true noisy divergence from the original, from which Equation~\eqref{eq:power} predicts its rejection probability.

\paragraph{Injected faults (RQ4).}
Eight deterministic fault injectors model common programming errors: removing an entangling gate (\texttt{remove\_entanglement}), applying the wrong measurement basis (\texttt{wrong\_basis}), swapping the operands of a two-qubit gate (\texttt{swap\_qubits}), inserting a spurious X gate (\texttt{extra\_x}), dropping the last gate before measurement (\texttt{drop\_last\_gate}), inserting a Z gate (\texttt{phase\_flip}), perturbing a rotation angle (\texttt{wrong\_angle}) and retargeting a two-qubit gate (\texttt{wrong\_target}). Of the $66$ applicable injections on the nine benchmarks, $16$ are equivalent on $|0\cdots0\rangle$ and are excluded from recall, leaving $50$ non-equivalent faults. Every fault is executed in $10$ calibration windows at each of the five noise levels ($2{,}500$ fault tests), and each window also judges three fresh correct runs ($1{,}350$ matched negatives). The fixed thresholds are evaluated at $\tau \in \{0.01, 0.02, 0.05, 0.1\}$. To validate the power model, each fault's true noisy divergence is estimated from $200{,}000$-shot runs of the program and the fault, and, independently of the benchmarks, synthetic distributions with $d \in \{4, 16, 64, 256\}$ outcomes are perturbed by $14$ fault sizes and tested $300$ times each.

\paragraph{Real bugs (RQ4).}
We use real bugs from Bugs4Q~\cite{zhao2023bugs4q}, reproducible bugs in Qiskit programs collected from GitHub issues, Stack Overflow and Quantum Computing Stack Exchange. We include every bug whose buggy and fixed programs differ in the circuit itself (gates, operands, measurement mapping or initial state) and run as static circuits ending in measurement, and exclude API, version, visualization and simulator-configuration bugs, bugs whose fix replaces the whole program, and dynamic circuits. Ten bugs qualify (Table~\ref{tab:rq4-realbugs}); we transcribe them to the current Qiskit version without changing their circuits. The fixed program is the reference $P$ and the buggy program $P'$. Each bug is evaluated at the five noise levels and on FakeManilaV2 and FakeLagosV2, with ten windows per environment, each judging three runs of the fixed and one run of the buggy program; rejections in the first three windows of each environment are attributed by linear ZNE.

\paragraph{Attribution cases (RQ3).}
Each of the 27 cells contributes ten noise-amplifying and five noise-neutral mutants and the unmodified program (ground truth \textsc{Noise}), and eight additive and eight subtractive semantic mutants and every non-equivalent injected fault (ground truth \textsc{Bug}), $1{,}062$ cases in total ($582$ \textsc{Bug}, $480$ \textsc{Noise}). To compare extrapolation methods on identical data, the same cases are executed once more at $s = 1, 3, 5$ with the counts stored, and every method is applied to these counts; this comparison is repeated on two device snapshots (FakeManilaV2 and FakeLagosV2, five small benchmarks, $392$ cases), whose noise is not depolarizing.

\subsection{Execution Environments}
\label{subsec:noise-models}

\paragraph{Synthetic noise.}
Synthetic noise is parameterized by a single error level $\eta$: single-qubit gates receive a depolarizing channel with parameter $\eta$, two-qubit gates a depolarizing channel with parameter $\min(10\eta, 0.5)$, and readout a symmetric bit-flip channel with flip probability $\min(2\eta, 0.5)$. Five levels are evaluated: very low ($\eta = 0.0005$), low ($0.001$), medium ($0.005$), high ($0.02$) and very high ($0.05$, where the two-qubit channel saturates at the cap). The levels form a controlled stress scale rather than models of particular devices; realistic device noise is represented by the device snapshots and the IBM device below.

\paragraph{Coherent noise.}
Depolarizing noise is incoherent: it moves every output distribution towards the uniform one. Real devices also have coherent errors, systematic over-rotations that can shift output probabilities in either direction. To test the oracle under them, a second synthetic model follows every single-qubit gate by $R_X(\epsilon)$ and every two-qubit gate by $R_{ZZ}(5\epsilon)$ on the qubits of the gate, composed with depolarizing error at $\eta = 0.001$ and symmetric readout error $2\eta$, for $\epsilon \in \{0.02, 0.05\}$ rad. On the nine core benchmarks, $20$ calibration windows per benchmark and $\epsilon$ each judge five correct runs and one run of every applicable injected fault ($1{,}800$ correct runs and $2{,}000$ non-equivalent fault tests), with the oracles of the fault-detection design on identical executions.

\paragraph{Device snapshots.}
Three Qiskit fake backends are calibration snapshots of IBM Quantum devices: FakeManilaV2 (5 qubits), FakeLagosV2 (7 qubits) and FakeKolkataV2 (27 qubits). Each snapshot includes measured relaxation and dephasing times, gate error rates, readout assignment errors and the coupling map. The small benchmarks run on all three backends and the 8-qubit benchmarks additionally on FakeKolkataV2, with $15$ calibration windows of five correct runs per (benchmark, backend) pair. For detection (RQ4), eight mutants per family and every applicable injected fault are judged under the fixed policy. Because the transpiler's layout and routing choices change the noise a circuit experiences, two compilation policies are evaluated. All executions are transpiled by the Qiskit transpiler at optimization level 1. Under the \emph{fixed} policy, every circuit is transpiled once with a fixed seed and every program under test is placed on the initial layout of the reference; this is the recommended practice and the default of the framework. Under the \emph{random} policy, every execution is transpiled afresh, so that layout and routing may change between runs; this emulates a pipeline that recompiles on every submission.

\begin{table}[t]
\centering
\caption{Sources of evidence. Executions: runs of a circuit with $4096$ shots. Windows: calibrations from which a reference and a threshold are built. Simulated experiments are listed by research question (over $199{,}000$ executions in total).}
\label{tab:evidence}
\footnotesize
\setlength{\tabcolsep}{3pt}
\begin{tabular}{P{2.7cm}P{2.9cm}P{3.3cm}P{1.3cm}rr}
\toprule
\textbf{Source} & \textbf{Backends} & \textbf{Programs} & \textbf{RQs} & \textbf{Executions} & \textbf{Windows} \\
\midrule
Synthetic noise (simulation) & Aer, 5 depolarizing levels & 9 core, 32 QASMBench, 10 real bugs, scaling to 16 qubits & RQ1--RQ5 & \multicolumn{2}{r}{see caption} \\
Coherent noise (simulation) & Aer, 2 coherent-error levels & 9 core & RQ1, RQ4 & \multicolumn{2}{r}{see caption} \\
Device snapshots (simulation) & FakeManilaV2, FakeLagosV2, FakeKolkataV2 & core benchmarks, real bugs & RQ1--RQ4 & \multicolumn{2}{r}{see caption} \\
\midrule
Hardware: probe & \texttt{ibm\_fez}, 1 session & Bell, GHZ-3 & RQ1 & $42$ & $2$ \\
Hardware: HM, first job\textsuperscript{a} & \texttt{ibm\_kingston}, 2 jobs & Bell, GHZ-3, QFT-3, QAOA-3, VQE-3 & RQ1--RQ4 & $132$ & $5$ \\
Hardware: HM, second job & \texttt{ibm\_kingston}, 1 job & same & RQ5 & $34$ & $5$\textsuperscript{b} \\
Hardware: HM, third job & \texttt{ibm\_kingston}, 1 job & Bell-8, GHZ-8, QAOA-8, HWE-6; 8 real bugs & RQ1, RQ4 & $164$ & $12$ \\
Hardware: HM, fourth job & \texttt{ibm\_kingston}, 1 job & as first job & RQ5 & $74$ & $5$ \\
Hardware: H1 and H3 & \texttt{ibm\_kingston}, 2 jobs & GHZ-3, QAOA-3, Bell-8 & RQ1--RQ4 & $206$ & $3$ \\
Hardware: H2 & \texttt{ibm\_kingston}, 8 jobs & GHZ-3, QAOA-3 & RQ5 & $128$ & $16$ \\
\bottomrule
\multicolumn{6}{l}{\footnotesize \textsuperscript{a}Including the four-circuit probe that measured the QPU time per circuit. \textsuperscript{b}Judged against the calibrations of the first job.}
\end{tabular}
\end{table}

\paragraph{IBM Quantum hardware.}
\label{subsec:live-device}
The hardware evaluation consists of a probe, a mini campaign (HM) and three campaigns (H1 to H3), executed through the Qiskit Runtime sampler with $4096$ shots under the fixed compilation policy (Table~\ref{tab:evidence}). In every job, the circuits are interleaved in a seeded random order and placed on one initial layout per program, chosen by the success probability estimated from the device's reported errors and kept for every later job on that program. Raw counts, transpiled circuits, job identifiers and a backend snapshot are archived. The \emph{probe} executed Bell and GHZ-3 on \texttt{ibm\_fez} (IBM Heron) on 16 June 2026: per circuit, eight calibration runs, one reference run, three test runs and three runs for each of three fixed thresholds.

\emph{HM} covers all five research questions and is sized to fit the free IBM Quantum allowance, so that the hardware evaluation can be reproduced without paid access. Its first job executes, per benchmark (Bell, GHZ-3, QFT-3, QAOA-3, VQE-3), eight calibration runs, one single-reference run, four correct runs, up to three non-equivalent faults, one additive and one subtractive semantic mutant, two noise-amplifying \texttt{cx\_pair} mutants, one noise-neutral mutant, and the reference and the four non-neutral mutants at noise scale $3$. The second job, $24.6$ hours later, executes two sentinel runs, three correct runs and two faults per benchmark, judged against the first job's calibrations. The third job extends HM to the four larger core benchmarks (eight calibration runs, four correct runs, three faults and two semantic mutants each) and to the eight observable real bugs (eight calibration runs and two runs of the fixed program, and two runs of the buggy program, each). The fourth job, $48.1$ hours after the first, repeats the second with eight new calibration runs, so that the sentinel policy is evaluated with a full recalibration. The SPC envelope is calibrated on the nine runs of $P$ of the first job. HM ran on \texttt{ibm\_kingston} (IBM Heron) between 21 and 23 October 2026 and used $446$ seconds of QPU time.

\emph{H1 to H3} use GHZ-3, QAOA-3 and Bell-8 with more runs per benchmark. \emph{H1} (detection) executes, per benchmark, eight calibration runs, ten correct runs, twenty semantic and ten semantics-preserving mutants and every non-equivalent fault. \emph{H2} (drift) repeats eight calibration runs of GHZ-3 and QAOA-3 in eight sessions over two calendar days. \emph{H3} (attribution) executes five noise-amplifying mutants (physical pairs only, since $Z$, $S$ and $T$ are virtual on IBM devices) and five semantic mutants of GHZ-3 and QAOA-3, with the reference, at noise scales 1 and 3. They ran on \texttt{ibm\_kingston} on 22 and 23 October 2026 and used $378$ seconds. Simulation provides the rates and intervals on which the answers to the research questions rest; the hardware runs test whether the same behavior appears on a physical device.

\subsection{Drift Scenarios (RQ5)}
\label{subsec:design-drift}

Drift is modeled as a change of the synthetic noise level, on GHZ-3, QAOA-3, Bell-8 and QAOA-8. \emph{Drift after calibration}: the oracle is calibrated at $\eta_0 = 0.005$ and tested on correct runs at $\eta_0(1 + r)$, $r \in \{0, 0.1, 0.25, 0.5, 1.0\}$, with $25$ windows of five test runs. \emph{Recalibration policies} are compared on paired executions for $r \in \{0, 0.1, 0.25, 0.5\}$: no recalibration; a sentinel with $m \in \{1, 2, 4\}$ runs tested at $\alpha_s \in \{10^{-3}, 10^{-2}\}$, followed by recalibration ($K = 8$ runs at the new noise level) if it rejects; and co-scheduled calibration, in which the reference is always built in the same session as the tests. Each window executes one calibration at $\eta_0$, four sentinel runs, one recalibration set and five test runs, and every policy uses the subset it needs. \emph{Drift during calibration}: the noise level of the eight calibration runs increases linearly from $\eta_0$ to $\eta_0(1 + r)$, and the rate at which the homogeneity test flags drift is recorded over $25$ windows. Execution cost is computed per campaign of $M$ programs under test.

\subsection{Baselines}
\label{subsec:baselines}

\paragraph{SPC envelope.}
The envelope oracle accepts a run if its divergence from a single reference run is at most $\tau = \max(\mu + k\sigma, \tau_{\min})$, where $\mu$ and $\sigma$ are the mean and standard deviation of the $\binom{K}{2}$ pairwise divergences among $K$ calibration runs of $P$. This is the classical statistical-process-control envelope~\cite{montgomery2020introduction} applied to the self-baseline, with $K = 12$, $k = 4$ and $\tau_{\min} = 0.02$. It shares the self-baseline reference with NAO but sets its threshold heuristically, which isolates the contribution of the null-quantile threshold.

\paragraph{Fixed threshold.}
The fixed-threshold oracle rejects when the divergence between the test run and the \emph{ideal} output distribution, estimated by a noiseless simulated execution of $P$ with $N$ shots, exceeds a constant $\tau \in \{0.005, 0.01, 0.02, 0.05, 0.10\}$ (subsets of these values in the experiments with many oracles). This is the decision rule that noise-oblivious practice, and tools such as QMutPy and Muskit, effectively implement.

\paragraph{QOIN-style filter.}
A reproducible stand-in for the noise-filtering strategy of QOIN~\cite{muqeet2024qoin}: a per-outcome multiplicative noise factor is estimated from $K_c = 5$ executions of the correct circuit (additive smoothing $10^{-3}$), its inverse is applied to each test run, and the corrected distribution is compared with the ideal output at $\tau = 0.05$. We do not reproduce QOIN's learned filter (a neural network fine-tuned per circuit by transfer learning); the comparison is at the level of the strategy, not the implementation.

\paragraph{Re-execution triage (RQ3).}
The attribution baseline compares $P$ and $P'$ under three re-executions: with four times the shots, under the noise model scaled by $0.5$, and on a noiseless simulator. It labels the case \textsc{Bug} if the noiseless divergence exceeds $0.1$, \textsc{Noise} if it does not and at least one noisy re-execution also stays within that tolerance, and \textsc{Inconclusive} otherwise (which did not occur). It needs a scalable noise model and a noiseless simulator, so it cannot run on hardware; it quantifies what the hardware-executable ZNE procedure gains or loses against simulator-dependent triage.

\paragraph{Noise-calibrated global threshold (RQ4).}
Following Fortz et al.~\cite{fortz2026robust}, this baseline compares the Jensen--Shannon distance between a run and the \emph{exact} ideal output with one threshold per noise model, $T = Q_{87.5}(\mu_i) + Q_{87.5}(\mathit{SE}_i)$, where $\mu_i$ and $\mathit{SE}_i$ are the mean and standard error of the distance over $30$ noisy executions of correct program $i$ and $Q_{87.5}$ is the $87.5$th percentile over programs. The thresholds are computed from executions separate from the evaluation, which follows the fault-detection design with NAO judging the same executions, at $4096$ shots for both oracles (Fortz et al.\ use $10{,}000$). Because the percentile is taken over the programs to which the threshold is applied, a separate experiment recomputes it without each program in turn and applies it to fresh runs of that program.

\paragraph{NAO-1 and marginal NAO.}
NAO-1 (Section~\ref{subsec:oracle}) is the low-cost variant of NAO; it differs only in pooling and the drift test. For the scaling study, \emph{marginal NAO} applies the threshold of Equation~\eqref{eq:chi2-threshold} to every two-qubit marginal at level $\alpha / \binom{n}{2}$ (Bonferroni) and rejects if any marginal exceeds its threshold.

\paragraph{Ablation (RQ4).}
The comparisons above change more than one component at a time, so the ablation builds six oracles that add one component each: (A) ideal reference with fixed $\tau = 0.02$; (B) one self-baseline run with the same threshold; (C) one self-baseline run with the SPC envelope; (D) one self-baseline run with the null-quantile threshold (NAO-1); (E) the pooled reference without the drift test; and (F) NAO. All six judge the same executions in two settings: the nine core benchmarks at five noise levels ($10$ windows per cell), and the five small benchmarks on the three snapshots under the random compilation policy ($10$ windows each), where the drift test is designed to matter. Each window judges five correct runs and one run of every applicable fault, on separate seeds.

\paragraph{Information and executions available to each oracle.}
Table~\ref{tab:oracle-resources} lists what each oracle uses. The self-baseline oracles need no simulation of the program but spend executions of the reference on the backend; the oracles that compare with the ideal output need no calibration executions but need a noiseless simulation, whose cost grows exponentially with the qubit count, and the global threshold also a device noise model and runs of a set of programs. All oracles judge identical test executions, but not at equal calibration budget (NAO $8$, SPC envelope $13$); NAO-1, with one calibration execution, and the ablation control for this. Execution costs per campaign are in Table~\ref{tab:rq5-cost}.

\begin{table}[t]
\centering
\caption{Information and executions used by each oracle for one (program, backend) pair. Calibration: executions of the reference program on the backend before testing; per test: executions of the program under test.}
\label{tab:oracle-resources}
\footnotesize
\setlength{\tabcolsep}{4pt}
\begin{tabular}{P{3.2cm}P{3.4cm}P{4.2cm}rr}
\toprule
\textbf{Oracle} & \textbf{Reference} & \textbf{Classical computation or models needed} & \textbf{Calibration} & \textbf{Per test} \\
\midrule
NAO                       & pool of $8$ runs of $P$ & none                                                     & $8$  & $1$ \\
NAO-1                     & $1$ run of $P$          & none                                                     & $1$  & $1$ \\
SPC envelope              & $1$ run of $P$          & none; threshold from $12$ further runs of $P$            & $13$ & $1$ \\
Fixed threshold           & ideal output of $P$     & noiseless simulation of $P$                              & $0$  & $1$ \\
QOIN-style filter         & ideal output of $P$     & noiseless simulation of $P$; correction from $5$ runs of $P$ & $5$ & $1$ \\
Global threshold          & exact ideal output of $P$ & statevector simulation of $P$; device noise model; $30$ noisy executions of each calibration program & $0$ & $1$ \\
\midrule
ZNE attribution (linear)  & runs of $P$ and $P'$ at scales $1$, $3$ & none                                              & \multicolumn{2}{r}{$4$ per attributed case} \\
Re-execution triage       & outputs of $P$ and $P'$ & noiseless simulator; scalable noise model              & \multicolumn{2}{r}{$3$ settings for $P$ and $P'$} \\
\bottomrule
\end{tabular}
\end{table}

\subsection{Calibration Windows and Independence}
\label{subsec:windows}

A \emph{calibration window} is one set of calibration executions of $P$ from which every oracle builds its reference and threshold. In RQ1 and RQ4, a window has $13$ executions of $P$: the first eight form NAO's pooled reference, the thirteenth is the single reference of NAO-1 and the SPC envelope, and the first twelve calibrate the envelope; the QOIN-style filter uses the first five. Within a window, all oracles judge the same test executions (paired design). Test executions are independent multinomial samples given the window, but tests of one window share its calibration, so if the calibration underestimates the variability, all of them are affected together. False-positive rates are therefore reported with 95\% cluster-bootstrap intervals that resample whole windows ($2000$ resamples)~\cite{field2007bootstrapping}, and RQ1 draws $40$ independent windows per cell. In RQ2, all mutants of a cell share one window, so the class contrasts are also tested conditionally on the window and at the level of cells.

\subsection{Statistical Procedures}
\label{subsec:stats}

Table~\ref{tab:stat-tests} lists each test with its hypotheses and interpretation. All tests are two-sided; when a family of tests is reported, the Bonferroni-corrected level is stated with the result.

\begin{table}[t]
\centering
\caption{Statistical procedures. $p$: probability of a result at least as extreme as the observed one if $H_0$ holds; a small $p$ is evidence against $H_0$.}
\label{tab:stat-tests}
\footnotesize
\setlength{\tabcolsep}{3pt}
\begin{tabular}{@{}P{2.6cm}P{2.0cm}P{4.6cm}P{4.8cm}@{}}
\toprule
\textbf{Procedure} & \textbf{Used in} & \textbf{Hypotheses} & \textbf{Interpretation} \\
\midrule
Wilson interval~\cite{wilson1927probable} & all proportions & (interval estimate) & 95\% interval for a rate; valid near 0 and 1 \\
Cluster bootstrap~\cite{efron1993introduction,field2007bootstrapping} & RQ1, RQ4 & (interval estimate) & 95\% interval that accounts for tests sharing a calibration window \\
Calibration check & RQ1 & $H_0$: FPR $= \alpha$ & the oracle is calibrated if $\alpha$ lies in the cluster-bootstrap interval \\
Fisher exact test~\cite{fisher1922interpretation} & RQ2 & $H_0$: two classes have equal rejection probability; $H_1$: they differ & per-mutant test; Bonferroni over the 3 class pairs ($p < 0.0167$) \\
Cochran--Mantel--Haenszel test~\cite{agresti2013categorical} & RQ2 & $H_0$: equal rejection probability within every cell (common odds ratio 1); $H_1$: common odds ratio $\neq 1$ & stratified by cell, so it conditions on the shared calibration window \\
Cell-level sign-flip permutation test & RQ2 & $H_0$: the per-cell difference of class rejection rates is symmetric around 0 & treats the cell as the unit of analysis; the most conservative test (exact up to 16 cells, $200{,}000$ random sign patterns otherwise) \\
Brier score~\cite{brier1950verification} & RQ2 & (accuracy of predicted probabilities) & mean squared difference between predicted rejection probability and outcome; lower is better, compared with a constant predictor \\
Mann--Whitney U~\cite{mann1947test} with Cliff's $\delta$~\cite{cliff1993dominance} & RQ2 & $H_0$: the normalized divergences $D/\tau$ of two classes have the same distribution & $|\delta| < 0.147$ negligible, $< 0.33$ small, $< 0.474$ medium, otherwise large~\cite{romano2006appropriate} \\
Cohen's $\kappa$~\cite{cohen1960coefficient} & RQ3 & (agreement beyond chance) & $0.61$--$0.80$ substantial, $> 0.80$ almost perfect agreement~\cite{landis1977measurement} \\
Sobol indices~\cite{sobol2001global,saltelli2008global} & parameters & (variance decomposition) & $S_1$: share of output variance due to one factor alone; $S_T$: share including all its interactions \\
\bottomrule
\end{tabular}
\end{table}

The Sobol indices are computed exactly by functional ANOVA on complete factorial grids, so they carry no Monte Carlo error.

\subsection{Parameter Selection}
\label{subsec:params}

The parameters of NAO fall into four groups with different roles, and only the first is chosen to control false positives. The \emph{statistical} parameter, the level $\alpha$, sets the nominal false-positive rate. The \emph{resource} parameters, the calibration size $K$ and the shot count $N$, set sensitivity and execution cost. The \emph{numerical} parameter, the bootstrap size $B$, sets how accurately the quantile is estimated; it costs classical computation only ($B = 4000$ in the main false-positive evaluation of RQ1 and in the hardware analyses, $3000$ in most other experiments, and $2000$ in the sparsity, scaling, attribution and sentinel experiments; every value is recorded with its results, and RQ1 measures its effect). The \emph{drift-handling} parameters, the homogeneity level $\alpha_{\mathrm{drift}} = 0.01$ with the widening factor $k = 4$ and the sentinel's $m$ and $\alpha_s$, trade unnecessary recalibration against protection from drift and are evaluated in RQ5. This section selects $\alpha$, $K$ and $N$. Their roles follow from Sections~\ref{subsec:oracle} and~\ref{subsec:power}: $\alpha$ sets the false-positive rate, while $K$ and $N$ enter only through $1/n_{\mathrm{ref}} + 1/n_{\mathrm{test}}$ and therefore govern sensitivity. To check this before fixing values, a factorial study was run on separate seeds from all main experiments: $\alpha \in \{10^{-2}, 10^{-3}, 10^{-4}\}$ $\times$ $K \in \{1, 4, 8, 16\}$ $\times$ $N \in \{1024, 4096, 16384\}$ ($36$ configurations) on five benchmarks (Bell, GHZ-3, QAOA-3, Bell-8, QAOA-8) at medium noise, eight windows each, with the non-equivalent injected faults as positives. For comparison, the same analysis was run for the SPC envelope on a $5 \times 5 \times 5$ grid ($K \in \{4, 8, 12, 16, 20\}$, $k \in \{2, \dots, 6\}$, $\tau_{\min} \in \{0.005, 0.01, 0.02, 0.03, 0.05\}$; $125$ configurations) on all nine benchmarks and five noise levels with five windows per cell and paired executions. Table~\ref{tab:sobol} reports the indices.

\begin{table}[t]
\centering
\caption{Variance-based sensitivity of the false-positive rate and recall to oracle parameters (configuration level, exact functional ANOVA on the complete grid). $S_1$: first-order index; $S_T$: total index. For NAO, the FPR is governed by $\alpha$ and recall by $N$, as the null and power models predict; for the SPC envelope, recall is governed almost entirely by the floor $\tau_{\min}$.}
\label{tab:sobol}
\small
\begin{tabular}{llrrrr}
\toprule
& & \multicolumn{2}{c}{\textbf{FPR}} & \multicolumn{2}{c}{\textbf{Recall}} \\
\cmidrule(lr){3-4}\cmidrule(lr){5-6}
\textbf{Oracle} & \textbf{Factor} & $S_1$ & $S_T$ & $S_1$ & $S_T$ \\
\midrule
\multirow{3}{*}{NAO} & $\alpha$ & 0.558 & 0.895 & 0.043 & 0.175 \\
& $K$ & 0.018 & 0.295 & 0.062 & 0.231 \\
& $N$ (shots) & 0.024 & 0.327 & 0.639 & 0.880 \\
\midrule
\multirow{3}{*}{SPC envelope} & $K$ & 0.083 & 0.308 & 0.000 & 0.000 \\
& $k$ & 0.291 & 0.700 & 0.006 & 0.008 \\
& $\tau_{\min}$ & 0.154 & 0.532 & 0.992 & 0.994 \\
\bottomrule
\end{tabular}
\end{table}

For NAO the analysis confirms the analytic roles. The FPR is dominated by $\alpha$ ($S_T = 0.895$); the remaining variance of the FPR is interaction with sampling noise of the rate estimate itself, since every configuration's FPR lies between $0$ and $0.019$. Recall is dominated by $N$ ($S_T = 0.880$). No parameter dominates both responses, which is the property a tester needs: the false-positive rate can be chosen without affecting sensitivity, and sensitivity can be bought with shots without affecting the false-positive rate. For the SPC envelope, by contrast, recall depends almost only on the floor $\tau_{\min}$ ($S_1 = 0.992$): the envelope's sensitivity is set by a constant chosen in advance, and only $75$ of its $125$ configurations have zero false positives. False positives occur in $76\%$ of the configurations with $\tau_{\min} \le 0.01$, in $48\%$ of those with $\tau_{\min} = 0.02$ and in none with $\tau_{\min} \ge 0.03$, where soundness is bought with sensitivity: the mean recall of sound configurations falls from $0.801$ at $\tau_{\min} = 0.02$ to $0.691$ at $\tau_{\min} = 0.05$. The full NAO grid is listed in Table~\ref{tab:app-grid}.

The values used in all main experiments follow from this analysis.
\begin{itemize}
\item $\alpha = 10^{-3}$ for the primary oracle. The level is a design choice whose consequences are measured. A campaign that judges $T$ correct programs raises $\alpha T$ false alarms in expectation, each costing an investigation or an attribution run: ten per $1000$ tests at $10^{-2}$, one at $10^{-3}$. The price of a stricter level is sensitivity: at $N = 4096$ and $K = 8$, the minimum divergence detected with $80\%$ power is $0.00077$--$0.00396$ at $10^{-2}$, $0.00107$--$0.00501$ at $10^{-3}$ and $0.00136$--$0.00591$ at $10^{-4}$ for $4$ to $256$ outcomes, $27$--$39\%$ larger from $10^{-2}$ to $10^{-3}$ and $18$--$27\%$ from $10^{-3}$ to $10^{-4}$; in the grid, recall at $4096$ shots is $1.000$ at all three levels. At $10^{-2}$, the single-reference variant is liberal ($0.0140$ in RQ1). At $10^{-4}$, the threshold is the largest of $B = 4000$ null draws, so the tail quantile is poorly resolved, and the rate is hard to assess: $9{,}000$ correct runs would show $0.9$ expected false positives, and observing none would only bound the rate below about $3.3 \times 10^{-4}$ for independent runs. $10^{-3}$ is therefore the strictest evaluated level whose threshold is adequately estimated and whose rate can be assessed within this study's budget. $\alpha = 10^{-2}$ is reported alongside in RQ1.
\item $K = 8$: at fixed power, the minimum detectable divergence is proportional to $1/(KN) + 1/N$, so $K = 8$ obtains $87.5\%$ of the reduction that an unlimited calibration would obtain over $K = 1$, whereas $K = 16$ would add about $6$ percentage points for twice the calibration cost. In the grid, $K = 4$ and $K = 8$ have zero false positives at all three shot counts at $\alpha = 10^{-3}$, and $K = 8$ has the higher recall at $1024$ shots ($0.978$ versus $0.966$).
\item $N = 4096$: the FPR does not depend on $N$ (Table~\ref{tab:sobol}), so $N$ is chosen for sensitivity. In the grid, recall at $\alpha = 10^{-3}$ and $K = 8$ rises from $0.978$ at $1024$ shots to $1.000$ at $4096$ and stays there at $16{,}384$, which costs four times as much. The FPR does depend on $N$ through the shots per observed outcome: $4096$ shots provide $16$ per outcome for a full $8$-qubit register, the width of the core benchmarks, which is where the sparsity study shows the nominal level to hold.
\end{itemize}

The SPC envelope is evaluated at $K = 12$, $k = 4$ and $\tau_{\min} = 0.02$, a sound configuration of its grid whose recall ($0.804$) is within $0.002$ of the best sound configuration with $k = 4$. The best sound configuration of the whole grid ($K = 20$, $k = 5$, $\tau_{\min} = 0.005$) reaches recall $0.858$ on the grid's executions; it is selected in hindsight on the evaluation data and is reported only for comparison.

\section{Results}
\label{sec:results}

Each subsection reports the core benchmarks first, then the additional analyses and the QASMBench programs, and the IBM Quantum hardware last. Unless stated otherwise, NAO pools $K = 8$ calibration runs at $\alpha = 10^{-3}$, and all executions use $4096$ shots. Secondary tables and a figure are in Appendix~\ref{app:tables}.

\subsection{RQ1: Soundness of the Noise-Aware Oracle}
\label{subsec:rq1-results}

\begin{figure}[t]
\centering
\includegraphics[width=0.85\linewidth]{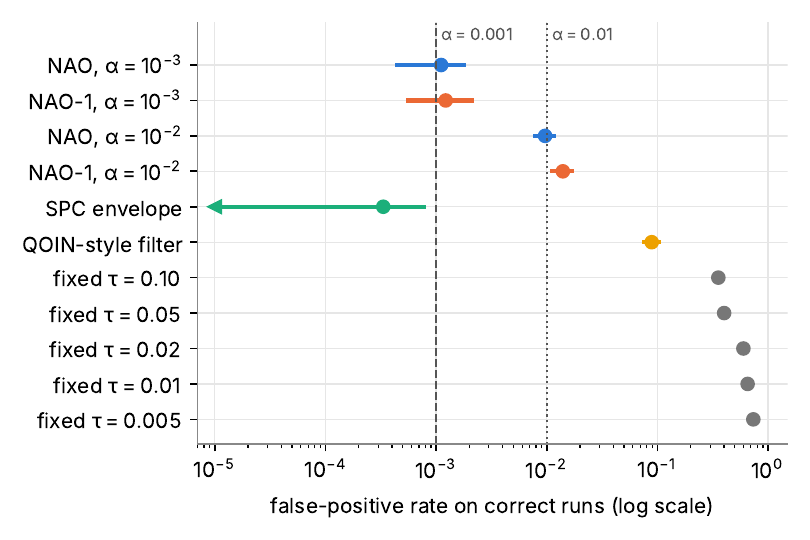}
\Description{Horizontal interval plot on a logarithmic axis of the false-positive rate of each oracle on correct programs. NAO and NAO-1 lie near the nominal levels 0.001 and 0.01, the SPC envelope lies below them, the QOIN-style filter near 0.09, and the fixed thresholds between 0.36 and 0.74.}
\caption{RQ1: false-positive rate of each oracle on $9{,}000$ correct runs (point: observed rate; bar: 95\% cluster-bootstrap interval over calibration windows; log scale). The dashed and dotted lines mark the nominal levels $\alpha = 10^{-3}$ and $10^{-2}$. An arrow marks an interval whose lower bound is $0$, which a log scale cannot show (SPC envelope: $3$ false positives, interval $[0, 0.0008]$).}
\label{fig:soundness}
\end{figure}

\paragraph{Synthetic noise.}
Table~\ref{tab:rq1-soundness} and Figure~\ref{fig:soundness} report the false-positive rate of every oracle over $9{,}000$ executions of correct programs (nine benchmarks, five noise levels, $40$ windows of five runs per cell). NAO's rate is $0.0011$ at $\alpha = 10^{-3}$ and $0.0097$ at $10^{-2}$, and both cluster-bootstrap intervals contain the nominal level. NAO-1 is calibrated at $10^{-3}$ and slightly liberal at $10^{-2}$ ($0.0140$), because one reference run underestimates the support of the output distribution and therefore the threshold.

Neither programs nor windows hide liberal behavior. Per program, NAO's rate lies between $0$ and $0.003$ at $\alpha = 10^{-3}$ and between $0.005$ and $0.013$ at $10^{-2}$, and no exact binomial test against $\alpha$ is significant (smallest $p = 0.080$; Table~\ref{tab:app-fpr-program}). Per window, false positives do not cluster: at $\alpha = 10^{-3}$, $10$ of $1{,}800$ windows contain one and none contains two ($8.96$ and $0.02$ expected for independent runs), and at $10^{-2}$, $85$ contain one and one contains two ($86.5$ and $1.8$ expected). NAO-1 at $10^{-2}$ is the exception: it is liberal on the three eight-qubit benchmarks ($0.018$--$0.027$, $p \le 0.017$), and $19$ windows contain two or more false positives, because one run of a wider distribution misestimates the threshold for the whole window. Pooling removes this error.

Every fixed threshold is unsound, from $35.6\%$ false positives at $\tau = 0.1$ to $73.7\%$ at $\tau = 0.005$, because the divergence of a correct run from the ideal output grows with the noise and the circuit. The thresholds $0.05$ and $0.1$ are sound at very low and low noise but reject $22$--$34\%$ of the correct runs at medium noise and $67$--$89\%$ at the two highest levels (Table~\ref{tab:app-fpr-noise}): a fixed threshold is sound only above a noise-dependent floor that is not known in advance. The QOIN-style filter reduces the rate to $8.9\%$; it is sound up to medium noise and rejects $22\%$ at the two highest levels, where a per-outcome multiplicative correction no longer undoes the noise. The SPC envelope has the lowest rate ($0.0003$), the expected behavior of a conservative threshold whose cost in sensitivity RQ2 and RQ4 quantify.

\begin{table}[t]
\centering
\caption{RQ1: false-positive rate on correct programs ($9$ benchmarks, $5$ noise levels, $40$ calibration windows of $5$ runs per cell). CI: 95\% cluster bootstrap over windows. For NAO and NAO-1, the target is the nominal level $\alpha$; for the baselines, lower is better.}
\label{tab:rq1-soundness}
\small
\begin{tabular}{lrrr}
\toprule
\textbf{Oracle} & \textbf{FP / tests} & \textbf{FPR} & \textbf{95\% CI} \\
\midrule
NAO ($\alpha = 10^{-3}$)   & 10/9000   & 0.0011 & [0.0004, 0.0018] \\
NAO ($\alpha = 10^{-2}$)   & 87/9000   & 0.0097 & [0.0078, 0.0117] \\
NAO-1 ($\alpha = 10^{-3}$) & 11/9000   & 0.0012 & [0.0006, 0.0021] \\
NAO-1 ($\alpha = 10^{-2}$) & 126/9000  & 0.0140 & [0.0112, 0.0170] \\
\midrule
SPC envelope               & 3/9000    & 0.0003 & [0.0000, 0.0008] \\
QOIN-style filter          & 800/9000  & 0.0889 & [0.0761, 0.1028] \\
Fixed $\tau = 0.1$         & 3200/9000 & 0.3556 & [0.3333, 0.3778] \\
Fixed $\tau = 0.05$        & 3611/9000 & 0.4012 & [0.3784, 0.4238] \\
Fixed $\tau = 0.02$        & 5391/9000 & 0.5990 & [0.5760, 0.6213] \\
Fixed $\tau = 0.01$        & 5886/9000 & 0.6540 & [0.6317, 0.6750] \\
Fixed $\tau = 0.005$       & 6630/9000 & 0.7367 & [0.7158, 0.7558] \\
\bottomrule
\end{tabular}
\end{table}

\paragraph{Thresholds.}
The bootstrap threshold of NAO-1 agrees with the closed form of Equation~\eqref{eq:chi2-threshold} to within $12\%$ in all 45 cells, and pooling lowers it by $28$--$46\%$ (Table~\ref{tab:rq1-thresholds}). The SPC envelope sits at its floor $\tau_{\min} = 0.02$ in $36$ of the $45$ cells, $14$ to $25$ times above the NAO threshold on the five small benchmarks, which explains both its low false-positive rate and its lower sensitivity. The \emph{dispersion ratio}, the observed mean pairwise divergence among calibration runs divided by its null expectation, is $0.986$ (median over cells; range $0.825$--$1.084$): on a stationary simulator, run-to-run variability is the finite-shot variability of the null model.

\paragraph{Extended suite.}
On the $32$ QASMBench programs (Table~\ref{tab:ext-suite}), NAO's rate is $0.0013$ over $8{,}000$ correct runs (interval $[0.0005, 0.0021]$), and no program exceeds $5\%$. The fixed thresholds reject $50$--$82\%$ of the correct runs and are unsound on $28$ to $30$ programs; the QOIN-style filter has $5.4\%$ false positives and the noise-calibrated global threshold $12.7\%$, each concentrated on a few programs. Soundness therefore holds on programs not used to design the oracle, including programs with partial measurement and up to $1024$ possible outcomes.

\begin{table}[t]
\centering
\caption{Extended suite of $32$ QASMBench programs (RQ1 and RQ4): $8{,}000$ correct runs and $163$ non-equivalent injected faults, five noise levels, ten windows per program and level; all oracles judge the same executions. Unsound programs: programs with FPR $> 0.05$. $\uparrow$: higher is better; $\downarrow$: lower is better; for NAO and NAO-1 the FPR target is $\alpha = 10^{-3}$.}
\label{tab:ext-suite}
\small
\setlength{\tabcolsep}{4pt}
\begin{tabular}{lrrrrrr}
\toprule
\textbf{Oracle} & \textbf{FPR} & \textbf{Unsound programs $\downarrow$} & \textbf{Recall $\uparrow$} & \textbf{Precision $\uparrow$} & \textbf{F1 $\uparrow$} \\
\midrule
NAO                               & 0.0013 & 0/32  & 0.776 & 0.998 & \textbf{0.873} \\
NAO-1                             & 0.0009 & 0/32  & 0.760 & 0.999 & 0.863 \\
SPC envelope                      & 0.0000 & 0/32  & 0.660 & 1.000 & 0.795 \\
Fixed $\tau = 0.01$               & 0.8181 & 30/32 & 0.995 & 0.553 & 0.711 \\
Fixed $\tau = 0.05$               & 0.6230 & 30/32 & 0.976 & 0.615 & 0.754 \\
Fixed $\tau = 0.1$                & 0.5031 & 28/32 & 0.905 & 0.647 & 0.754 \\
QOIN-style filter                 & 0.0544 & 3/32  & 0.549 & 0.911 & 0.685 \\
Noise-calibrated global threshold & 0.1265 & 6/32  & 0.529 & 0.810 & 0.640 \\
\bottomrule
\end{tabular}
\end{table}

\paragraph{Device snapshots.}
Under the fixed compilation policy, NAO is sound on all three snapshots (Table~\ref{tab:rq1-snapshots}; FPR $\le 0.003$). Under the random policy, layout and routing change between runs, so correct runs are no longer samples of one distribution: NAO-1 then reaches $10$--$11\%$ false positives, whereas NAO stays at $\le 0.005$ because its homogeneity test flags the over-dispersed calibrations ($20$--$25\%$ of the windows, against $1\%$ under the fixed policy; $73$--$93\%$ on QAOA-3, whose layouts produce visibly different outputs) and widens the threshold. The fixed thresholds reach $14$--$80\%$ false positives.

\begin{table}[t]
\centering
\caption{RQ1: false-positive rate on device-snapshot backends ($15$ windows of $5$ correct runs per benchmark; Bell, GHZ-3, QFT-3, QAOA-3 and VQE-3 on all backends, and Bell-8, GHZ-8 and QAOA-8 on FakeKolkataV2). Drift: fraction of windows in which the homogeneity test of NAO rejects. Lower FPR is better.}
\label{tab:rq1-snapshots}
\small
\setlength{\tabcolsep}{4pt}
\begin{tabular}{llrrrrrrr}
\toprule
\textbf{Backend} & \textbf{Compilation} & \textbf{NAO} & \textbf{NAO-1} & \textbf{SPC} & \textbf{$\tau{=}0.005$} & \textbf{$\tau{=}0.05$} & \textbf{$\tau{=}0.1$} & \textbf{Drift} \\
\midrule
FakeManilaV2  & fixed  & 0.000 & 0.000 & 0.000 & 0.800 & 0.200 & 0.000 & 0.01 \\
              & random & 0.005 & 0.107 & 0.000 & 0.800 & 0.200 & 0.000 & 0.24 \\
FakeLagosV2   & fixed  & 0.003 & 0.000 & 0.000 & 0.800 & 0.800 & 0.600 & 0.01 \\
              & random & 0.005 & 0.107 & 0.000 & 0.800 & 0.800 & 0.600 & 0.20 \\
FakeKolkataV2 & fixed  & 0.002 & 0.000 & 0.000 & 0.628 & 0.137 & 0.000 & 0.01 \\
              & random & 0.002 & 0.100 & 0.000 & 0.650 & 0.142 & 0.012 & 0.25 \\
\bottomrule
\end{tabular}
\end{table}

\paragraph{Coherent noise.}
Under coherent over-rotation errors, the reference experiences the same errors as the program under test. NAO rejects $1$ of $1{,}800$ correct runs (interval $[0, 0.0017]$), and the homogeneity test flags at most $5\%$ of the windows of any benchmark. The fixed thresholds reject $46.6\%$ ($\tau = 0.01$), $38.8\%$ ($0.02$) and $5.5\%$ ($0.05$) of the correct runs, and $\tau = 0.1$ is sound but detects only $82\%$ of the faults. On $2{,}000$ non-equivalent fault tests, NAO has recall $0.981$, precision $0.999$ and F1 $0.990$, against $0.979$ for the SPC envelope and at most $0.922$ for the fixed thresholds. NAO also rejects $59$ of $640$ runs of equivalent faults, because an inserted gate brings its own over-rotation: the noise-amplifying effect of RQ2 in coherent form.

\paragraph{Monte Carlo stability of the bootstrap threshold.}
At $\alpha = 10^{-3}$ and $B = 4000$, only about four bootstrap draws lie beyond the threshold. For one window of each core benchmark at low, medium and high noise ($27$ cells), the threshold was re-estimated $30$ times for each $B$ from a fixed plug-in estimate and compared with a reference null of $400{,}000$ draws (Table~\ref{tab:rq1-bootstrap}). At $B = 4000$, the implied false-positive rate averages $0.00101$, but its 5th--95th percentile range is $0.17$--$2.37$ times the nominal level: the Monte Carlo error averages out over windows, which is why RQ1 observes $0.0011$, but a single window's rate can be well below or more than twice the nominal level. At $B = 40{,}000$, the range narrows to $0.66$--$1.44$ times $\alpha$ at about three seconds per window for $256$ outcomes, negligible against the QPU time of a calibration; testers who need the level per window should use the larger $B$. Experiments with $B = 2000$ or $3000$ lie between the rows for $1000$ and $4000$.

\begin{table}[t]
\centering
\caption{RQ1: Monte Carlo stability of the bootstrap threshold at $\alpha = 10^{-3}$ ($27$ cells: nine core benchmarks at low, medium and high noise; $30$ re-estimations per cell and $B$; reference null of $400{,}000$ draws). Implied FPR: the false-positive rate of an estimated threshold under the reference null. Target: $\alpha = 0.001$.}
\label{tab:rq1-bootstrap}
\small
\begin{tabular}{rrrr}
\toprule
$B$ & \textbf{Relative s.d.\ of threshold, median (max)} & \textbf{Implied FPR, mean} & \textbf{Implied FPR, 5th--95th percentile range} \\
\midrule
$1000$     & $0.107$ ($0.204$) & $0.00101$ & $0.00001$--$0.00447$ \\
$4000$     & $0.047$ ($0.079$) & $0.00101$ & $0.00017$--$0.00237$ \\
$40{,}000$ & $0.014$ ($0.025$) & $0.00102$ & $0.00066$--$0.00144$ \\
\bottomrule
\end{tabular}
\end{table}

\paragraph{Calibration under sparse supports.}
We distinguish the \emph{possible} outcomes of an $m$-qubit register ($2^m$), the \emph{occupied} outcomes ($\pi_x > 0$) and the outcomes \emph{observed} in a run. Table~\ref{tab:rq1-sparsity} tests the threshold where shots per outcome are few, by sampling only: test and reference runs are multinomial samples of one distribution with $d = 8$ to $16{,}384$ occupied outcomes and three shapes (flat, skewed and head-tail), so every rejection is a false positive. Down to $16$ shots per occupied outcome, the rate stays close to the nominal level for all shapes: at $\alpha = 10^{-2}$, $0.005$--$0.017$ for NAO and $0.002$--$0.023$ for NAO-1. At $4$ shots per occupied outcome it becomes liberal, depending on the shape: $0.015$ (flat), $0.098$ (skewed) and $0.264$ (head-tail) at $\alpha = 10^{-2}$, and $0.062$ for head-tail at $10^{-3}$. With many outcomes, the null distribution is narrow relative to its mean, so a small bias of the plug-in estimate moves the rate by several times $\alpha$. With fewer shots than outcomes, the unseen-outcome correction overshoots and the oracle becomes conservative for the skewed and head-tail shapes but stays liberal for the flat shape ($0.062$ at $4096$ outcomes).

A tester does not know the occupied support, so the boundary must be stated in a quantity computable from the reference. The Good--Turing unseen mass of the reference does not separate calibrated from liberal windows; the \emph{shots per observed outcome} of one run of $P$ (shots divided by the distinct outcomes observed) does. In every window with $16$ or more, across all three shapes, NAO's rate is $0.0003$--$0.0005$ at $\alpha = 10^{-3}$, $0.009$ at $10^{-2}$ and $0.046$--$0.052$ at $0.05$ ($1{,}800$ windows); in windows with fewer than $8$, it is $0.024$--$0.063$ at $10^{-2}$, liberal or, at the largest supports, conservative ($1{,}800$ windows); no window fell between $8$ and $16$. In these experiments the nominal level therefore held with at least about $16$ shots per observed outcome and was not reliably controlled below $8$. This is an empirical operating criterion for the evaluated shapes, shot budget ($N = 4096$, $K = 8$) and bootstrap size ($B = 2000$), not a certificate of calibration; a tester can check it after the first calibration run. For a full register it means $8$ qubits at $4096$ shots, but concentrated outputs meet it at larger widths: the three QASMBench programs with $9$ or $10$ qubits produced one false positive in $750$ correct runs, and GHZ circuits remain sound up to $16$ qubits (RQ4), whereas QAOA-12, which occupies almost all $4096$ outcomes, does not. Below the boundary, more shots or the marginal comparison of RQ4 are needed.

\begin{table}[t]
\centering
\caption{RQ1: observed false-positive rate under sparse supports ($N = 4096$ shots; $1000$ correct test runs per cell; NAO pools $K = 8$ calibration runs, NAO-1 uses one). Flat: $\pi \sim$ Dirichlet$(1, \dots, 1)$; skewed: Zipf with exponent $1.1$; head-tail: four outcomes with $80\%$ of the mass (weights $1, \frac{1}{2}, \frac{1}{3}, \frac{1}{4}$) and a uniform tail. N1: NAO-1. Target: the nominal level $\alpha$.}
\label{tab:rq1-sparsity}
\footnotesize
\setlength{\tabcolsep}{3pt}
\begin{tabular}{rrrrrrrrrrr}
\toprule
& & \multicolumn{3}{c}{\textbf{Flat}} & \multicolumn{3}{c}{\textbf{Skewed}} & \multicolumn{3}{c}{\textbf{Head-tail}} \\
\cmidrule(lr){3-5}\cmidrule(lr){6-8}\cmidrule(lr){9-11}
& & \multicolumn{2}{c}{$\alpha{=}0.01$} & $0.05$ & \multicolumn{2}{c}{$\alpha{=}0.01$} & $0.05$ & \multicolumn{2}{c}{$\alpha{=}0.01$} & $0.05$ \\
\cmidrule(lr){3-4}\cmidrule(lr){5-5}\cmidrule(lr){6-7}\cmidrule(lr){8-8}\cmidrule(lr){9-10}\cmidrule(lr){11-11}
\textbf{$d$} & \textbf{$N/d$} & NAO & N1 & NAO & NAO & N1 & NAO & NAO & N1 & NAO \\
\midrule
8      & 512  & 0.005 & 0.014 & 0.042 & 0.006 & 0.013 & 0.048 & 0.009 & 0.011 & 0.051 \\
64     & 64   & 0.008 & 0.009 & 0.046 & 0.017 & 0.017 & 0.061 & 0.010 & 0.002 & 0.062 \\
256    & 16   & 0.010 & 0.018 & 0.046 & 0.008 & 0.014 & 0.049 & 0.008 & 0.023 & 0.042 \\
1024   & 4    & 0.015 & 0.054 & 0.075 & 0.098 & 0.058 & 0.281 & 0.264 & 0.006 & 0.514 \\
4096   & 1    & 0.062 & 0.178 & 0.214 & 0.000 & 0.000 & 0.000 & 0.000 & 0.000 & 0.000 \\
16384  & 0.25 & 0.043 & 0.067 & 0.150 & 0.000 & 0.000 & 0.000 & 0.000 & 0.000 & 0.000 \\
\bottomrule
\end{tabular}
\end{table}

\paragraph{IBM Quantum hardware.}
In the probe on \texttt{ibm\_fez}, the divergences of the test runs from the reference ($\le 0.0015$) stayed below the closed-form NAO-1 thresholds at both levels, whereas a fixed threshold of $0.005$ against the ideal output rejected all six runs, because a correct run lay $0.019$--$0.035$ from the ideal output. In H1 on \texttt{ibm\_kingston}, NAO and NAO-1 rejected none of the $30$ correct runs of GHZ-3, QAOA-3 and Bell-8 (Wilson $95\%$ upper bound $0.11$). In HM (Table~\ref{tab:hm}), NAO rejected none of the $20$ correct runs of the small programs, none of the $16$ of the larger programs and none of the $16$ runs of the fixed real-bug programs, whereas $\tau = 0.005$ and $0.02$ rejected $10$ and $4$, $16$ and $12$, and $13$ and $4$; on Bell-8 and GHZ-8, a correct run lay $0.061$--$0.085$ from the ideal output. The threshold $\tau = 0.05$ was sound on the small programs but not on the larger ones. QAOA-8 occupied all $256$ outcomes, the boundary of $16$ shots per observed outcome, and NAO remained sound on it. The homogeneity test flagged no calibration on hardware. These counts cannot estimate a rate of order $10^{-3}$; they test whether the contrast between the self-baseline and the ideal reference observed in simulation holds on the device.

\begin{table}[t]
\centering
\caption{Mini campaign HM on \texttt{ibm\_kingston}: rejected / tested runs per oracle. Small programs: Bell, GHZ-3, QFT-3, QAOA-3 and VQE-3 (first job). Larger programs: Bell-8, GHZ-8, QAOA-8 and HWE-6. Real bugs: the eight observable Bugs4Q bugs (runs of the fixed and of the buggy programs). Faults: non-equivalent injected faults and semantic mutants. NA, NN: noise-amplifying and noise-neutral mutants; NA rejections have no preferred direction (Section~\ref{subsec:rq2-results}). $\uparrow$: higher is better; $\downarrow$: lower is better; --: not evaluated.}
\label{tab:hm}
\footnotesize
\setlength{\tabcolsep}{3pt}
\begin{tabular}{lrrrrrrrr}
\toprule
& \multicolumn{4}{c}{\textbf{Small programs}} & \multicolumn{2}{c}{\textbf{Larger programs}} & \multicolumn{2}{c}{\textbf{Real bugs}} \\
\cmidrule(lr){2-5}\cmidrule(lr){6-7}\cmidrule(lr){8-9}
\textbf{Oracle} & \textbf{Correct $\downarrow$} & \textbf{Faults $\uparrow$} & \textbf{NA} & \textbf{NN $\downarrow$} & \textbf{Correct $\downarrow$} & \textbf{Faults $\uparrow$} & \textbf{Fixed $\downarrow$} & \textbf{Buggy $\uparrow$} \\
\midrule
NAO                   & 0/20  & 23/23 & 4/10 & 0/5 & 0/16  & 19/20 & 0/16  & 16/16 \\
NAO-1                 & 1/20  & 23/23 & 4/10 & 1/5 & 0/16  & 19/20 & 0/16  & 16/16 \\
SPC envelope          & 0/20  & 23/23 & 0/10 & 0/5 & --    & --    & --    & -- \\
Fixed $\tau = 0.005$  & 10/20 & 23/23 & 5/10 & 3/5 & 16/16 & 20/20 & 13/16 & 16/16 \\
Fixed $\tau = 0.02$   & 4/20  & 23/23 & 2/10 & 1/5 & 12/16 & 19/20 & 4/16  & 16/16 \\
Fixed $\tau = 0.05$   & 0/20  & 21/23 & 1/10 & 0/5 & 8/16  & 18/20 & 0/16  & 16/16 \\
\bottomrule
\end{tabular}
\end{table}

\begin{tcolorbox}[colback=gray!8, colframe=gray!50, title=Answer to RQ1]
Under the evaluated conditions, the noise-aware oracle attains its nominal false-positive rate on average over windows: $0.0011$ at $\alpha = 10^{-3}$ and $0.0097$ at $10^{-2}$ over $9{,}000$ correct runs, and $0.0013$ over $8{,}000$ correct runs of $32$ QASMBench programs, none above $5\%$. It remains sound under coherent errors ($1$ of $1{,}800$) and on device snapshots, including under recompilation, where its homogeneity test prevents the $10\%$ rate of the single-reference variant. Fixed thresholds produce $36$--$74\%$ false positives, a QOIN-style filter $8.9\%$ and, on the QASMBench programs, a globally calibrated threshold $12.7\%$. The level held whenever one reference run had about $16$ or more shots per observed outcome, an empirical criterion a tester can check; below $8$ it was not reliably controlled (up to $0.264$ at $\alpha = 10^{-2}$), and the range between was not sampled. On an IBM Heron device, NAO rejected none of $82$ correct runs of $17$ programs, while fixed thresholds rejected up to all correct runs of the eight-qubit programs.
\end{tcolorbox}

\subsection{RQ2: Equivalent Mutants, Taxonomy and Mutation Score}
\label{subsec:rq2-results}

\paragraph{Equivalent mutants.}
Table~\ref{tab:rq2-census} reports the fraction of raw operator draws that are equivalent to the original. On $|0\cdots0\rangle$, $69.7\%$ of T-gate, $48.1\%$ of CNOT, $28.2\%$ of rotation and $24.3\%$ of X insertions are equivalent: a T gate before measurement changes only a relative phase unless later gates convert it into a population change, and a CNOT whose control is $|0\rangle$ is the identity. Over four inputs, the CNOT rate falls to $20.9\%$; the other rates barely change. Subtractive operators rarely produce equivalent mutants ($0$--$9.7\%$). Placement is decisive: the same insertions placed after the final measurement are equivalent in $1350/1350$ draws, against $636/1350$ before it. Among the injected faults, $16$ of $66$ applicable injections are equivalent on $|0\cdots0\rangle$, $12$ of them because they change the unitary but not the measured output.

\begin{table}[t]
\centering
\caption{RQ2: fraction of raw operator draws that are equivalent mutants (exact statevector labels; 100 draws per operator and applicable benchmark). $|0\cdots0\rangle$: fraction of draws whose ideal output equals the original's on the all-zero input state, the input of all noisy executions; 4 inputs: fraction equal on all four tested inputs (the all-zero state and three random computational-basis states). The noise-amplifying and noise-neutral operators are equivalent by design; for semantic operators, a lower rate means fewer draws are wasted.}
\label{tab:rq2-census}
\small
\begin{tabular}{llrrr}
\toprule
\textbf{Class (family)} & \textbf{Operator} & \textbf{Draws} & \textbf{$|0\cdots0\rangle$} & \textbf{4 inputs} \\
\midrule
Semantic (additive)    & \texttt{insert\_t}             & 900 & 0.697 & 0.697 \\
                       & \texttt{insert\_cnot}          & 900 & 0.481 & 0.209 \\
                       & \texttt{insert\_rotation}      & 900 & 0.282 & 0.267 \\
                       & \texttt{insert\_x}             & 900 & 0.243 & 0.243 \\
Semantic (subtractive) & \texttt{remove\_gate}          & 900 & 0.097 & 0.066 \\
                       & \texttt{remove\_1q\_gate}      & 900 & 0.000 & 0.000 \\
                       & \texttt{rotation\_to\_identity}& 400 & 0.000 & 0.000 \\
\midrule
Noise-amplifying       & \texttt{inverse\_pair\_1q}, \texttt{cx\_pair}, \texttt{rotation\_pair} & 2700 & 1.000 & 1.000 \\
Noise-neutral          & \texttt{commute\_swap}, \texttt{cx\_to\_h\_cz\_h} & 1300 & 1.000 & 1.000 \\
\bottomrule
\end{tabular}
\end{table}

\paragraph{Mutation score.}
Table~\ref{tab:rq2-mutation} reports the rejection rate of each class (with Wilson intervals in Figure~\ref{fig:mutation-by-class}). NAO rejects $93.8\%$ of the semantic mutants (NAMS, Definition~\ref{def:nams}), NAO-1 $91.9\%$ and the SPC envelope $78.9\%$. Every semantic mutant is non-equivalent, so the remaining $6.2\%$ are genuine misses; they concentrate at high noise (NAMS $0.978$, $0.974$ and $0.863$ at low, medium and high noise) and on small changes such as \texttt{insert\_t} ($0.738$) and \texttt{rotation\_to\_identity} ($0.826$; Table~\ref{tab:app-operator}). The original program passed in every cell.

Noise-neutral mutants are almost never rejected ($2/405$). Noise-amplifying mutants are rejected in $20.7\%$ of cases, increasingly with noise ($0.148$, $0.222$, $0.252$) and almost only through \texttt{cx\_pair} ($53.8\%$, against $0\%$ for single-qubit and rotation pairs). These rejections are not statistical errors: a CNOT pair adds two-qubit gate error, so the mutant's noisy distribution genuinely differs, and a sensitive differential oracle detects it. They are, however, false alarms with respect to the program's semantics, the case that attribution (RQ3) must resolve. The SPC envelope rejects fewer ($6.4\%$) only because it is less sensitive to any difference.

\begin{table}[t]
\centering
\caption{RQ2: rejection rate by taxonomy class ($27$ cells; $810$ semantic mutants and $405$ of each semantics-preserving class). For the semantic class this is the noise-aware mutation score (NAMS, higher is better); for the two semantics-preserving classes it is a false-rejection rate (lower is better).}
\label{tab:rq2-mutation}
\small
\begin{tabular}{lrrr}
\toprule
\textbf{Oracle} & \textbf{Semantic (NAMS $\uparrow$)} & \textbf{Noise-amplifying $\downarrow$} & \textbf{Noise-neutral $\downarrow$} \\
\midrule
NAO          & 760/810 = 0.938 & 84/405 = 0.207 & 2/405 = 0.005 \\
NAO-1        & 744/810 = 0.919 & 75/405 = 0.185 & 5/405 = 0.012 \\
SPC envelope & 639/810 = 0.789 & 26/405 = 0.064 & 0/405 = 0.000 \\
\bottomrule
\end{tabular}
\end{table}

\paragraph{Taxonomy.}
The taxonomy is supported if the three classes have different rejection probabilities under the oracle. Table~\ref{tab:rq2-taxonomy} tests every pair at three levels of conservativeness (Section~\ref{subsec:stats}): per-mutant Fisher tests, the CMH test stratified by cell, which conditions on the shared calibration window, and the cell-level sign-flip test, which treats each cell as one observation.

The semantic class differs from both semantics-preserving classes in all $27$ cells, with large effects ($\delta \ge 0.75$), CMH $p < 10^{-30}$ at every noise level, and the minimum attainable cell-level $p$ ($0.0039$). The noise-amplifying and noise-neutral classes differ under the CMH test at every level ($p \le 4.3 \times 10^{-6}$) and at the cell level when the $27$ cells are pooled ($p = 2.5 \times 10^{-5}$; higher rate in $17$ cells, equal in $10$, lower in none). Per level, the cell-level test is significant only at high noise ($p = 0.016$), because at low noise only $4$ of $9$ cells reject any noise-amplifying mutant. The effect is small ($\delta = 0.23$--$0.27$, intervals excluding zero and reaching at most $0.38$), as expected of a second-order effect that becomes visible only where the added gates contribute enough error. The three classes are therefore distinguishable, with the separation between the two semantics-preserving classes weaker and noise-dependent.

\begin{table}[t]
\centering
\caption{RQ2: tests of the taxonomy under NAO. Rates: rejection rates of the two classes compared. CMH: Cochran--Mantel--Haenszel test stratified by cell. Cells: cells in which the first class has a higher / equal / lower rejection rate; sign-flip: cell-level permutation test. $\delta$: Cliff's delta on $D/\tau$ with 95\% cluster-bootstrap interval over cells. Bonferroni level for three pairs: $0.0167$.}
\label{tab:rq2-taxonomy}
\footnotesize
\setlength{\tabcolsep}{3pt}
\begin{tabular}{llrrrrrr}
\toprule
\textbf{Noise} & \textbf{Contrast} & \textbf{Rates} & \textbf{Fisher $p$} & \textbf{CMH $p$} & \textbf{Cells $>$/$=$/$<$} & \textbf{Sign-flip $p$} & \textbf{$\delta$} \\
\midrule
low    & semantic vs amplifying & 0.978 / 0.148 & $2.5{\times}10^{-71}$  & $1.1{\times}10^{-65}$  & 9/0/0 & 0.004 & 0.98 [0.96, 1.00] \\
       & semantic vs neutral    & 0.978 / 0.000 & $2.6{\times}10^{-101}$ & $5.2{\times}10^{-82}$  & 9/0/0 & 0.004 & 0.99 [0.98, 1.00] \\
       & amplifying vs neutral  & 0.148 / 0.000 & $8.9{\times}10^{-7}$   & $4.3{\times}10^{-6}$   & 4/5/0 & 0.125 & 0.23 [0.15, 0.34] \\
\addlinespace
medium & semantic vs amplifying & 0.974 / 0.222 & $7.9{\times}10^{-60}$  & $1.2{\times}10^{-56}$  & 9/0/0 & 0.004 & 0.94 [0.90, 0.99] \\
       & semantic vs neutral    & 0.974 / 0.000 & $5.2{\times}10^{-100}$ & $3.3{\times}10^{-81}$  & 9/0/0 & 0.004 & 0.99 [0.97, 1.00] \\
       & amplifying vs neutral  & 0.222 / 0.000 & $3.0{\times}10^{-10}$  & $8.6{\times}10^{-9}$   & 6/3/0 & 0.031 & 0.27 [0.15, 0.38] \\
\addlinespace
high   & semantic vs amplifying & 0.863 / 0.252 & $1.8{\times}10^{-34}$  & $3.1{\times}10^{-38}$  & 9/0/0 & 0.004 & 0.75 [0.67, 0.84] \\
       & semantic vs neutral    & 0.863 / 0.015 & $2.6{\times}10^{-69}$  & $7.4{\times}10^{-63}$  & 9/0/0 & 0.004 & 0.94 [0.88, 0.98] \\
       & amplifying vs neutral  & 0.252 / 0.015 & $2.3{\times}10^{-9}$   & $1.4{\times}10^{-8}$   & 7/2/0 & 0.016 & 0.25 [0.13, 0.36] \\
\addlinespace
pooled & amplifying vs neutral  & 0.207 / 0.005 & $9.8{\times}10^{-25}$  & $1.7{\times}10^{-21}$  & 17/10/0 & $2.5{\times}10^{-5}$ & 0.24 [0.18, 0.31] \\
\bottomrule
\end{tabular}
\end{table}

\paragraph{Physical criterion after compilation.}
The tests above use the classes as assigned by operator. Table~\ref{tab:rq2-compiled} checks the physical criterion of Definition~\ref{def:taxonomy} for the $615$ noise-amplifying and $615$ noise-neutral mutants of the core and QASMBench programs, compiled as in the experiments for three targets: the synthetic-noise backend, a CNOT-native IBM snapshot (FakeKolkataV2) and a CZ-native IBM Heron snapshot (FakeKingston). The check counts compiled operations; it establishes exposure to additional gate error, which the rejection rates and the dose study measure. On the synthetic backend, every noise-amplifying mutant adds operations and no noise-neutral mutant adds a two-qubit gate, but \texttt{cx\_to\_h\_cz\_h} keeps its Hadamards in $173$ of $283$ mutants, which are therefore weakly noise-amplifying. They do not drive the contrasts: none of the $159$ core noise-neutral executions that add no operation is rejected, and with the class restricted to verified mutants the contrast with the noise-amplifying class remains significant on the QASMBench programs at every level (CMH $p \le 8.8 \times 10^{-11}$, sign-flip $p \le 2.7 \times 10^{-4}$) and on the core benchmarks when pooled (CMH $p = 1.4 \times 10^{-4}$). On the IBM targets, virtual $R_z$ gates make $Z$, $S$ and $T$ pairs add operations only where barriers prevent merging, and a few noise-neutral reorderings are routed differently. Class membership is therefore a property of the mutant and the target and should be verified on the target. In the archived hardware circuits, $16$ of the $18$ noise-amplifying mutants of H1 add physical operations and none of its $12$ noise-neutral mutants does; HM and H3 used only physical pairs. Because every engine is seeded identically, the single-qubit pairs drawn in the mutation analysis are $Y$, $Z$, $S$ and $T$ pairs.

\begin{table}[t]
\centering
\caption{RQ2: semantics-preserving mutants whose compiled circuit contains more physical operations than the compiled original (core benchmarks and QASMBench programs; in parentheses: of these, mutants with additional two-qubit gates). Synthetic: the Aer backend of the synthetic noise model, on which every gate is noisy. CX-native: FakeKolkataV2; CZ-native: FakeKingston; on both, $R_z$ is virtual. Expected: all for noise-amplifying operators, none for noise-neutral operators.}
\label{tab:rq2-compiled}
\footnotesize
\setlength{\tabcolsep}{4pt}
\begin{tabular}{llrrr}
\toprule
\textbf{Class} & \textbf{Operator} & \textbf{Synthetic} & \textbf{CX-native} & \textbf{CZ-native} \\
\midrule
Noise-amplifying & \texttt{cx\_pair}                      & 234/234 & 228/234 & 234/234 \\
                 & \texttt{inverse\_pair\_1q}, $Y$        & 76/76   & 76/76   & 76/76 \\
                 & \texttt{inverse\_pair\_1q}, $Z, S, T$  & 129/129 & 15/129  & 36/129 \\
                 & \texttt{rotation\_pair}                & 176/176 & 176/176 & 176/176 \\
\midrule
Noise-neutral    & \texttt{commute\_swap}                 & 0/228   & 6/228 (6)  & 10/228 (6) \\
                 & \texttt{cx\_to\_h\_cz\_h}              & 173/283 (0) & 28/283 (9) & 13/283 (9) \\
                 & \texttt{x\_to\_h\_z\_h}                & 0/104   & 0/104   & 7/104 (0) \\
\bottomrule
\end{tabular}
\end{table}

The two semantic families do not separate: their rejection rates are $0.978$ and $0.978$ at low noise, $0.970$ and $0.978$ at medium noise, and $0.815$ and $0.911$ at high noise ($p = 0.033$, not significant after correction; $\delta = -0.11$). Additive and subtractive operators are families within one class.

\paragraph{Noise-amplification dose.}
Whether a noise-amplifying mutant is rejected should depend on how much it changes the noisy output distribution, which depends on the error it adds and on the program. Table~\ref{tab:rq2-dose} confirms this: rejection rises with the number of inserted pairs and with noise, from $0.000$ for one single-qubit pair to $0.789$ for four CNOT pairs. The power model of Equation~\eqref{eq:power}, applied to each mutant's true noisy divergence, predicts these rejections (Brier score $0.028$ over $1{,}620$ mutants, against $0.246$ for a constant predictor; mutants predicted below $0.1$ are rejected in $0.9\%$ of cases, those above $0.9$ in $98.5\%$). The program matters as well: CNOT pairs are rejected in $98$--$100\%$ of cases on Bell, GHZ-3, Bell-8 and GHZ-8, whose concentrated outputs are visibly blurred by added error, but never on QFT-3 and in $2\%$ on QAOA-8, whose nearly uniform outputs depolarizing error barely changes. The weak separation between the semantics-preserving classes is therefore a property of noise amplification, not a defect of the taxonomy: the oracle rejects a noise-amplifying mutant as often as its added error is detectable at the shot budget.

\begin{table}[t]
\centering
\caption{RQ2: rejection of noise-amplifying mutants by dose ($k$ inserted inverse pairs; $90$ mutants per row, nine benchmarks). Observed: rejection rate under NAO. Predicted: mean rejection probability from the power model applied to each mutant's noisy divergence (estimated from $200{,}000$ shots). Lower observed rates are better from the tester's view, but the oracle is correct to detect a real distributional change; the question is whether its rejections are predictable.}
\label{tab:rq2-dose}
\small
\begin{tabular}{llrrrrrr}
\toprule
& & \multicolumn{2}{c}{\textbf{Low}} & \multicolumn{2}{c}{\textbf{Medium}} & \multicolumn{2}{c}{\textbf{High}} \\
\cmidrule(lr){3-4}\cmidrule(lr){5-6}\cmidrule(lr){7-8}
\textbf{Pairs} & $k$ & Observed & Predicted & Observed & Predicted & Observed & Predicted \\
\midrule
CNOT         & 1  & 0.422 & 0.359 & 0.633 & 0.660 & 0.722 & 0.708 \\
             & 2  & 0.489 & 0.501 & 0.733 & 0.753 & 0.756 & 0.769 \\
             & 4  & 0.644 & 0.671 & 0.789 & 0.817 & 0.789 & 0.805 \\
Single-qubit & 1  & 0.000 & 0.002 & 0.000 & 0.005 & 0.011 & 0.009 \\
             & 4  & 0.011 & 0.010 & 0.144 & 0.112 & 0.244 & 0.251 \\
             & 16 & 0.167 & 0.199 & 0.611 & 0.565 & 0.656 & 0.671 \\
\bottomrule
\end{tabular}
\end{table}

\paragraph{Extended suite.}
On the $32$ QASMBench programs, the original program failed NAO in one of $96$ cells (Linear-solver at high noise), which is excluded. NAO rejects $77.2\%$ of the $2{,}850$ semantic mutants ($93.5\%$, $80.5\%$ and $56.8\%$ at low, medium and high noise), $15.7\%$ of the noise-amplifying and $0.3\%$ of the noise-neutral mutants; the SPC envelope $60.8\%$, $1.3\%$ and $0\%$. With $32$ cells per level, every contrast is significant at every noise level under both the CMH test ($p \le 8 \times 10^{-14}$) and the cell-level test ($p \le 1.5 \times 10^{-4}$). The contrast between the semantics-preserving classes is again small ($\delta = 0.21$--$0.32$), whereas the semantic class separates from the noise-amplifying class with $\delta = 0.95$, $0.78$ and $0.51$ at low, medium and high noise.

\paragraph{IBM Quantum hardware.}
On \texttt{ibm\_kingston}, NAO rejected $4$ of $18$ noise-amplifying and none of $12$ noise-neutral mutants in H1, and $4$ of $10$ and none of $5$ in HM: the ordering of the simulation.

\paragraph{What the evidence supports.}
The evidence supports three claims to different degrees. \emph{The classes are well defined}: semantic membership is decided exactly for every mutant, and physical membership is checkable on any target by compiled gate counts. \emph{The classes behave differently in the evaluated data}: strongly for semantic against semantics-preserving mutants, and weakly but consistently for noise-amplifying against noise-neutral mutants, also with verified classes. \emph{The classes predict behavior on programs not used to design them}: the QASMBench programs reproduce every contrast, the power model predicts individual rejections, and the ordering recurs on hardware. This third claim rests on depolarizing noise, IBM device snapshots and one IBM device.

\begin{tcolorbox}[colback=gray!8, colframe=gray!50, title=Answer to RQ2]
Equivalent mutants are frequent: $24$--$70\%$ of raw draws of the gate-insertion operators are equivalent on $|0\cdots0\rangle$, and every insertion after the final measurement is. With exact labels, NAO kills $93.8\%$ of the semantic mutants, against $78.9\%$ for the SPC envelope. The three classes are distinguishable: semantic mutants separate from both semantics-preserving classes in every cell with large effects, and noise-amplifying from noise-neutral mutants conditionally on the window at every noise level, more weakly because the separation depends on the dose of added error and on the program, as the power model predicts (Brier score $0.028$). The separation persists with classes verified on compiled circuits, but membership depends on the target: on IBM devices, virtual $Z$, $S$ and $T$ pairs add no physical operation. Additive and subtractive operators form one class. On an IBM Heron device, NAO rejected $8$ of $28$ noise-amplifying and none of $17$ noise-neutral mutants.
\end{tcolorbox}

\subsection{RQ3: Bug--Noise Attribution}
\label{subsec:rq3-results}

Attribution is invoked only after the oracle rejects a program, so the operationally relevant cases are the $597$ that NAO rejects: $543$ bugs and $54$ noise effects, all noise-amplifying mutants. Table~\ref{tab:rq3-by-noise} gives their confusion matrices by noise level. At low and medium noise, linear ZNE labels $372$ of $377$ rejected bugs \textsc{Bug} (recall $0.98$--$0.99$, precision $0.984$--$1.000$) and $27$ of $30$ rejected noise effects \textsc{Noise} ($90\%$); quadratic ZNE labels $28$ correctly ($93\%$) at slightly lower bug recall. At high noise, noise recall falls to $3/24$ (linear) and $12/24$ (quadratic): the folded circuits at scales 3 and 5 approach the maximally mixed distribution, the extrapolation leaves a residual difference between the programs, and the null of the decision step, which contains sampling variation only, does not cover it (Section~\ref{subsec:attribution}). Over all $597$ cases, $\kappa$ is $0.564$ (linear) and $0.599$ (quadratic; Table~\ref{tab:rq3-attribution}), reflecting both the high-noise failures and the imbalance of the subset.

\begin{table}[!htb]
\centering
\caption{RQ3: attribution of the cases rejected by NAO, by noise level (positive class: \textsc{Bug}). TP/FN: bugs labeled \textsc{Bug}/\textsc{Noise}; FP/TN: noise effects labeled \textsc{Bug}/\textsc{Noise}. Higher is better in the last three columns.}
\label{tab:rq3-by-noise}
\small
\begin{tabular}{llrrrrrrr}
\toprule
\textbf{Method} & \textbf{Noise} & \textbf{TP} & \textbf{FN} & \textbf{FP} & \textbf{TN} & \textbf{Bug rec.} & \textbf{Bug prec.} & \textbf{Noise rec.} \\
\midrule
ZNE linear    & low    & 186 & 3  & 0  & 9  & 0.984 & 1.000 & 1.000 \\
              & medium & 186 & 2  & 3  & 18 & 0.989 & 0.984 & 0.857 \\
              & high   & 155 & 11 & 21 & 3  & 0.934 & 0.881 & 0.125 \\
\addlinespace
ZNE quadratic & low    & 182 & 7  & 0  & 9  & 0.963 & 1.000 & 1.000 \\
              & medium & 182 & 6  & 2  & 19 & 0.968 & 0.989 & 0.905 \\
              & high   & 148 & 18 & 12 & 12 & 0.892 & 0.925 & 0.500 \\
\bottomrule
\end{tabular}
\end{table}

For comparison with procedures evaluated on every case, Table~\ref{tab:rq3-attribution} also reports all $1{,}062$ cases regardless of the oracle verdict: linear ZNE reaches accuracy $0.920$ and $\kappa = 0.839$, and the quadratic variant is equivalent ($\kappa = 0.830$) at $50\%$ more executions. Re-execution triage never labels a noise case as a bug but misses $31.6\%$ of the bugs ($\kappa = 0.662$), because its fixed tolerance exceeds the divergence of many semantic changes, and it needs a noise model and a noiseless simulator.

\begin{table}[t]
\centering
\caption{RQ3: attribution against exact labels. Triggered: cases rejected by NAO ($543$ bugs, $54$ noise), on which attribution is used in practice. All cases: every case regardless of the oracle verdict ($582$ bugs, $480$ noise). Higher is better in every column. Executions: per attributed case.}
\label{tab:rq3-attribution}
\small
\setlength{\tabcolsep}{4pt}
\begin{tabular}{llrrrrrr}
\toprule
\textbf{Cases} & \textbf{Method} & \textbf{Accuracy} & \textbf{$\kappa$} & \textbf{Bug prec.} & \textbf{Bug rec.} & \textbf{Noise rec.} & \textbf{Executions} \\
\midrule
All       & ZNE linear         & 0.920 & 0.839 & 0.941 & 0.911 & 0.931 & 4 \\
          & ZNE quadratic      & 0.915 & 0.830 & 0.956 & 0.887 & 0.950 & 6 \\
          & Re-execution triage & 0.827 & 0.662 & 1.000 & 0.684 & 1.000 & 6\textsuperscript{a} \\
\midrule
Triggered & ZNE linear         & 0.933 & 0.564 & 0.956 & 0.971 & 0.556 & 4 \\
          & ZNE quadratic      & 0.925 & 0.599 & 0.973 & 0.943 & 0.741 & 6 \\
          & Re-execution triage & 0.754 & 0.328 & 1.000 & 0.729 & 1.000 & 6\textsuperscript{a} \\
\bottomrule
\multicolumn{8}{l}{\footnotesize \textsuperscript{a}Requires a scalable noise model and a noiseless simulator; not executable on hardware.}
\end{tabular}
\end{table}

\paragraph{Choice of extrapolation.}
On identical executions (Table~\ref{tab:rq3-extrapolation}), linear extrapolation has the highest agreement on all cases ($\kappa = 0.854$) and quadratic extrapolation the highest noise recall on triggered cases ($0.759$). Exponential extrapolation, which should remove the bias between circuits that decay at different rates, does not help: dividing by the estimated decay factor amplifies sampling noise and widens the null. On the device snapshots, all three reach $\kappa \ge 0.883$. We keep linear extrapolation as the default and recommend the quadratic variant when false bug reports cost more than six executions per case.

\paragraph{IBM Quantum hardware.}
In H3 and in HM on \texttt{ibm\_kingston}, linear ZNE attributed $10$ semantic and $10$ noise-amplifying mutants each, with accuracy $0.900$ and $\kappa = 0.800$ in both: every semantic mutant was labeled \textsc{Bug}, and $8$ of $10$ noise-amplifying mutants \textsc{Noise}; the four errors are noise-amplifying mutants of Bell and GHZ-3 whose extrapolated distance ($0.086$--$0.152$) exceeded the null quantile. To check the folding, the archived transpiled circuits of every attribution pair were compared at $s = 3$ and $s = 1$: in $44$ of $45$ pairs with two-qubit gates, the folded circuit has exactly three times the two-qubit gates and $2.98$--$3.03$ times the first-order error budget $\sum_g n_g e_g$. The exception, a GHZ-3 mutant of HM, was routed differently ($2.40$ times the two-qubit gates) and has the largest extrapolated distance of all HM cases. Matching counts validate the folding construction, not the effective noise scale, which idle time, crosstalk and coherent errors can change; on hardware, attribution depends on the transpiler preserving the folded structure, which a tester can check before trusting a verdict.

\begin{tcolorbox}[colback=gray!8, colframe=gray!50, title=Answer to RQ3]
Zero-noise extrapolation attributes the deviations the oracle reports without a noiseless simulator, at four executions per case, but reliably only where its extrapolation model holds. At low and medium noise, it labels $98$--$99\%$ of the rejected bugs \textsc{Bug} and $90\%$ (quadratic: $93\%$) of the rejected noise effects \textsc{Noise}. At high noise, where folding saturates the output, it labels most rejected noise effects \textsc{Bug} (noise recall $0.125$; quadratic $0.500$), and exponential extrapolation does not remedy this. Over all cases, $\kappa = 0.839$ (device snapshots: $0.918$); on an IBM Heron device, $\kappa = 0.800$ in H3 and HM, and the one folded circuit whose compiled structure fell short of the nominal factor produced the largest misattribution. ZNE attribution is a conditional diagnostic, not a general identifier of causes.
\end{tcolorbox}

\subsection{RQ4: Fault Detection, Power and Scaling}
\label{subsec:rq4-results}

\paragraph{Recall, precision and F1.}
Table~\ref{tab:rq4-faults} reports the confusion matrix on the $50$ non-equivalent injected faults ($2{,}500$ tests) and $1{,}350$ matched correct runs. NAO detects $88.8\%$ of the faults with precision $0.999$ and F1 $0.941$; NAO-1 detects $87.7\%$ and the SPC envelope $80.2\%$ (F1 $0.890$). The fixed thresholds reach higher recall (up to $0.985$) only by rejecting correct runs as well ($65.6\%$ at $\tau = 0.01$), and their F1 ($0.830$--$0.861$) is below NAO's. They also reject $25$--$47\%$ of the $800$ equivalent fault tests, against $1/800$ for NAO.

\begin{table}[t]
\centering
\caption{RQ4: detection of non-equivalent injected faults (positives, $2{,}500$ tests) and matched correct runs (negatives, $1{,}350$). TP/FN: detected/missed faults; FP/TN: rejected/accepted correct runs. $\uparrow$: higher is better; $\downarrow$: lower is better.}
\label{tab:rq4-faults}
\small
\begin{tabular}{lrrrrrrrr}
\toprule
\textbf{Oracle} & \textbf{TP} & \textbf{FN} & \textbf{FP} & \textbf{TN} & \textbf{Recall $\uparrow$} & \textbf{Precision $\uparrow$} & \textbf{F1 $\uparrow$} & \textbf{FPR $\downarrow$} \\
\midrule
NAO                 & 2221 & 279 & 2   & 1348 & 0.888 & 0.999 & \textbf{0.941} & 0.0015 \\
NAO-1               & 2192 & 308 & 1   & 1349 & 0.877 & 1.000 & 0.934 & 0.0007 \\
SPC envelope        & 2004 & 496 & 0   & 1350 & 0.802 & 1.000 & 0.890 & 0.0000 \\
Fixed $\tau = 0.01$ & 2462 & 38  & 886 & 464  & 0.985 & 0.735 & 0.842 & 0.6563 \\
Fixed $\tau = 0.02$ & 2431 & 69  & 803 & 547  & 0.972 & 0.752 & 0.848 & 0.5948 \\
Fixed $\tau = 0.05$ & 2303 & 197 & 544 & 806  & 0.921 & 0.809 & 0.861 & 0.4030 \\
Fixed $\tau = 0.1$  & 2115 & 385 & 480 & 870  & 0.846 & 0.815 & 0.830 & 0.3556 \\
\bottomrule
\end{tabular}
\end{table}

\begin{figure}[t]
\centering
\includegraphics[width=0.75\linewidth]{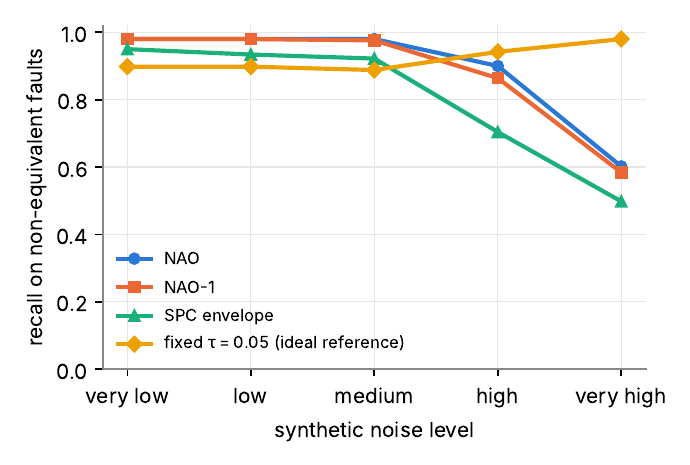}
\Description{Line chart of recall on non-equivalent injected faults against noise level for NAO, NAO-1, the SPC envelope and a fixed threshold. Recall stays near 0.98 up to medium noise and falls at high and very high noise, most steeply for the SPC envelope.}
\caption{RQ4: recall on non-equivalent injected faults by synthetic noise level ($\eta = 0.0005$, $0.001$, $0.005$, $0.02$, $0.05$; $500$ fault tests per level). The fixed threshold compares with the ideal output; its rising recall at high noise comes with rejecting most correct runs (Table~\ref{tab:rq4-faults}).}
\label{fig:recall-by-noise}
\end{figure}

\paragraph{Where faults are missed.}
NAO detects $98\%$ of the faults from very low to medium noise, $90.0\%$ at high and $60.2\%$ at very high noise (Figure~\ref{fig:recall-by-noise}). The misses at the two highest levels are faults whose effect noise has erased: both distributions approach the maximally mixed one, and their divergence falls below what $4096$ shots resolve. By injector, recall ranges from $0.77$ (\texttt{wrong\_angle}) to $0.97$ (\texttt{swap\_qubits}; Table~\ref{tab:app-injector}), so small changes are the hardest to detect. The SPC envelope loses most at high noise ($0.704$ and $0.498$), where its floor exceeds the divergence of many faults.

\begin{figure}[t]
\centering
\includegraphics[width=\linewidth]{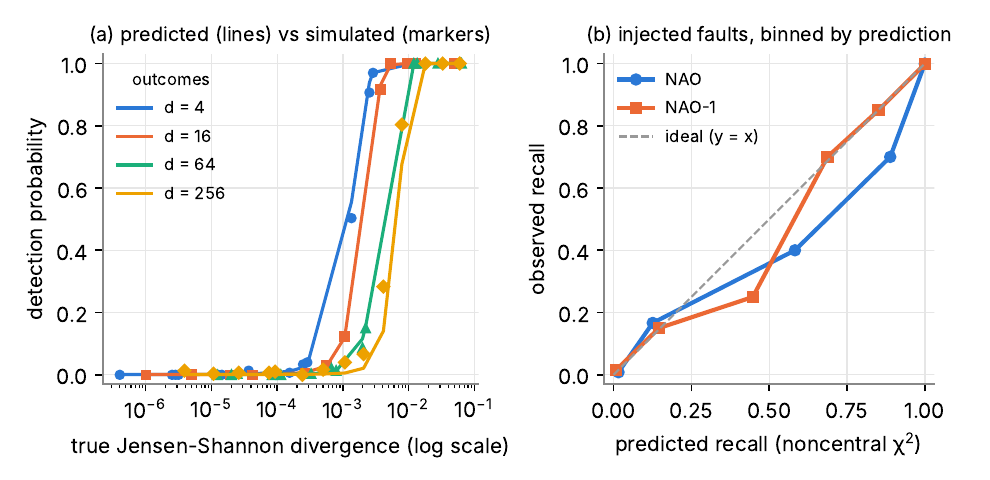}
\Description{Two panels. Left: detection probability against fault size for 4, 16, 64 and 256 outcomes, with predicted curves and observed markers close together. Right: observed recall against predicted power for injected faults, binned, lying close to the diagonal.}
\caption{RQ4: (a) detection probability predicted by the noncentral chi-square model (lines) and observed in simulation (markers) on synthetic distributions with $d$ outcomes; (b) predicted versus observed recall on the $2{,}500$ injected-fault tests, binned by predicted power; the dashed line is perfect agreement ($y = x$).}
\label{fig:power}
\end{figure}

\paragraph{Analytic power model.}
Figure~\ref{fig:power} compares the power model of Equation~\eqref{eq:power} with simulation. On synthetic distributions with $4$ to $256$ outcomes (panel a), predicted and observed detection probability differ by $0.011$ on average over $56$ points, at most $0.14$ at $256$ outcomes, where the chi-square approximation is least accurate. On the injected faults (panel b), faults predicted to be detected with probability above $0.9$ are detected in $99.8\%$ (NAO-1) and $100\%$ (NAO) of the tests, and those predicted below $0.1$ in $1.5\%$ and $0.8\%$. By noise level, NAO's mean predicted and observed recall differ by at most $0.013$, and by number of occupied outcomes by at most $0.007$; the Brier score is $0.006$, against $0.099$ for a constant predictor. The model has no fitted parameters, and each fault's divergence is estimated from separate $200{,}000$-shot executions, so this is an out-of-sample check, limited to the evaluated faults, programs and noise levels. The model predicts whether a fault of known size is detected, not which faults a program contains; it serves to plan shot budgets (Section~\ref{subsec:guidance}). Pooling $K = 8$ runs reduces the minimum detectable divergence at $80\%$ power by $44\%$, from $0.0019$ to $0.0011$ for four outcomes and from $0.0089$ to $0.0050$ for $256$ (Table~\ref{tab:rq4-mdd}).

\paragraph{Noise-calibrated global threshold.}
A threshold calibrated on a noise model of the device and shared by all programs, as proposed by Fortz et al.~\cite{fortz2026robust}, does not remove the problem of fixed thresholds. On the same executions in a separate run of the fault-detection design (Table~\ref{tab:rq4-global}), it has a false-positive rate of $0.108$, recall $0.602$ and F1 $0.725$, against $0.0007$, $0.888$ and $0.940$ for NAO. Its false positives concentrate on Bell-8 ($47\%$) and GHZ-8 ($50\%$), whose distance from the ideal output exceeds the $87.5$th percentile over programs, and its recall falls from $0.96$ at very low to $0.23$ at very high noise, because the threshold must grow to accommodate the noisiest programs. The threshold is also computed from the programs it is applied to. Leaving each program out of the calibration in turn, on fresh executions, raises its false-positive rate from $0.097$ to $0.221$ on the core benchmarks and from $0.124$ to $0.145$ on the QASMBench programs, with unchanged recall: Bell-8 and GHZ-8 are then rejected in $99$--$100\%$ of their correct runs. NAO uses no statistic across programs.

\paragraph{Ablation.}
Table~\ref{tab:rq4-ablation} adds the components of NAO one at a time on identical executions (a separate run, so row F differs from Table~\ref{tab:rq4-faults} by sampling variation). On synthetic noise, replacing the ideal reference by one self-baseline run (A $\to$ B) lowers the false-positive rate from $0.599$ to $0.192$, but a fixed threshold remains unsound, because the finite-shot variation between two runs depends on the program and the noise level. The SPC envelope (C) removes the false positives at the price of recall ($0.802$). The null-quantile threshold (D) keeps the rate nominal ($0.0013$) and raises F1 from $0.890$ to $0.934$, the largest gain of any component. Pooling (E) raises recall to $0.886$. The drift test (F) changes no verdict on synthetic noise, where calibrations are homogeneous, but is decisive under recompilation: without it, D and E reject $8.3\%$ and $12.1\%$ of the correct runs, because the variation between compilations is not part of the finite-shot null, and pooling makes this worse by narrowing the null. With it, the rate falls to $0.009$ at recall $0.961$, and F has the highest F1 ($0.976$). B and C are sound under recompilation, because the change between compilations of these small programs stays below $0.02$, but lose $4$--$5$ points of recall; whether a constant threshold is sound depends on the setting and cannot be known before it is measured.

\begin{table}[t]
\centering
\caption{RQ4: ablation of NAO. Each row adds one component to the row above, and all rows judge identical executions. Synthetic noise: nine core benchmarks at five noise levels ($450$ windows; $2{,}250$ correct runs, $2{,}500$ non-equivalent fault tests). Snapshots, recompilation: five small benchmarks on three device snapshots under the random compilation policy ($150$ windows; $750$ correct runs, $750$ non-equivalent fault tests). $\uparrow$: higher is better; $\downarrow$: lower is better.}
\label{tab:rq4-ablation}
\footnotesize
\setlength{\tabcolsep}{4pt}
\begin{tabular}{lllrrrrrr}
\toprule
& & & \multicolumn{3}{c}{\textbf{Synthetic noise}} & \multicolumn{3}{c}{\textbf{Snapshots, recompilation}} \\
\cmidrule(lr){4-6}\cmidrule(lr){7-9}
& \textbf{Reference} & \textbf{Threshold} & \textbf{FPR $\downarrow$} & \textbf{Recall $\uparrow$} & \textbf{F1 $\uparrow$} & \textbf{FPR $\downarrow$} & \textbf{Recall $\uparrow$} & \textbf{F1 $\uparrow$} \\
\midrule
A & Ideal (noiseless)   & Fixed $\tau = 0.02$                & 0.5987 & 0.970 & 0.773 & 0.4667 & 0.988 & 0.805 \\
B & 1 self-baseline run & Fixed $\tau = 0.02$                & 0.1916 & 0.838 & 0.834 & 0.0000 & 0.920 & 0.958 \\
C & 1 self-baseline run & SPC envelope                       & 0.0000 & 0.802 & 0.890 & 0.0000 & 0.915 & 0.955 \\
D & 1 self-baseline run & Null quantile (NAO-1)              & 0.0013 & 0.878 & 0.934 & 0.0827 & 0.977 & 0.949 \\
E & 8 pooled runs       & Null quantile                      & 0.0009 & 0.886 & 0.939 & 0.1213 & 0.979 & 0.932 \\
F & 8 pooled runs       & Null quantile + drift test (NAO)   & 0.0009 & 0.886 & 0.939 & 0.0093 & 0.961 & 0.976 \\
\bottomrule
\end{tabular}
\end{table}

\paragraph{Real bugs.}
Table~\ref{tab:rq4-realbugs} reports the ten real bugs from Bugs4Q. On the eight that change the ideal output on $|0\cdots0\rangle$, NAO detects $96.4\%$ of the buggy runs across five noise levels and two snapshots, with precision $0.994$, F1 $0.979$ and a false-positive rate of $0.0014$ on the fixed programs; all its misses occur at very high noise. The SPC envelope detects $89.6\%$ (F1 $0.945$), and the fixed thresholds detect $96$--$98\%$ but reject $42$--$77\%$ of the runs of the fixed programs (F1 $0.41$--$0.54$). ZNE attribution labels $160$ of $163$ attributed rejections of observable bugs \textsc{Bug}.

Two bugs do not change the ideal output on $|0\cdots0\rangle$: Terra \#5580 exchanges controls and targets of two controlled-$R_z$ gates, which changes only phases, and Terra \#5098 misspecifies a control state that does not fire on this input. NAO nevertheless rejects $82\%$ of their runs, because the buggy circuits compile to different physical gate sequences whose noisy outputs genuinely differ, the effect seen for noise-amplifying mutants. ZNE labels $16$ of the $36$ attributed rejections \textsc{Noise}; for \#5098, a \textsc{Bug} verdict is correct on other inputs.

\begin{table}[t]
\centering
\caption{RQ4: real bugs from Bugs4Q~\cite{zhao2023bugs4q}. JSD: ideal divergence between fixed and buggy program on $|0\cdots0\rangle$ (0: not observable on this input). Detection: fraction of buggy runs rejected over $70$ windows (five noise levels and two device snapshots). Higher detection is better for observable bugs.}
\label{tab:rq4-realbugs}
\footnotesize
\setlength{\tabcolsep}{4pt}
\begin{tabular}{llrrrrr}
\toprule
\textbf{Bug (Bugs4Q source)} & \textbf{Kind} & \textbf{Qubits} & \textbf{JSD} & \textbf{NAO} & \textbf{SPC} & \textbf{Fixed $0.05$} \\
\midrule
Aer \#664                       & measurement mapping & 3 & 1.000 & 1.00 & 1.00 & 1.00 \\
Stack Overflow 63283443         & qubit order         & 3 & 1.000 & 0.97 & 0.86 & 1.00 \\
Stack Overflow 62661255         & wrong gate          & 2 & 0.311 & 1.00 & 0.89 & 1.00 \\
Stack Overflow, inverse QFT     & wrong subroutine    & 3 & 1.000 & 0.91 & 0.86 & 1.00 \\
Stack Exchange 9                & measurement mapping & 3 & 1.000 & 1.00 & 1.00 & 1.00 \\
Stack Exchange 15               & wrong subroutine    & 4 & 0.581 & 0.86 & 0.71 & 1.00 \\
Stack Exchange (p.\,1--25) 9    & missing gates       & 4 & 0.829 & 0.97 & 0.86 & 0.86 \\
Stack Exchange (p.\,1--25) 19   & wrong initial state & 2 & 1.000 & 1.00 & 1.00 & 1.00 \\
Terra \#5580                    & qubit order         & 3 & 0     & 1.00 & 0.44 & 0.60 \\
Terra \#5098                    & control state       & 3 & 0     & 0.64 & 0.00 & 0.71 \\
\midrule
\multicolumn{4}{l}{Observable bugs: recall / precision / F1} & 0.964 / 0.994 / 0.979 & 0.896 / 1.000 / 0.945 & 0.982 / 0.330 / 0.494 \\
\multicolumn{4}{l}{FPR on runs of the fixed programs $\downarrow$} & 0.0014 & 0.0000 & 0.5314 \\
\bottomrule
\end{tabular}
\end{table}

\paragraph{Detection on device snapshots.}
On the device snapshots under the fixed policy, NAO detects $96.6\%$ of the semantic mutants and non-equivalent faults ($n = 381$) and rejects $2.2\%$ of the semantics-preserving mutants ($n = 325$); the SPC envelope detects $85.8\%$ ($1.5\%$), and $\tau = 0.05$ detects $89.8\%$ but rejects $33.5\%$.

\paragraph{Scaling.}
Table~\ref{tab:rq4-scaling} extends the evaluation to GHZ circuits of up to 16 qubits and QAOA circuits of up to 12 qubits at medium noise. Full-distribution NAO is sound on every GHZ circuit, whose output concentrates on few outcomes at any width, and detects every fault. On QAOA-12, whose output occupies almost all $4096$ outcomes with about one shot each, it rejects $2$ of $32$ correct runs ($0.062$), because the unseen-outcome correction and the chi-square approximation are no longer accurate. Marginal NAO, which compares all two-qubit marginals, is sound on every circuit at recall $1.000$; the SPC envelope is sound but loses recall on QAOA-12 ($0.700$). Full-distribution comparison is therefore limited by the ratio of shots to occupied outcomes rather than by the qubit count.

\begin{table}[t]
\centering
\caption{RQ4: scaling at $\eta = 0.005$ and $4096$ shots. Shots/outcome: $4096 / 2^n$. Full: NAO on the full distribution; marginal: NAO on all two-qubit marginals (Bonferroni). FPR: lower is better; recall: higher is better.}
\label{tab:rq4-scaling}
\small
\begin{tabular}{lrrrrrrr}
\toprule
& & & \multicolumn{3}{c}{\textbf{FPR $\downarrow$}} & \multicolumn{2}{c}{\textbf{Recall $\uparrow$}} \\
\cmidrule(lr){4-6}\cmidrule(lr){7-8}
\textbf{Circuit} & \textbf{Qubits} & \textbf{Shots/outcome} & \textbf{Full} & \textbf{Marginal} & \textbf{SPC} & \textbf{Full} & \textbf{SPC} \\
\midrule
GHZ     & 8  & 16    & 0.000 & 0.000 & 0.000 & 1.000 & 1.000 \\
GHZ     & 10 & 4     & 0.000 & 0.000 & 0.000 & 1.000 & 1.000 \\
GHZ     & 12 & 1     & 0.000 & 0.000 & 0.000 & 1.000 & 1.000 \\
GHZ     & 14 & 0.25  & 0.000 & 0.000 & 0.000 & 1.000 & 1.000 \\
GHZ     & 16 & 0.06  & 0.000 & 0.000 & 0.000 & 1.000 & 1.000 \\
QAOA    & 8  & 16    & 0.000 & 0.000 & 0.000 & 1.000 & 1.000 \\
QAOA    & 10 & 4     & 0.000 & 0.000 & 0.000 & 1.000 & 0.975 \\
QAOA    & 12 & 1     & 0.062 & 0.000 & 0.000 & 1.000 & 0.700 \\
\bottomrule
\end{tabular}
\end{table}

\paragraph{Extended suite.}
On the $163$ non-equivalent faults of the QASMBench programs (Table~\ref{tab:ext-suite}), NAO has recall $0.776$, precision $0.998$ and the highest F1 ($0.873$). Recall is lower than on the core benchmarks because the suite contains deeper programs: it is $0.99$--$1.00$ at very low and low noise and $0.39$ at very high noise, and at medium noise NAO detects $6\%$ of the fault runs of HHL-7 (depth $551$) and none of DNN-8 ($192$ two-qubit gates), whose noisy outputs are nearly uniform, whereas shallow programs such as Cat-state-4 keep recall $0.89$--$1.00$ even at high noise. The SPC envelope reaches F1 $0.795$, the QOIN-style filter $0.685$ and the global threshold $0.640$.

\paragraph{IBM Quantum hardware.}
In H1, NAO and NAO-1 detected all $78$ semantic mutants and non-equivalent faults of GHZ-3, QAOA-3 and Bell-8 (Wilson $95\%$ lower bound $0.95$). In HM (Table~\ref{tab:hm}), NAO detected all $23$ faults and semantic mutants of the small programs, $19$ of $20$ of the larger programs and all $16$ runs of the buggy real-bug programs; every fixed threshold either detected fewer or rejected correct runs as well. The one miss is a mutant of HWE-6 whose ideal divergence, $8.5 \times 10^{-5}$, is $33$ times smaller than the minimum detectable divergence for $64$ outcomes ($0.0028$), so the power model predicts it.

\begin{tcolorbox}[colback=gray!8, colframe=gray!50, title=Answer to RQ4]
Soundness does not cost sensitivity: NAO detects $88.8\%$ of non-equivalent faults with precision $0.999$ and the highest F1 of all oracles ($0.941$). A threshold calibrated globally on the device's noise model reaches F1 $0.725$ with $10.8\%$ false positives, and more on programs held out from its calibration. The ablation attributes the result to the oracle's components: the self-baseline reference removes most false positives, the null-quantile threshold gives the largest gain in F1, and the drift test keeps the oracle sound under recompilation ($0.009$ against $0.121$). The ranking holds on the QASMBench programs (F1 $0.873$) and on real bugs ($96.4\%$ of observable runs detected, F1 $0.979$). Misses concentrate at the two highest noise levels and on small changes, and the power model predicts detection accurately enough to plan shot budgets. Full-distribution comparison was sound on GHZ circuits up to 16 qubits and failed near one shot per occupied outcome (QAOA-12), where marginal comparison remained sound. On an IBM Heron device, NAO detected $120$ of $121$ faults and semantic mutants, missing one the power model predicts to be undetectable, and all runs of eight real bugs.
\end{tcolorbox}

\subsection{RQ5: Drift and Cost}
\label{subsec:rq5-results}

\begin{figure}[t]
\centering
\includegraphics[width=\linewidth]{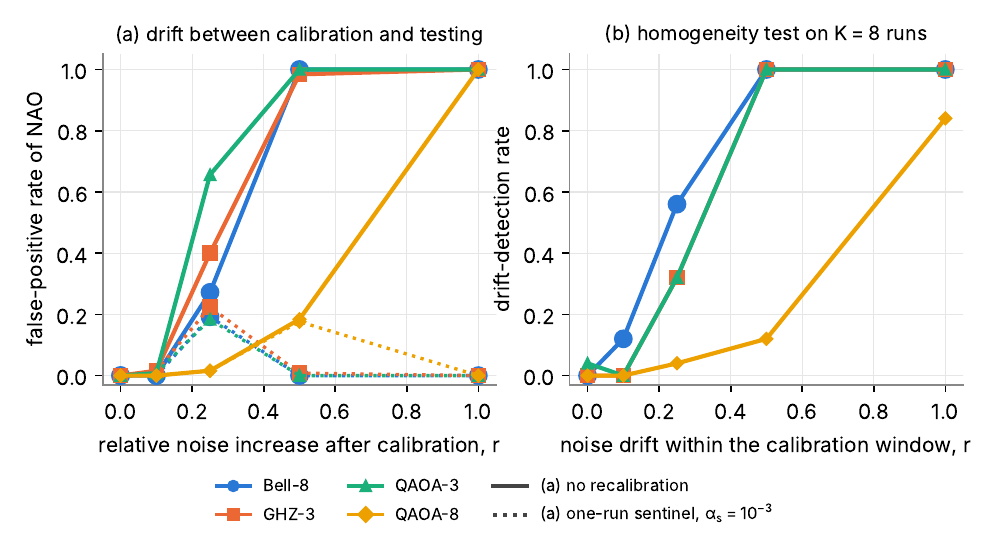}
\Description{Two panels. Left: false-positive rate of NAO after the noise level drifts by a relative amount r between calibration and testing, without recalibration and with a one-run sentinel, for four benchmarks. Right: fraction of calibration windows in which the homogeneity test detects drift that occurs within the window, against r.}
\caption{RQ5: (a) false-positive rate of NAO when the noise level at test time is $\eta_0(1+r)$ after calibration at $\eta_0 = 0.005$, without recalibration (solid) and with a one-run sentinel at $\alpha_s = 10^{-3}$ followed by recalibration (dotted); (b) fraction of calibration windows in which the homogeneity test detects drift that occurs within the eight calibration runs. $25$ windows per point. Panel (a) uses executions separate from those of Table~\ref{tab:rq5-drift}, so its values differ from the table by sampling variation.}
\label{fig:drift}
\end{figure}

\paragraph{Drift after calibration.}
Figure~\ref{fig:drift}a reports the false-positive rate when the noise level at test time is $\eta_0(1 + r)$ after calibration at $\eta_0 = 0.005$. A $10\%$ increase is tolerated (FPR $\le 0.016$); a $25\%$ increase raises the rate of NAO to $0.27$--$0.66$ on GHZ-3, QAOA-3 and Bell-8, and a $50\%$ increase to $0.98$--$1.00$. This is expected of an oracle calibrated to a nominal level: after drift, a correct program no longer has the reference's output distribution, and the oracle detects the difference. The SPC envelope is insensitive to drift on the small benchmarks only because its floor is far above their divergences, and QAOA-8 is least affected because its threshold is large relative to the change.

\paragraph{Recalibration policies.}
Table~\ref{tab:rq5-drift} compares recalibration policies on paired executions. A one-run sentinel at $\alpha_s = 10^{-3}$ restores soundness after a $50\%$ drift but not after a $25\%$ drift (FPR $0.17$--$0.26$), because one run at a strict level has limited power. Two sentinel runs pooled and tested at $\alpha_s = 10^{-2}$ detect every $50\%$ drift and, on GHZ-3, QAOA-3 and Bell-8, every $25\%$ drift, and keep the false-positive rate of the subsequent tests at or below $0.008$ for every benchmark and drift level. The price is unnecessary recalibration: in at most $4\%$ of drift-free windows, and in up to $24\%$ under a $10\%$ drift, which does not yet harm the tests. Every policy with $m \ge 2$ keeps the rate at or below $0.04$. Co-scheduled calibration is sound by construction and costs $K$ executions per session.

\begin{table}[t]
\centering
\caption{RQ5: false-positive rate of NAO after the noise level changes from $\eta_0 = 0.005$ to $\eta_0(1 + r)$ between calibration and testing, under recalibration policies evaluated on the same executions ($25$ windows of $5$ test runs per row). Sentinel $(m, \alpha_s)$: $m$ runs of $P$ pooled and tested at level $\alpha_s$ before testing; recalibration rate in parentheses. Co-scheduled: calibration in the same session as the tests. Lower FPR is better.}
\label{tab:rq5-drift}
\small
\begin{tabular}{lrrrrr}
\toprule
\textbf{Benchmark} & \textbf{$r$} & \textbf{None} & \textbf{Sentinel $(1, 10^{-3})$} & \textbf{Sentinel $(2, 10^{-2})$} & \textbf{Co-scheduled} \\
\midrule
GHZ-3  & 0.00 & 0.000 & 0.000 (0.00) & 0.000 (0.04) & 0.000 \\
       & 0.10 & 0.000 & 0.000 (0.00) & 0.000 (0.16) & 0.000 \\
       & 0.25 & 0.464 & 0.264 (0.44) & 0.000 (1.00) & 0.000 \\
       & 0.50 & 0.992 & 0.008 (1.00) & 0.008 (1.00) & 0.008 \\
QAOA-3 & 0.00 & 0.000 & 0.000 (0.00) & 0.000 (0.00) & 0.000 \\
       & 0.10 & 0.024 & 0.008 (0.04) & 0.008 (0.24) & 0.000 \\
       & 0.25 & 0.624 & 0.168 (0.76) & 0.000 (1.00) & 0.000 \\
       & 0.50 & 1.000 & 0.000 (1.00) & 0.000 (1.00) & 0.000 \\
Bell-8 & 0.00 & 0.008 & 0.008 (0.00) & 0.008 (0.00) & 0.000 \\
       & 0.10 & 0.000 & 0.000 (0.00) & 0.000 (0.04) & 0.000 \\
       & 0.25 & 0.320 & 0.176 (0.32) & 0.000 (1.00) & 0.000 \\
       & 0.50 & 1.000 & 0.000 (1.00) & 0.000 (1.00) & 0.000 \\
QAOA-8 & 0.00 & 0.000 & 0.000 (0.00) & 0.000 (0.00) & 0.000 \\
       & 0.10 & 0.000 & 0.000 (0.00) & 0.000 (0.00) & 0.000 \\
       & 0.25 & 0.000 & 0.000 (0.00) & 0.000 (0.04) & 0.000 \\
       & 0.50 & 0.128 & 0.112 (0.12) & 0.000 (1.00) & 0.000 \\
\bottomrule
\end{tabular}
\end{table}

\paragraph{Drift during calibration.}
When the noise drifts linearly within the eight calibration runs (Figure~\ref{fig:drift}b), the homogeneity test detects $r = 0.5$ in every window on GHZ-3, QAOA-3 and Bell-8 and $r = 0.25$ in $32$--$56\%$ of windows; on QAOA-8, it detects $r = 1.0$ in $84\%$. At $r = 0$, it flags $0$--$4\%$ of windows, consistent with $\alpha_{\mathrm{drift}} = 0.01$.

\paragraph{Cost.}
Calibration is a fixed cost per (program, backend) pair (Table~\ref{tab:rq5-cost}): NAO needs $K = 8$ executions, NAO-1 one, the SPC envelope $13$ and the QOIN-style filter $K_c = 5$ plus a noiseless simulation. A mutation campaign of $100$ mutants costs $108$ executions with NAO, $8\%$ more than an (unsound) fixed threshold, and $112$ with a two-run sentinel every $50$ tests, plus eight executions per recalibration. On \texttt{ibm\_kingston}, the $128$ executions of the first HM job used $140$ seconds of QPU time, about $1.1$ seconds per $4096$-shot execution including overhead (H1: $1.09$ seconds), so calibration costs about $9$ seconds per (program, backend) pair, a sentinel about $2$ seconds, and the $100$-mutant campaign about two minutes. Attribution adds four executions per rejection to be explained.

\paragraph{IBM Quantum hardware.}
Campaign H2 executed eight calibration sessions of GHZ-3 and QAOA-3 on \texttt{ibm\_kingston} over $26.3$ hours. Within a session, the runs were homogeneous: the homogeneity test flagged none of the $16$ calibrations. Between sessions, they were not. Judging each later session against each earlier calibration ($28$ ordered pairs per program), a stale calibration of GHZ-3 rejected $34\%$ of the correct runs less than six hours later, $57\%$ at twelve to twenty-four hours and all of them after a day, and the divergence between sessions reached $3.5$ times the NAO threshold. QAOA-3, executed on other qubits, moved by at most $0.54$ times its threshold, and its stale calibrations rejected $4\%$ of the correct runs. To evaluate the sentinel, the unit of analysis is an ordered pair of sessions ($56$ pairs): the first two runs of the later session form the sentinel, tested at $\alpha_s = 10^{-2}$ against the earlier calibration, and its remaining six runs are the protected test runs ($336$). In $27$ pairs the stale calibration rejected at least one protected run, and the sentinel fired in $22$ of them; in the other $29$, it fired in $9$, causing unnecessary recalibration. Without the sentinel, $92$ of the $336$ protected runs ($27.4\%$) are rejected; with it, and recalibration in the same session when it fires, $6$ ($1.8\%$). Pairs that share a session are not independent, so these counts describe this campaign rather than estimate rates.

In HM, the second job ($24.6$ hours after the first) was judged against the first job's calibrations. GHZ-3 had drifted (divergence $0.0019$ against a threshold of $0.0013$), and its stale calibration rejected all three correct runs ($3$ of $15$ overall), while the other benchmarks moved by $0.0004$ or less. The two-run sentinel fired on GHZ-3 and on QFT-3 and VQE-3, which had barely moved. Since the allowance did not permit full recalibration, the two sentinel runs served as the new reference, and one correct run was rejected (VQE-3). The fourth job ($48.1$ hours) applied the policy as specified, with an eight-run recalibration: the stale calibration again rejected all three runs of GHZ-3, the sentinel fired on GHZ-3 and on three benchmarks that had moved less than their thresholds, and with the recalibrated references no correct run was rejected ($0$ of $15$). All $9$ faults were detected under every policy in both jobs. Drift between days is therefore real on this device, invalidated one calibration in five on both days, and is removed by the sentinel policy at the cost of unnecessary recalibrations, the trade-off of the lenient level $\alpha_s = 10^{-2}$ measured in simulation.

\begin{tcolorbox}[colback=gray!8, colframe=gray!50, title=Answer to RQ5]
Without recalibration, NAO tolerates a $10\%$ change of the noise level between calibration and testing but not a $25\%$ change, because it is sensitive enough to detect the changed output distribution. A two-run sentinel at $\alpha_s = 10^{-2}$ before each batch detects such drift and keeps the false-positive rate at or below $0.008$ under every evaluated drift, at the cost of recalibrating in up to $4\%$ of drift-free batches; co-scheduled calibration is sound by construction. The homogeneity test detects strong drift within the calibration window. Calibration adds $K = 8$ executions per program and backend, $8\%$ of a $100$-mutant campaign. On an IBM device, drift is real and program-dependent: calibrations of GHZ-3 went stale within hours, whereas QAOA-3 was barely affected. The sentinel caught $22$ of $27$ stale calibrations that produced false positives, reducing them from $27.4\%$ to $1.8\%$ of the protected runs, and in HM the full policy removed all false positives, at the cost of some unnecessary recalibrations.
\end{tcolorbox}

\section{Discussion}
\label{sec:discussion}

\subsection{Self-Baseline Calibration as an Oracle Reformulation}
\label{subsec:reformulation}

Conventional NISQ correctness oracles retain the noiseless ideal distribution as their reference and pursue robustness by acting on the \emph{comparison}: a fixed-threshold oracle widens the tolerance, and a noise-filtering oracle in the style of QOIN estimates and removes the hardware noise before comparing with the ideal. Both continue to ask how far a run lies from an output the device cannot produce. Self-baseline calibration changes the \emph{reference}: the oracle asks whether a run lies within the variability that the reference program itself exhibits under repeated execution on the same backend.

The evaluation separates the design decisions this involves, and the ablation (Table~\ref{tab:rq4-ablation}) measures each on identical executions. The choice of reference removes most false positives: every oracle that compares against the ideal output produces false positives that grow with the noise or concentrate on particular programs, whereas the oracles that compare against the self-baseline are sound in the evaluated settings. The reference alone is not sufficient, since a fixed threshold against one self-baseline run still rejects a fifth of the correct runs on synthetic noise. The choice of threshold decides whether the false-positive rate is known in advance and how sensitive the oracle is: the SPC envelope shares the reference but is set by its floor, which costs recall and mutation score, whereas the null-quantile threshold turns the false-positive rate into a nominal level and makes recall predictable from the power model. Pooling adds sensitivity and stabilizes the threshold across windows, and the drift test is inactive on homogeneous calibration runs but indispensable under recompilation. Table~\ref{tab:oracle-paradigms} summarizes the three paradigms.

\begin{table*}[t]
\centering
\caption{Three oracle paradigms for NISQ correctness testing. The noise-filtering column is a reproducible, ML-free stand-in for QOIN's paradigm; QOIN itself learns the filter through transfer learning~\cite{muqeet2024qoin}. FPR: false-positive rate over $9{,}000$ correct runs (RQ1); recall and F1 on non-equivalent injected faults (RQ4).}
\label{tab:oracle-paradigms}
\small
\begin{tabular}{@{}P{2.4cm}P{3.3cm}P{3.5cm}P{4.6cm}@{}}
\toprule
 & \textbf{Fixed or global threshold} & \textbf{Noise filtering (QOIN-style)} & \textbf{Self-baseline calibration (NAO)} \\
\midrule
Reference & noiseless ideal output & ideal output, after noise removal & pooled executions of the reference program on the same backend \\
\addlinespace
Requires & classical simulation of the ideal output; a noise model for global calibration & $K_c$ correct runs and the ideal output & $K$ runs of the reference program; no ideal output, noise model or learned model \\
\addlinespace
Threshold & chosen by the tester, or one per noise model for all programs & fixed on the corrected output & $(1-\alpha)$ quantile of the finite-shot null distribution \\
\addlinespace
FPR known in advance & no & no & yes ($\alpha$) \\
\addlinespace
Observed & FPR $0.36$--$0.74$, F1 $0.830$--$0.861$ (fixed); FPR $0.108$, F1 $0.725$ (global) & FPR $0.089$ & FPR $0.0011$ at $\alpha = 10^{-3}$; recall $0.888$; F1 $0.941$ \\
\bottomrule
\end{tabular}
\end{table*}

\subsection{What the Oracle Decides, and What It Cannot}
\label{subsec:what-decides}

The oracle decides whether $P'$ behaves like $P$ on the backend; it does not decide whether $P$ is correct. The expected output is not needed, because the reference's own noisy behavior on the same backend is the specification. If the reference is faulty, the oracle accepts versions that share its fault and rejects versions that fix it, so a faulty reference is detected only by tests of the reference itself, such as property-based or specification-based tests on small instances~\cite{pontolillo2025ideal,barrera2021state}, or by comparison with the ideal output where it can be computed. Within the differential setting, the oracle distinguishes noise from a fault only in a statistical sense. A rejection can still be caused by noise, because the noise changed (RQ5) or because the program under test interacts differently with it (RQ2); these are the cases for which the sentinel policy and the attribution procedure are intended.

\subsection{Sensitivity and Operating Regimes}
\label{subsec:tradeoff}
\label{subsec:applicability}

An oracle that accepts every run has no false positives and detects nothing, so soundness alone is not evidence of usefulness. The trade-off of NAO is explicit: lowering $\alpha$ raises the threshold, and Equation~\eqref{eq:power} gives the recall that is lost. Its misses are not caused by a conservative threshold: at the two highest noise levels, noise has removed most of the faults' effect from the output distribution, and the fixed thresholds that still detect these faults do so by rejecting most correct runs as well. Depth reduces sensitivity as width does, because noise accumulates over more gates, as the deeper QASMBench programs show. The applicability of full-distribution comparison is governed by the ratio between the shot budget and the outcomes a program actually occupies rather than by its qubit count: concentrated outputs remain testable at $16$ qubits, whereas outputs spread over all outcomes leave the evaluated regime near one shot per outcome, where marginal comparison remains an option. The evaluated range, up to $16$ qubits and two circuit families in the scaling study, is too narrow to establish this as a general law.

\subsection{Beyond Quantum}
\label{subsec:beyond-quantum}

The principle underlying the oracle, to judge correctness against a baseline that the system establishes about itself rather than against a fixed external tolerance, is not specific to quantum execution. Flaky tests in continuous integration~\cite{luo2014empirical,parry2021survey}, machine-learning inference affected by numerical nondeterminism~\cite{zhang2020machine} and performance-regression testing all judge a noisy signal against a tolerance fixed in advance. The three ingredients developed here, a repeated-execution baseline, the null distribution of the comparison statistic under ``same behavior'', and a homogeneity check of the baseline, have natural analogues wherever the sampling distribution of the correctness signal can be modeled or bootstrapped. We state this as a structural observation, not a validated result: we have not evaluated the mechanism outside the quantum setting.

\subsection{Implications for Practice}
\label{subsec:guidance}

\begin{enumerate}
\item \textbf{Use the noise-aware oracle for regression and mutation testing against a trusted reference.} Choose $\alpha$ from the number of tests and the acceptable number of false alarms: at $\alpha = 10^{-3}$, a campaign of $1000$ tests of correct programs is expected to raise one false alarm.
\item \textbf{Use NAO by default; use NAO-1 only on a stable, compiled-once backend.} NAO-1 saves seven executions per program and backend, but it cannot detect drift or recompilation and is less sensitive.
\item \textbf{Compile once.} Transpile the reference and every program under test with a fixed seed onto one initial layout; recompilation between runs changes the noise the program experiences.
\item \textbf{Plan the shot budget with the power model.} When the size of the faults to detect is not known, fix the smallest divergence whose detection matters, for example the ideal divergence of the smallest change considered a regression, computed by statevector simulation on a reduced instance, and solve Equation~\eqref{eq:power} for $N$ at $K = 8$ and the desired power. Where no such divergence can be named, report the minimum detectable divergence of the available budget as the campaign's sensitivity, so that a passing test is read as ``no change larger than this''. After the first calibration run, check that it has about $16$ or more shots per observed outcome; otherwise add shots or compare marginals.
\item \textbf{Check the baseline before each batch.} The need to recalibrate cannot be predicted from the elapsed time alone. Before each batch of tests, execute the reference twice and test the pooled runs at $\alpha_s = 10^{-2}$; recalibrate if the test rejects. Where the platform allows it, submit the calibration runs in the same job as the tests.
\item \textbf{Label mutants before running them.} Exact statevector labels are cheap at the widths at which mutation analysis is usually performed, and they make the mutation score interpretable.
\item \textbf{Attribute rejections of semantics-preserving changes, at low and medium noise.} When a refactoring that should preserve semantics is rejected, ZNE attribution (four executions) indicates whether the rejection is a noise effect or a fault; at high noise, treat a \textsc{Bug} verdict with caution, check that the folded circuits keep their structure, or use the quadratic variant.
\item \textbf{Develop on device snapshots.} Under fixed compilation, the snapshot results agree with the synthetic-noise results, so test suites can be developed on free, reproducible fake backends before hardware time is committed.
\end{enumerate}

\subsection{Implications for Research}
\label{subsec:research-implications}

\paragraph{State the false-positive rate of an oracle.}
A test oracle for stochastic programs should state the false-positive rate it is designed for and show that it attains it. Reporting only that an oracle produced no false positives does not distinguish a calibrated oracle from an insensitive one; reporting recall, precision and F1 alongside does.

\paragraph{Label mutants exactly and place them before measurement.}
Mutation scores of quantum programs are interpretable only if equivalent mutants are excluded, which exact labels make possible at moderate widths.

\paragraph{Validate taxonomies conditionally on shared calibrations.}
When mutants share a calibration, tests that treat them as independent overstate significance; stratified and cell-level tests give the honest strength of a taxonomy's support. Classes defined by the mechanism of a change should also be validated against a quantitative account of that mechanism, such as the dose-response model used here, and, where the mechanism depends on compilation, membership should be checked on the compiled circuits of the target.

\section{Threats to Validity}
\label{sec:threats}

We organize threats following the classification of Wohlin et al.~\cite{wohlin2012experimentation}: internal, construct, external and conclusion validity.

\subsection{Internal Validity}
\label{subsec:internal-threats}

\paragraph{Stochastic execution.}
Finite shot counts introduce sampling variance in every reported quantity. All experiments are driven by recorded seeds (mutant generation, simulator sampling, transpilation and bootstrap), so rerunning the experiments with these seeds reproduces every reported number exactly; the main rates are reported with confidence intervals, and the package reports intervals for all of them.

\paragraph{Parameter selection.}
The parameters of NAO were chosen from the null and power models and a factorial study on separate seeds (Section~\ref{subsec:params}) that used five of the nine core benchmarks at one noise level, so it is not independent of them; the $32$ QASMBench programs, which played no role in any design decision, reproduce the results with the same parameters. Moreover, the false-positive rate is set by $\alpha$ through the null model rather than by tuning, and every grid configuration with $4096$ or more shots has recall $1.000$, so the chosen configuration is not a selected optimum.

\paragraph{Shared calibrations.}
Tests in the same calibration window share a reference and a threshold. RQ1 and RQ4 use $40$ and $10$ independent windows per cell, and the false-positive rates of RQ1 carry cluster-bootstrap intervals over windows; RQ2 conditions on the window with a stratified test and additionally treats cells as units (Section~\ref{subsec:windows}).

\paragraph{Compilation.}
The transpiler can cancel inserted gates, change the layout between runs, or route programs under test differently from the reference. Inserted gate pairs are separated by barriers, programs under test are placed on the layout of the reference, and the effect of recompilation is measured explicitly (RQ1, random policy).

\paragraph{Implementation.}
The implementation includes $23$ unit tests covering the threshold computation, the equivalence labels, the mutation operators, the folding and extrapolation used for attribution and the statistical procedures. The bootstrap threshold is checked against the closed-form chi-square threshold in every cell of RQ1.

\subsection{Construct Validity}
\label{subsec:construct-threats}

\paragraph{Computational-basis observability.}
The oracle compares measured output distributions, so a change that leaves the measured distribution unchanged on the tested input cannot be detected, whatever the distance or threshold (Section~\ref{subsec:framework-overview}). Twelve of the injected faults change the unitary but not the output on $|0\cdots0\rangle$; they are reported separately and excluded from recall. Measurement in additional bases~\cite{oldfield2025faster} would make such faults observable and is compatible with the threshold construction, which only requires a multinomial count vector.

\paragraph{Choice of distance.}
We use the Jensen--Shannon divergence~\cite{lin1991divergence}, which is symmetric, bounded and well defined with zero counts. The null-quantile construction applies to any statistic whose null distribution can be bootstrapped from the reference counts; we do not claim that the Jensen--Shannon divergence is the most powerful choice.

\paragraph{Correctness of references.}
The differential setting assumes that the reference is correct. For injected faults and mutants this holds by construction. For the real bugs, the fixed program from Bugs4Q is the reference; a fixed program can itself be imperfect (the fixed phase-estimation program of Stack Exchange 15, for example, does not produce a sharp phase peak), but the oracle's verdict concerns only whether the buggy and fixed programs behave alike.

\paragraph{Equivalence labels.}
Equivalence is defined with respect to the measured output distribution on the tested inputs. A mutant that is equivalent on these inputs may be non-equivalent on others; the four-input census shows that this matters for CNOT insertions and rarely for other operators. The labels require statevector simulation and are therefore available only at widths that can be simulated.

\paragraph{Attribution labels and noise scaling.}
The ground truth of attribution is the exact ideal output on the tested input, so a change of the unitary invisible on that input counts as \textsc{Noise}; such changes are kept out of the attribution cases. The procedure assumes that folding realizes the nominal scale factors and that extrapolation removes the difference noise causes. On \texttt{ibm\_kingston}, the folded circuits had the nominal gate counts for all but one archived pair, which was misattributed, but gate counts do not capture idle errors, crosstalk or coherent errors. Saturation at high noise violates the second assumption, and other noise structures, such as coherent errors, which folding amplifies differently, may violate it as well.

\paragraph{Mutants and faults as proxies for real faults.}
Mutation scores and detection rates on injected faults are proxies for the detection of real faults. In classical software, mutant detection correlates with real fault detection, but not perfectly~\cite{papadakis2018mutation}; whether the same holds for quantum programs is not established, and we do not claim that the reported detection rates transfer unchanged to real defects. The ten real bugs from Bugs4Q are consistent with the injected-fault results (Section~\ref{subsec:rq4-results}), but they are too few to estimate the correlation between mutation scores and real-fault detection.

\paragraph{Taxonomy operationalization.}
The classes are operationalized by the operators of Table~\ref{tab:mutation-operators}. On the core benchmarks, the separation between the semantics-preserving classes is driven by \texttt{cx\_pair}, since a single pair of single-qubit gates adds too little error to be detected; operators that add more physical operations may separate more strongly. Operators are proxies for the physical criterion, which depends on the target (Table~\ref{tab:rq2-compiled}); the contrasts hold with classes verified on the synthetic backend, but a study on another target should verify membership there. Every engine used the same seed, so operator draws are correlated across programs, and only $Y$, $Z$, $S$ and $T$ single-qubit pairs were drawn; $H$ and $X$ pairs were not evaluated. The dose study uses depolarizing noise, under which uniform outputs are immune to added error; under non-unital noise such as amplitude damping, the programs that react to noise amplification may differ.

\paragraph{Baselines.}
The QOIN-style filter is a stand-in for the noise-filtering strategy, not a reimplementation of QOIN's transfer-learning pipeline~\cite{muqeet2024qoin}; a learned filter may achieve a lower false-positive rate. The noise-calibrated global threshold follows Fortz et al.~\cite{fortz2026robust} on our program sets at $4096$ shots; its percentile is computed over the programs it is applied to, which favors it (on held-out programs its rate is higher), and a different program set or shot count changes its threshold, though not the fact that one threshold must serve all programs. Other attribution methods, such as noise fingerprinting~\cite{bensoussan2026toward}, were not compared.

\subsection{External Validity}
\label{subsec:external-threats}

\paragraph{Benchmarks.}
The nine core benchmarks cover state preparation, an arithmetic kernel and variational algorithms with fixed parameters, up to 8 qubits; the extended suite adds $32$ QASMBench programs from $2$ to $10$ qubits across arithmetic, error correction, oracle algorithms, phase estimation, linear solvers, simulation, machine learning and variational circuits, and the scaling study extends two families to 16 qubits. Neither covers programs with mid-circuit measurement, reset or classical control, or parameterized circuits whose parameters change between versions; QASMBench programs with these features were excluded. Only ten real bugs from Bugs4Q met the inclusion criteria, so the real-bug results are evidence for the injected-fault results rather than a separate estimate of detection rates.

\paragraph{Noise.}
The main synthetic model is depolarizing with symmetric readout error; the coherent model reproduces the main results at two error strengths with a fixed error axis only. Device snapshots add realistic error rates, relaxation and coupling maps, but not drift, crosstalk or non-Markovian effects, and RQ5 models drift as a change of the error level only. Hardware drift was measured on one device for a few programs over up to two days and was strongly program-dependent (GHZ-3 against QAOA-3), so it cannot establish how often recalibration is needed in general.

\paragraph{Hardware.}
The hardware evaluation covers IBM Heron devices of one technology. The campaigns are sized to fit free IBM Quantum allowances: with $82$ correct runs and $121$ faults and semantic mutants in total, they confirm or refute the direction of the simulation results, but their rates carry wide intervals and are not estimates of the device's false-positive rate or recall. Devices of other vendors and modalities (trapped ions, neutral atoms) have different noise structures.

\paragraph{Scope of the evidence.}
Taken together, the results establish the oracle's behavior within the evaluated program families, shot budgets, noise models (depolarizing, coherent over-rotation and device snapshots), calibration procedures and IBM Heron devices. They do not establish universal false-positive control or detection performance across quantum workloads and devices; generalization to programs with mid-circuit measurement or classical control, to other hardware modalities and to noise processes with drift, crosstalk or non-Markovian memory remains an open empirical question.

\paragraph{Scope of the generality claim.}
Every experiment concerns quantum programs. The relevance of self-baseline calibration to other non-deterministic software (Section~\ref{subsec:beyond-quantum}) is a motivated hypothesis, not an evaluated claim.

\subsection{Conclusion Validity}
\label{subsec:conclusion-threats}

\paragraph{Statistical power.}
The soundness result rests on $9{,}000$ correct runs in $1{,}800$ independent calibration windows, enough to distinguish an FPR of $10^{-3}$ from $2 \times 10^{-3}$ only roughly; the cluster-bootstrap interval at $\alpha = 10^{-3}$, $[0.0004, 0.0018]$, quantifies this. On the core benchmarks, the per-noise-level cell-level taxonomy tests have nine cells each and therefore limited power; the extended suite, with $32$ cells per level, resolves this for all class contrasts.

\paragraph{Approximate level.}
The level $\alpha$ is met only approximately (Section~\ref{subsubsec:level}). The boundary of about $16$ shots per observed outcome held for the three evaluated shapes, but the size of the violation below it depends strongly on the shape, other shapes may move it, no window fell between $8$ and $16$, and the study used one shot budget ($N = 4096$, $K = 8$) and one bootstrap size ($B = 2000$). At $B = 4000$, the rate of an individual window lies between about $0.17$ and $2.37$ times the nominal level, and smaller $B$ varies more; every reported rate is an average over windows, and claims about a single window require a larger $B$.

\paragraph{Multiple testing.}
The taxonomy is tested on three class pairs at three noise levels and pooled; Bonferroni correction over the three pairs is applied within each level. With correction over all twelve tests, the CMH results remain significant ($p \le 4.3 \times 10^{-6}$), and the cell-level result for the semantics-preserving classes is significant only when pooled.

\paragraph{Single research team.}
All experiments were designed, executed and analyzed by the authors. To mitigate this, all results are produced by scripts from recorded seeds, every number reported in the tables is regenerated from the raw per-execution records, and the raw hardware data, including job identifiers and transpiled circuits, are archived.

\section{Conclusion}
\label{sec:conclusion}

We presented a noise-aware testing framework for NISQ quantum software and evaluated it on nine core benchmarks, 32 QASMBench programs and ten real bugs under five depolarizing and two coherent-error noise levels, three device-snapshot backends, circuits of up to 16 qubits and IBM Quantum hardware, against fixed-threshold, envelope-based, QOIN-style, noise-model-calibrated and re-execution baselines.

The results support four claims. First, the noise-aware oracle, built on self-baseline calibration with a null-quantile threshold, approximately attains its nominal false-positive rate in the evaluated settings in which calibration and test runs are samples of one distribution and one run of the reference provides about $16$ or more shots per observed outcome; this criterion is empirical, not a universal guarantee. Under these conditions it controls false positives where fixed thresholds against the ideal output are unsound, without a loss of sensitivity: it has the highest F1 of all evaluated oracles on the core and QASMBench programs, detects most runs of observable real bugs, and its power model predicts its recall. Second, equivalent mutants are frequent among quantum mutation operators, and exact statevector labels make mutation scores interpretable. Third, the three-class taxonomy is distinguishable under the oracle, strongly between semantic and semantics-preserving mutants and more weakly between noise-amplifying and noise-neutral ones, a difference the power model explains and that persists on compiled circuits. Fourth, zero-noise extrapolation attributes rejections to faults or noise without a noiseless simulator, reliably at low and medium noise. An ablation measures the contribution of each design decision, and the hardware campaigns on an IBM Heron device confirm soundness, detection and the need for a drift sentinel.

The evaluation also shows where the approach stops working: drift between calibration and testing causes false positives unless it is detected before testing, full-distribution comparison is not reliably calibrated with too few shots per observed outcome, and attribution of noise effects fails at high noise.

The contribution beyond the specific framework is a change in how correctness oracles are built for software whose behavior is visible only through noise: instead of selecting a tolerance, derive the threshold from the system's own observed variability and state the false-positive rate it implies. We conjecture, but do not demonstrate, that the same principle is useful for other non-deterministic software.

\subsection{Future Work}
\label{subsec:future}

\begin{enumerate}
\item \textbf{Hardware at scale and across vendors.} Extend the hardware campaigns to more devices, larger programs and other qubit modalities (trapped ions, neutral atoms), and characterize how often recalibration is needed in practice.
\item \textbf{Continuous drift monitoring.} Complement the sentinel by continuous monitoring, for example by sequential tests on reference executions interleaved with the tests or by live noise fingerprinting of the device~\cite{bensoussan2026toward}, to reduce unnecessary recalibration.
\item \textbf{Sample-efficient representations.} Generalize the null-quantile threshold from two-qubit marginals to observables and classical-shadow estimators, whose estimation cost does not grow with the output support, to extend sound testing beyond the shots-per-outcome limit.
\item \textbf{Phase-sensitive oracles.} Combine the threshold construction with measurement in multiple bases~\cite{oldfield2025faster} to detect faults that do not change the computational-basis distribution.
\item \textbf{Mutation scores and real faults.} Combine exact labels with the wider program set and distance metrics of robust mutation analysis under noise~\cite{fortz2026robust}, and study whether quantum mutation scores correlate with the detection of real quantum faults.
\item \textbf{Direct comparison with QOIN.} Reproduce QOIN's learned noise filter on the same benchmarks for a like-for-like comparison.
\end{enumerate}

\section*{Declarations}
\paragraph{Competing interests.} The authors declare no competing interests.
\paragraph{Funding.} This research received no specific grant from any funding agency in the public, commercial or not-for-profit sectors.

\begin{acks}
We acknowledge the use of IBM Quantum services for this work. The views expressed are those of the authors and do not reflect the official policy or position of IBM or the IBM Quantum team.
\end{acks}

\bibliographystyle{ACM-Reference-Format-unsrt}
\bibliography{references}

\appendix
\section{Derivation of the Null Distribution}
\label{app:derivation}

Let $\hat p$ and $\hat q$ be the empirical distributions of $n_1$ and $n_2$ independent multinomial draws from $\pi$ over $d$ outcomes with $\pi_x > 0$, and $m = (\hat p + \hat q)/2$. With natural logarithms, a second-order Taylor expansion of $f(t) = t \ln t$ around $m_x$ gives
\begin{equation*}
\mathrm{JS}_e(\hat p, \hat q) = \frac{1}{2}\sum_x \left[ \hat p_x \ln \frac{\hat p_x}{m_x} + \hat q_x \ln \frac{\hat q_x}{m_x} \right] \approx \frac{1}{8} \sum_x \frac{(\hat p_x - \hat q_x)^2}{m_x} \approx \frac{1}{8} \sum_x \frac{(\hat p_x - \hat q_x)^2}{\pi_x},
\end{equation*}
since $m \to \pi$. Under $H_0$, $\sqrt{n_1 n_2/(n_1 + n_2)}\,(\hat p - \hat q)$ converges to a Gaussian vector with covariance $\mathrm{diag}(\pi) - \pi\pi^\top$, so the Pearson-type sum $\frac{n_1 n_2}{n_1 + n_2}\sum_x (\hat p_x - \hat q_x)^2/\pi_x$ converges to $\chi^2_{d-1}$~\cite{agresti2013categorical,read1988goodness}. With $D = \mathrm{JS}_e/\ln 2$ (base 2) and $n_1 n_2/(n_1 + n_2) = 1/(1/n_1 + 1/n_2)$, this yields Equation~\eqref{eq:null}. Under a fixed alternative with true divergence $\delta$ (base 2), the same expansion gives a noncentral $\chi^2_{d-1}(\lambda)$ with $\lambda = 8 \ln 2\, \delta / (1/n_1 + 1/n_2)$, which yields Equation~\eqref{eq:power}. The approximation requires that most outcomes have several expected counts; RQ1 (sparse supports) and RQ4 (scaling) show where it breaks down.

\section{Supplementary Results}
\label{app:tables}

The main text summarizes the following tables and figure.

\begin{table}[htbp]
\centering
\caption{RQ5: executions per test campaign of $M$ programs under test, for one (program, backend) pair. Sentinel $(m, \alpha_s) = (2, 10^{-2})$: before every $50$ tests, $m = 2$ runs of the reference program are pooled and tested against the reference at level $\alpha_s = 10^{-2}$. Lower is better.}
\label{tab:rq5-cost}
\small
\begin{tabular}{lrrr}
\toprule
\textbf{Oracle} & \textbf{$M = 10$} & \textbf{$M = 100$} & \textbf{$M = 1000$} \\
\midrule
Fixed threshold (unsound)              & 10 & 100 & 1000 \\
NAO-1                                  & 11 & 101 & 1001 \\
QOIN-style filter ($K_c = 5$)          & 16 & 106 & 1006 \\
NAO ($K = 8$)                          & 18 & 108 & 1008 \\
NAO ($K = 8$), sentinel $(2, 10^{-2})$ every 50 tests\textsuperscript{a} & 20 & 112 & 1048 \\
SPC envelope ($K = 12$)                & 23 & 113 & 1013 \\
\bottomrule
\multicolumn{4}{l}{\footnotesize \textsuperscript{a}Plus $K = 8$ executions per recalibration.}
\end{tabular}
\end{table}

\begin{table}[htbp]
\centering
\caption{Parameter study of NAO (Section~\ref{subsec:params}): false-positive rate (160 correct runs per configuration) and recall on non-equivalent faults (232 fault tests per configuration), five benchmarks at medium noise, eight windows. $K = 1$ is NAO-1.}
\label{tab:app-grid}
\small
\begin{tabular}{llrrrrrr}
\toprule
& & \multicolumn{2}{c}{$N = 1024$} & \multicolumn{2}{c}{$N = 4096$} & \multicolumn{2}{c}{$N = 16384$} \\
\cmidrule(lr){3-4}\cmidrule(lr){5-6}\cmidrule(lr){7-8}
$\alpha$ & $K$ & FPR & Recall & FPR & Recall & FPR & Recall \\
\midrule
$10^{-2}$ & 1 & 0.0125 & 0.970 & 0.0000 & 1.000 & 0.0125 & 1.000 \\
$10^{-2}$ & 4 & 0.0187 & 0.996 & 0.0125 & 1.000 & 0.0125 & 1.000 \\
$10^{-2}$ & 8 & 0.0063 & 0.991 & 0.0063 & 1.000 & 0.0063 & 1.000 \\
$10^{-2}$ & 16 & 0.0125 & 0.991 & 0.0000 & 1.000 & 0.0187 & 1.000 \\
$10^{-3}$ & 1 & 0.0000 & 0.940 & 0.0063 & 1.000 & 0.0000 & 1.000 \\
$10^{-3}$ & 4 & 0.0000 & 0.966 & 0.0000 & 1.000 & 0.0000 & 1.000 \\
$10^{-3}$ & 8 & 0.0000 & 0.978 & 0.0000 & 1.000 & 0.0000 & 1.000 \\
$10^{-3}$ & 16 & 0.0063 & 0.983 & 0.0000 & 1.000 & 0.0000 & 1.000 \\
$10^{-4}$ & 1 & 0.0000 & 0.948 & 0.0000 & 1.000 & 0.0000 & 1.000 \\
$10^{-4}$ & 4 & 0.0000 & 0.978 & 0.0000 & 1.000 & 0.0000 & 1.000 \\
$10^{-4}$ & 8 & 0.0000 & 0.974 & 0.0063 & 1.000 & 0.0000 & 1.000 \\
$10^{-4}$ & 16 & 0.0000 & 0.961 & 0.0000 & 1.000 & 0.0000 & 1.000 \\
\bottomrule
\end{tabular}
\end{table}

\begin{table}[htbp]
\centering
\caption{False positives per core benchmark over $1{,}000$ correct runs (five noise levels, $200$ calibration windows; RQ1), with 95\% Wilson intervals. For NAO and NAO-1 the target is $\alpha$.}
\label{tab:app-fpr-program}
\footnotesize
\setlength{\tabcolsep}{4pt}
\begin{tabular}{lrrrr}
\toprule
\textbf{Benchmark} & \textbf{NAO, $\alpha = 10^{-3}$} & \textbf{NAO-1, $\alpha = 10^{-3}$} & \textbf{NAO, $\alpha = 10^{-2}$} & \textbf{NAO-1, $\alpha = 10^{-2}$} \\
\midrule
Bell    & 0 [0.0000, 0.0038] & 2 [0.0005, 0.0073] & 13 [0.0076, 0.0221] & 12 [0.0069, 0.0209] \\
GHZ-3   & 2 [0.0005, 0.0073] & 1 [0.0002, 0.0056] & 12 [0.0069, 0.0209] & 8 [0.0041, 0.0157] \\
QFT-3   & 0 [0.0000, 0.0038] & 3 [0.0010, 0.0088] & 10 [0.0054, 0.0183] & 11 [0.0062, 0.0196] \\
QAOA-3  & 1 [0.0002, 0.0056] & 0 [0.0000, 0.0038] & 9 [0.0047, 0.0170] & 14 [0.0084, 0.0234] \\
VQE-3   & 1 [0.0002, 0.0056] & 1 [0.0002, 0.0056] & 12 [0.0069, 0.0209] & 9 [0.0047, 0.0170] \\
Bell-8  & 0 [0.0000, 0.0038] & 1 [0.0002, 0.0056] & 5 [0.0021, 0.0117] & 20 [0.0130, 0.0307] \\
GHZ-8   & 2 [0.0005, 0.0073] & 3 [0.0010, 0.0088] & 11 [0.0062, 0.0196] & 27 [0.0186, 0.0390] \\
QAOA-8  & 3 [0.0010, 0.0088] & 0 [0.0000, 0.0038] & 10 [0.0054, 0.0183] & 18 [0.0114, 0.0283] \\
HWE-6   & 1 [0.0002, 0.0056] & 0 [0.0000, 0.0038] & 5 [0.0021, 0.0117] & 7 [0.0034, 0.0144] \\
\bottomrule
\end{tabular}
\end{table}

\begin{table}[htbp]
\centering
\caption{False-positive rate on correct programs by noise level ($1{,}800$ runs per level; RQ1). For NAO and NAO-1 the target is $\alpha$; for the baselines, lower is better.}
\label{tab:app-fpr-noise}
\small
\begin{tabular}{lrrrrr}
\toprule
\textbf{Oracle} & \textbf{Very low} & \textbf{Low} & \textbf{Medium} & \textbf{High} & \textbf{Very high} \\
\midrule
NAO ($\alpha = 10^{-3}$)   & 0.0011 & 0.0006 & 0.0022 & 0.0006 & 0.0011 \\
NAO ($\alpha = 10^{-2}$)   & 0.0100 & 0.0083 & 0.0100 & 0.0122 & 0.0078 \\
NAO-1 ($\alpha = 10^{-3}$) & 0.0006 & 0.0017 & 0.0028 & 0.0011 & 0.0000 \\
NAO-1 ($\alpha = 10^{-2}$) & 0.0078 & 0.0139 & 0.0211 & 0.0167 & 0.0106 \\
SPC envelope               & 0.0000 & 0.0000 & 0.0006 & 0.0006 & 0.0006 \\
QOIN-style filter          & 0.0000 & 0.0000 & 0.0000 & 0.2222 & 0.2222 \\
Fixed $\tau = 0.005$       & 0.4828 & 0.5817 & 0.8411 & 0.8889 & 0.8889 \\
Fixed $\tau = 0.01$        & 0.3344 & 0.3800 & 0.7778 & 0.8889 & 0.8889 \\
Fixed $\tau = 0.02$        & 0.1128 & 0.3333 & 0.7711 & 0.8889 & 0.8889 \\
Fixed $\tau = 0.05$        & 0.0000 & 0.0000 & 0.3372 & 0.7800 & 0.8889 \\
Fixed $\tau = 0.1$         & 0.0000 & 0.0000 & 0.2222 & 0.6667 & 0.8889 \\
\bottomrule
\end{tabular}
\end{table}

\begin{table}[htbp]
\centering
\caption{RQ1: median acceptance thresholds at $\alpha = 10^{-3}$ for selected cells (divergence scale). $\chi^2$: closed form of Equation~\eqref{eq:chi2-threshold} for NAO-1.}
\label{tab:rq1-thresholds}
\small
\begin{tabular}{lrrrr}
\toprule
\textbf{Cell} & \textbf{NAO-1} & \textbf{$\chi^2$ form} & \textbf{NAO} & \textbf{SPC envelope} \\
\midrule
Bell / medium   & 0.0014 & 0.0014 & 0.0008 & 0.0200 \\
GHZ-3 / medium  & 0.0022 & 0.0021 & 0.0012 & 0.0200 \\
HWE-6 / medium  & 0.0092 & 0.0091 & 0.0052 & 0.0200 \\
Bell-8 / medium & 0.0191 & 0.0176 & 0.0120 & 0.0200 \\
QAOA-8 / medium & 0.0302 & 0.0291 & 0.0172 & 0.0308 \\
GHZ-8 / high    & 0.0306 & 0.0284 & 0.0182 & 0.0323 \\
\bottomrule
\end{tabular}
\end{table}

\begin{table}[htbp]
\centering
\caption{Rejection rate of mutants by operator under NAO (RQ2). For semantic operators higher is better; for semantics-preserving operators lower is better.}
\label{tab:app-operator}
\small
\begin{tabular}{llrr}
\toprule
\textbf{Class} & \textbf{Operator} & $n$ & \textbf{Rejected} \\
\midrule
Semantic         & \texttt{insert\_cnot}          & 108 & 0.935 \\
                 & \texttt{insert\_rotation}      & 159 & 0.925 \\
                 & \texttt{insert\_t}             & 42  & 0.738 \\
                 & \texttt{insert\_x}             & 96  & 0.979 \\
                 & \texttt{remove\_gate}          & 165 & 1.000 \\
                 & \texttt{remove\_1q\_gate}      & 171 & 0.965 \\
                 & \texttt{rotation\_to\_identity}& 69  & 0.826 \\
Noise-amplifying & \texttt{cx\_pair}              & 156 & 0.538 \\
                 & \texttt{inverse\_pair\_1q}     & 135 & 0.000 \\
                 & \texttt{rotation\_pair}        & 114 & 0.000 \\
Noise-neutral    & \texttt{commute\_swap}         & 117 & 0.000 \\
                 & \texttt{cx\_to\_h\_cz\_h}      & 288 & 0.007 \\
\bottomrule
\end{tabular}
\end{table}

\begin{table}[htbp]
\centering
\caption{RQ3: attribution by extrapolation method on identical executions (counts at $s = 1, 3, 5$ stored once per case). Synthetic: the $1{,}062$ cases of Table~\ref{tab:rq3-attribution}, executed again so that every method judges the same counts; NAO's trigger is re-evaluated on these executions, so $603$ cases are triggered here ($58$ noise effects) against $597$ ($54$) in Table~\ref{tab:rq3-attribution}. Snapshots: FakeManilaV2 and FakeLagosV2, $392$ cases. Higher is better.}
\label{tab:rq3-extrapolation}
\small
\begin{tabular}{llrrrr}
\toprule
\textbf{Data} & \textbf{Extrapolation} & \textbf{Accuracy} & \textbf{$\kappa$} & \textbf{Bug rec.} & \textbf{Noise rec.} \\
\midrule
Synthetic, all       & linear      & 0.927 & 0.854 & 0.914 & 0.944 \\
                     & quadratic   & 0.917 & 0.834 & 0.887 & 0.954 \\
                     & exponential & 0.912 & 0.824 & 0.912 & 0.912 \\
Synthetic, triggered & linear      & 0.932 & 0.593 & 0.967 & 0.603 \\
                     & quadratic   & 0.924 & 0.615 & 0.941 & 0.759 \\
                     & exponential & 0.915 & 0.477 & 0.961 & 0.483 \\
Snapshots, all       & linear      & 0.959 & 0.918 & 0.952 & 0.967 \\
                     & quadratic   & 0.941 & 0.883 & 0.905 & 0.984 \\
                     & exponential & 0.962 & 0.923 & 0.957 & 0.967 \\
\bottomrule
\end{tabular}
\end{table}

\begin{table}[htbp]
\centering
\caption{Recall on non-equivalent injected faults by injector (RQ4). $n$: fault tests. Higher is better.}
\label{tab:app-injector}
\small
\begin{tabular}{lrrrrr}
\toprule
\textbf{Injector} & $n$ & \textbf{NAO} & \textbf{NAO-1} & \textbf{SPC} & \textbf{Fixed $\tau{=}0.05$} \\
\midrule
\texttt{drop\_last\_gate}      & 400 & 0.79 & 0.78 & 0.69 & 0.83 \\
\texttt{extra\_x}              & 400 & 0.93 & 0.91 & 0.86 & 1.00 \\
\texttt{phase\_flip}           & 200 & 0.78 & 0.73 & 0.54 & 0.66 \\
\texttt{remove\_entanglement}  & 400 & 0.93 & 0.93 & 0.93 & 1.00 \\
\texttt{swap\_qubits}          & 300 & 0.97 & 0.97 & 0.90 & 0.97 \\
\texttt{wrong\_angle}          & 200 & 0.77 & 0.76 & 0.55 & 0.85 \\
\texttt{wrong\_basis}          & 350 & 0.92 & 0.91 & 0.86 & 0.94 \\
\texttt{wrong\_target}         & 250 & 0.96 & 0.96 & 0.91 & 1.00 \\
\bottomrule
\end{tabular}
\end{table}

\begin{table}[htbp]
\centering
\caption{RQ4: minimum detectable Jensen--Shannon divergence at $4096$ shots, $80\%$ power and $\alpha = 10^{-3}$ (Equation~\eqref{eq:power}). Lower is better.}
\label{tab:rq4-mdd}
\small
\begin{tabular}{rrr}
\toprule
\textbf{Outcomes $d$} & \textbf{NAO-1} & \textbf{NAO ($K = 8$)} \\
\midrule
4   & 0.00190 & 0.00107 \\
16  & 0.00297 & 0.00167 \\
64  & 0.00497 & 0.00280 \\
256 & 0.00891 & 0.00501 \\
\bottomrule
\end{tabular}
\end{table}

\begin{table}[htbp]
\centering
\caption{RQ4: NAO and a noise-calibrated global threshold (after Fortz et al.~\cite{fortz2026robust}) on identical executions ($2{,}500$ non-equivalent fault tests, $1{,}350$ correct runs, $800$ equivalent fault tests). This is a separate run of the design of Table~\ref{tab:rq4-faults} with its own seeds, so NAO's values differ from that table by sampling variation (here $1$ false positive, there $2$). $\uparrow$: higher is better; $\downarrow$: lower is better.}
\label{tab:rq4-global}
\small
\begin{tabular}{lrrrrr}
\toprule
\textbf{Oracle} & \textbf{Recall $\uparrow$} & \textbf{Precision $\uparrow$} & \textbf{F1 $\uparrow$} & \textbf{FPR $\downarrow$} & \textbf{Equivalent rejected $\downarrow$} \\
\midrule
NAO                              & 0.888 & 1.000 & 0.940 & 0.0007 & 0/800 \\
Noise-calibrated global threshold & 0.602 & 0.912 & 0.725 & 0.1081 & 49/800 \\
\bottomrule
\end{tabular}
\end{table}

\begin{figure}[htbp]
\centering
\includegraphics[width=0.75\linewidth]{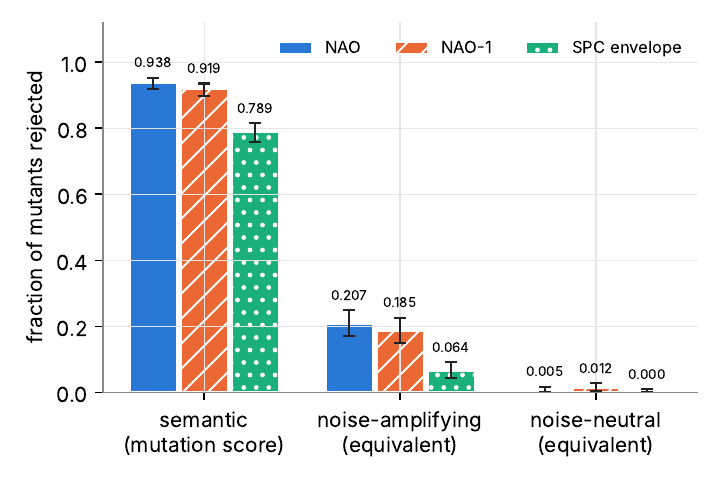}
\Description{Grouped bar chart of the fraction of mutants rejected per taxonomy class for NAO, NAO-1 and the SPC envelope. Semantic mutants are rejected in 79 to 94 percent of cases, noise-amplifying mutants in 6 to 21 percent and noise-neutral mutants in about 1 percent or fewer.}
\caption{RQ2: fraction of mutants rejected per taxonomy class and oracle ($810$ semantic mutants and $405$ of each semantics-preserving class; $27$ cells). Error bars: 95\% Wilson intervals; the value above each bar is the observed rate. For the semantic class the rate is the mutation score (higher is better); for the semantics-preserving classes it is a false-rejection rate.}
\label{fig:mutation-by-class}
\end{figure}

\end{document}